\documentclass{article}
\usepackage[utf8]{inputenc}
\usepackage{graphicx} % Required for inserting images
\usepackage{xcolor}
\usepackage{amsmath}
\usepackage{amsthm}
\usepackage{amsfonts}
\usepackage{enumerate}
\usepackage{makecell}
\usepackage{array}
\usepackage{changes}
\usepackage{url}
\usepackage{hyperref}
\usepackage{actuarialsymbol}
\newcommand{\cum}{\mathrm{cum}}
\newcommand{\CDC}{\mathrm{CDC}}

\newcommand{\DC}{\mathrm{IDC}}
\newcommand{\upper}{\mathrm{upper}}
\newcommand{\target}{\mathrm{target}}
\newcommand{\Ret}{\mathrm{Ret}}
\newcommand{\nom}{\mathrm{nom}}

\newcommand{\aitch}{\mathsf{h}}
\newcommand{\NRA}{\mathsf{NRA}}
\newcommand{\DCPI}{q}%{{q^{\mathrm{CPI}}}}
\newcommand{\DCPIt}{q_t}%{{q_t^{\mathrm{CPI}}}}
\newcommand{\CPI}{\mathsf{CPI}}
\newcommand{\constantindexrate}{q}

\newcommand{\age}{\mathsf{age}}

\newtheorem{theorem}{Theorem}
\theoremstyle{definition}

\AddToHook{cmd/added/before}{\def\Changes@AuthorColor{red}}
\AddToHook{cmd/deleted/before}{\def\Changes@AuthorColor{green}}
\AddToHook{cmd/replaced/before}{\def\Changes@AuthorColor{blue}}

\title{Intergenerational cross-subsidies in UK collective defined contribution funds}
\author{John Armstrong, James Dalby and Catherine Donnelly}

\begin{document}

\maketitle

\begin{abstract}
We evaluate the performance and level of intergenerational cross-subsidy in flat-accrual and dynamic-accrual collective
defined contribution (CDC) schemes which have been designed to be compatible with UK legislation. In the flat-accrual scheme, all members accrue the benefits at the same rate irrespective of age. This captures the most significant feature of the Royal Mail Collective Pension Plan, which is currently the only UK CDC scheme. The dynamic-accrual schemes seeks to reduce intergenerational cross-subsidies by varying the rate of benefit-accrual in accordance to the age of members and the current funding level.

We find that these CDC schemes can often be successful in smoothing pension outcomes post-retirement while outperforming a defined contribution scheme followed by annuity purchase at the point of retirement. However, this out-performance is not guaranteed in a flat-accrual scheme and there is little smoothing of projected pension outcomes before retirement. 

There are significant intergenerational cross-subsidies in the flat-accrual scheme. These qualitatively mirror the cross-subsidies seen in existing defined benefit schemes, but we find the magnitude of the cross-subsidies is much larger in flat-accrual CDC schemes. 

The dynamic-accrual scheme design is intended to reduce such cross-subsidies, but we find they still arise due to the approximate pricing methodology used to determine the benefits accrued by each contribution.  Although the cross-subsidies tend to cancel out over time, in any given year they can be large.  Thus, the benefits accrued by contributions should be calculated rigorously to reduce cross-subsidies.
\end{abstract}

\section{Introduction}

Collective defined contribution (CDC) pension schemes have arrived in the UK, after a long and fragile genesis \cite{wilkinson}. These are pension schemes which share risks, such as longevity, investment and inflation risk, collectively among the scheme members.  This is different to individual defined contribution (DC) pension schemes, where members bear only their own risk.   Unlike defined benefit (DB) pension schemes, the other main form of pension provision in the UK, benefit payments in CDC schemes are not guaranteed; indeed, they are expected to fluctuate.  It is important to understand what risks are being borne by which members in CDC schemes and to quantify them, to reduce and mitigate against unfairness between different members. 

We study two different whole-life CDC scheme designs which would comply with the current UK regulations. The collective risk-sharing is done through declaring the same increases on all members' benefits, a UK regulatory requirement for CDC schemes.  We refer to such schemes as {\em shared-indexation schemes} to highlight this common feature of their design.

The first shared-indexation scheme is a flat-accrual scheme, in which the contributions paid into the scheme are a fixed percentage of salaries.  In exchange, they receive a deferred retirement income linked to their current salary.  This scheme is closely modelled on that of the Royal Mail Collective Pension Plan (RMCPP), the first, and currently only, UK CDC scheme in existence.  The design of RMCPP has strong similarities with the defined benefit scheme it replaces.  The desire to launch the RMCPP has been the driving force behind the introduction of the legislation enabling CDC schemes in the UK \cite{wilkinson}.  Combined with the scale of the scheme, which is anticipated to have over 100,000 members, the importance and significance of the RMCPP is clear. It is the prototype for CDC plan design in the UK, for UK regulations and regulators, for actuaries, employers and workers' unions.  

The second shared-indexation scheme we consider is a dynamic-accrual CDC scheme.  In such a scheme, members can choose how much to contribute and, in return, receive a deferred retirement income that reflects financial market conditions at the time the contribution was paid.   While no such scheme exists in the UK at present, in 2024 the UK Department for Work and Pensions (DWP) held a consultation on multi-employer CDC and our dynamic-accrual scheme is intended to reflect a common understanding of UK pension professionals on how such a scheme might operate. To achieve this, we have worked with the Pensions Policy Institute, a not-for-profit UK research institute, who have conducted a series of interviews on CDC pensions with a broad range of stakeholders.  They have used these insights to publish two briefing notes on CDC schemes \cite{wilkinson,upton}. In addition, we have used the expertise of a project advisory group drawn from across the pension industry and academia.

The aim of this paper is to compare these two scheme designs
and evaluate the extent to which they succeed in design goals
such as providing an income for life, minimizing intergenerational unfairness and smoothing pension outcomes.

\medskip

Due to the novelty of UK CDC schemes, with the RMCPP launched in October 2024, there is little literature on their unique design. A 2023 report prepared for the ABI \cite{abi} summarises
numerous industry studies of similar designs \cite{governmentActuaries, aon, popat, abi, irri, wtw}, as well as providing its own additional analysis. However, there are important differences in details between the scheme designs in these studies and the RMCPP. Notable differences include the use of buffers and funding corridors.
Of these studies, only two perform stochastic modelling \cite{governmentActuaries,popat}. In the academic literature,
Owadally et al. \cite{owadally} discuss CDC plans in the UK with respect to policy, as well as their risks and the ensuing advantages and disadvantages, and model an earlier proposal stochastically. They also compare the outcomes for a single member of their CDC plan model against typical DC-based alternative pension schemes, when the scheme is mature. They find their CDC plan can pay higher and more stable pensions compared to the alternative schemes considered.

\medskip

This paper contributes to the literature in several ways. 
First, in Section \ref{sec:operationCDC}, we describe an explicit mathematical model for shared-indexation CDC schemes. This model describes the contract of
such a scheme. Once the realised contributions and investments are known, this model describes what
the resulting benefits will be. In the case of the
dynamic-accrual CDC scheme, this gives a
precise operationalisation of the broad shared understanding among pension professionals of what a dynamic-accrual CDC scheme should entail. 

In describing a flat-accrual scheme, we follow the key design features of the RMCPP, but there are differences which we now summarise.  Only a single-life pension in retirement is valued in our model -- spousal benefits, death benefits and the UK's tax-free cash lump sum payment made at the point of retirement are omitted. We include an upper bound on indexation which is not present in the RMCPP, but is required for dynamic-accrual and so aids comparison between the two scheme types. We assume that benefit cuts and bonuses that must occur when the fund is significantly over- or under-funded take effect immediately, rather than over time as is done for benefit cuts in the RMCPP. In our model, the relationship between contribution rate and level of benefits is determined by choosing a long-term target rate of indexation. In the RMCPP, these levels are fixed in the Trust Deed and Rules \cite{royalMailTrustDeeds}, but it is not specified how they are determined. Nevertheless, the member handbook does suggest that ``The aim is to provide increases to help keep up
with the cost of living'' \cite[p.\ 67]{royalMailHandbook}.

We endeavour to make a clear separation between the contract
of a CDC fund and our modelling assumptions. Section \ref{sec:operationCDC} describes the contract. Section
\ref{sec:modelling} describes modelling assumptions including demographic and economic factors.

Having established the contract of CDC funds, the second way our paper contributes to the literature is 
by studying the extent of intergenerational cross-subsidy in CDC
funds. This is the topic of Section \ref{sec:crossSubsidy}.

We find that flat-accrual schemes and dynamic-accrual schemes
contain very different levels of cross-subsidies. A dominant source of cross-subsidy in the flat-accrual scheme results from the combination of a fixed contribution rate and fixed accrual rate. This favours participants who are close to retirement as they receive the same benefit amount as younger members for the same contribution, despite there being less time for their investments to grow. This qualitatively mirrors the behaviour of the DB scheme that the RMCPP flat-accrual scheme is designed to replace. However, when examined quantitatively, we find that the level of cross-subsidy is considerably higher in a flat-accrual scheme than a DB scheme. 

Since the earliest generations in a flat-accrual scheme receive an income that is worth more than the value of their lifetime contributions, subsequent generations must receive an income that is worth less than their contributions. We find that the flat-accrual scheme broadly tends to a long-term steady-state situation where typical generations receive a retirement income worth less than their contributions. One can think of this as later generations paying interest on the debt built up by overpaying the earliest generations. We refer to this effect as ``drag''. It may appear counter-intuitive that a scheme can enter a steady-state which is not actuarially fair. This is possible because over an infinite time horizon one cannot argue that the total money paid into the scheme must equal the total money paid out, so we will also refer to this as an ``infinite-horizon effect''.

The dynamic-accrual scheme is designed to avoid the obvious cross-subsidies of a flat-accrual scheme. It is instead intended to be \emph{actuarially fair}, in the sense that members' contributions match their new benefit entitlements, at the time of each contribution. However, the implicit pricing methodology
that has been proposed in industry to do this is only approximate, and so cross-subsidies still appear. These cross-subsidies are stochastic in nature, meaning one cannot predict in advance which generations will lose out.

We briefly consider two ways to mitigate the cross-subsidies we have identified.
If a flat-accrual scheme is designed to replace a DB scheme, one might wish to design a scheme that has similar levels of cross-subsidy to the DB scheme it seeks to replace. This could be done using age-based-accrual rather than flat-accrual. In a dynamic-accrual scheme,  we show how to statistically calibrate a pricing formula to obtain fairer prices for benefits than the formula proposed by industry. These mitigations are able to manage the level of cross-subsidies without departing from a shared-indexation design, but do not eliminate them entirely. 

We use two distinct methodologies to examine the cross-subsidies. The cross-subsidies in a flat-accrual scheme and the notion of drag can be studied using a constant economic model. This allows us to find analytic results. However, the cross-subsidies in a dynamic-accrual scheme require stochastic modelling. In this case, we use a Black-Scholes economic model in order to ensure that any pension income can be priced unambiguously using the theory of risk-neutral pricing. The cross-subsidies then emerge from the difference between the true risk-neutral price and the ``small-noise'' approximation to the price used to determine how much members are charged for their pension benefits. In the case of flat-accrual schemes, stochastic modelling confirms the results obtained using a constant economic model.

 Cross-subsidies have been studied in other types of CDC scheme, and the high cross-subsidies and consequent drag are a specific consequence of the flat-accrual design. For example, \cite{haan} studies a Dutch-style CDC scheme and shows a greater financial fairness of those pension schemes.

The third way we contribute to the literature, is by
simulating the behaviour of these funds using
a stochastic
economic model and comparing the performance to other scheme
designs. This is done in Section \ref{sec:esgResults}, using the model described in \cite{armstrongMaffraPennanen}. We compare the performance with the alternatives of DC schemes and a pooled annuity fund.  Similar stochastic models have been used to study CDC designs that have been proposed for use in the UK e.g.\ \cite{popat, owadally}, but not for the current shared-indexation design of the RMCPP.

In all our modelling in this paper, we assume there is no systematic longevity risk and there are sufficient numbers of members so that individual longevity risk is eliminated through pooling. We do not allow for any heterogeneity in the schemes in our simulations.  However, we do show how heterogeneity can be incorporated into the scheme design if desired; results from simulating heterogeneous funds are discussed in \cite{upton}.

The fourth contribution to the literature is studying
the extent to which shared-indexation schemes are able to smooth
pension incomes. This is considered in Section \ref{sec:smoothing}.

The design of shared-indexation schemes is such that younger generations experience greater volatility in the value of their future benefits, while older generations experience a smoothing in this value. This allows the fund to invest in riskier assets while still providing a smooth income in retirement. Other CDC plans studied in the literature have the same feature and they can also out-perform alternative pension options by taking on more investment risk \cite{bonenkamp,cui,gollier}.

We find that the shared-indexation designs do provide some
smoothing of benefits after retirement, but they do not smooth
projected benefits before retirement by as much as has been suggested \cite{aon}.
The discrepancy between our results and the literature can be attributed to the fact that none of the studies we examined attempted to compute projected benefits using a stochastic model, and the deterministic approximations used to project benefits are not sufficiently accurate for this purpose.

Our fifth contribution is in Section \ref{sec:cap2}, where we consider how a scheme designer might control the outcomes in a CDC fund. We find that it is quite easy in a flat-accrual scheme to target a particular
growth rate for median incomes in retirement by adjusting the relationship between benefits and contributions. This flexibility is not available in a dynamic-accrual scheme, making it difficult to design a scheme so that it will produce a desired level of indexation.

Our final contribution, in Section \ref{sec:modelRisk}, is to briefly consider model risk by evaluating the cross-subsidies that occur when actual investment returns differ from their predictions. Since the level of pensions paid out depend on future predictions of returns, poor predictions of those returns lead to pension payments that are, in hindsight, either too low or too high. Consequently, future generations have higher or lower pensions. This type of cross-subsidy can go either way: it may benefit earlier or later generations to join. However, it is also true that once the plan begins its life, the cost due to previous predictions deviating from their observed expected values could be calculated.

\medskip

Our key findings are that:
\begin{itemize}
    \item dynamic-accrual schemes are not able to target a given level of annual pension increases as effectively as flat-accrual schemes; 
    \item dynamic-accrual schemes have substantially lower intergenerational cross-subsidies than flat-accrual schemes; 
    \item dynamic-accrual schemes still feature some level of intergenerational cross-subsidy as a result of the use of approximate pricing formulae in the design of CDC funds;
    \item flat-accrual schemes do not necessarily have higher pension outcomes than DC followed by life annuity purchase at retirement; 
    \item dynamic-accrual schemes can markedly outperform flat-accrual schemes because they do not experience drag effects;
    \item shared-indexation CDC schemes do not result in as much smoothing of projected pension benefits as has
    been suggested in the literature;
    \item nominal benefits are not an accurate means for estimating 
    projected benefits.
\end{itemize}

\section{The contract of shared-indexation CDC funds}
\label{sec:operationCDC}

In this section, we detail the contractual aspects of shared-indexation CDC funds.  The aspects of their operation which are common to both flat- and dynamic-accrual funds, namely the annual indexation of accrued pensions are described first. We then explain
the different approaches taken in the flat- and dynamic-accrual
CDC funds to the accrual of benefits. In Appendix \ref{sec:regulations}, we explain briefly how our description of CDC funds corresponds to the UK regulations.

\subsection{Elements common to all shared-indexation CDC funds}

\subsubsection{Scheme membership and contributions}

To describe the operation, we assume there are a total of $M$ different types of individual investing in the fund over time, indexed by $\xi \in \{0, 1,2,\ldots,M-1\}$. All individuals of type $\xi$ share the same demographic and financial characteristics. For the numerical examples in this paper, $\xi$ is simply a generation index (\cite{upton} studies the situation when individuals vary by both generation and sex). Each individual of type $\xi$ who is below the retirement age makes a contribution $C^\xi_t$ at time $t$ and receives, in exchange, an additional nominal benefit amount $B^\xi_t$ in retirement.

\subsubsection{Cumulative benefit and new benefit accrual amounts}

At any time $t$, each individual of type $\xi$ has accrued a cumulative nominal benefit of annual amount $B^{\xi,\cum}_{t-}$ before new contributions. At time $0$, this is equal to $0$ and we will describe inductively below how 
the value is computed at subsequent times.
Assuming the value $B^{\xi,\cum}_{t-}$ is known at time $t$,
after paying a contribution at time $t \in \mathbb{N}_{0}$, the individual accrues an additional annual nominal benefit of amount $B^\xi_t$.  This increases their cumulative nominal benefit to
\[
B^{\xi,\cum}_t := B^{\xi,\cum}_{t-}+B^\xi_t.
\]
Members do not accrue benefits until they begin contributing to the scheme (which may be at times $t>1$; the notation considers all members who are ever in the scheme, even those not currently in the scheme).  Retired individuals are paid $B^{\xi,\cum}_t$ at time $t$.  For individuals who have not yet retired, $B^{\xi,\cum}_t$ represents their currently accrued benefit entitlement before allowance for future pension increases and benefit adjustments.

\subsubsection{Pension increase mechanism}

Each year, all accrued pensions are increased by the same pension increase factor $\theta_t (1 + h^{\nom}_{t})$, in which $\theta_t$ is a \emph{bonus level} and $h^{\nom}_{t}$ is the {\em nominal indexation rate}.  It is currently a UK regulatory requirement that scheme members must receive identical adjustments to their accrued benefits.  This is the defining characteristic of a shared-indexation scheme. 

The procedure for determining the pair $(h^{\nom}_{t}, \theta_t)$ is described next.  Denote by $\DCPIt=\frac{\CPI_t}{\CPI_{t-1}}-1$ the annual effective rate of change in the Consumer Price Index, $\CPI_t$, from time $t-1$ to time $t$.  Define the {\em real indexation rate} $h_t$, at time $t$, as the solution to
\[
1 + h^{\nom}_{t} = (1 + h_t) \left( 1 + \DCPIt \right).
\] 
After the pension increase factor is determined at time $t$, the total benefit accrued by each individual of type $\xi$ becomes
\begin{equation}
B^{\xi,\cum}_{t-} := \theta_t \, (1 + \DCPIt)(1 + h_t) \, B^{\xi,\cum}_{t-1}, \quad \text{for }t\geq 1.
\label{eq:cdcDefining}
\end{equation}

\subsubsection{Calculation of pension increases}

Broadly, to calculate the real indexation rate $h_t$ at time $t$, first, the accrued benefits is projected from time $t$ onwards assuming that the future real indexation rate remains $h_t$ at all future times $s \geq t$.  Discounting the projected pension payments back to time $t$, their total value is equated with the market value of the scheme's assets.  The resulting expression can be solved for $h_t$.  

However, there are some critical elements necessary to complete the calculation.  The real indexation rate $h_t$ is constrained to lie in the range $[-\DCPIt, h^{\upper}_t]$, in which $h^{\upper}_t \geq -\DCPIt$ is chosen by the scheme manager.  Although in the RMCPP, a flat-accrual scheme, $h^{\upper}_t = \infty$, we will assume a finite upper limit for our studied flat-accrual scheme.  In the considered dynamic-accrual scheme, $h^{\upper}_t < \infty$ and the goal is to ensure that the nominal indexation rate awarded at time $t$ on accrued benefits is neither negative nor too high.

If the solution $h_t$ to the expression involving the scheme's asset value and discounted benefits would imply breaching either of these limits, then the real indexation level is set equal to the relevant limit and a bonus level $\theta_t$ is applied to the accrued benefit entitlements.

The intention is that, when the limits on $h_t$ are not breached, then $\theta_t=1$ and changes to benefits arise only from the indexation rates.  If, for example, the solution implied $h_t  < -\DCPIt$, then $h_t:=-\DCPIt$ and the bonus level falls below one, i.e. $\theta_t < 1$, corresponding to a nominal benefit cut.  The calculation of the value of $\theta_t$ is explained more fully below.

Let:
\begin{itemize}
\item $N^{\xi}_t$ be the number of surviving individuals of type $\xi$ in the fund at time $t$.

\item $p^{\xi}_{t,k}$ be the probability that an individual of type $\xi$ survives from time $t$ to time $t+k$, given that they are alive and age $x$ at time $t$.  It is assumed that future lifetimes are independent random variables.

\item $\nu^{\xi}_{t,k}$ be the cumulative discount factor for an individual of type $\xi$, that discounts 1 unit paid at time $t+k$ back to time $t$.  Let $\hat{R}^{\xi}_{t,k}$ be the annual effective rate of return predicted at time $t$ of the assets invested, in the time period $[\max\{t+k-1,t\}, t+k)$, in line with the investment strategy associated with individuals of type $\xi$.  Then, for $k \in \mathbb{N}_0$,
\[
\nu^{\xi}_{t,k} = \prod_{n=0}^{k} (1+\hat{R}^{\xi}_{t,n})^{-1}.
\]
Note that $\hat{R}^{\xi}_{t,0}=0$.

\item $\widehat{\DCPI}_{t,k}$ be the projection made at time $t$, of the annual change in CPI from time $t+k-1$ to time $t + k$, for $k=0,1,2,\ldots$.  Note that $\widehat{\DCPI}_{t,0}=\DCPIt$, the actual change in CPI over the year to time $t$.
\item $I_{t,k}(\aitch)$ be the projection made at time $t$ of the cumulative nominal indexation rate from time $t$ to time $t+k$, using a constant real indexation rate $\aitch$, for $k = 0,1,2,\ldots$, which is calculated as
\[ 
I_{t,k}(\aitch) := \prod_{n=0}^{k} (1 + \widehat{\DCPI}_{t,n})(1 + \aitch).
\]

\item ${\mathbf 1}^{\xi, \Ret}_t$ be the deterministic indicator function taking the value $1$ if individuals of type $\xi$ have retired at time $t$, and $0$ otherwise.

\item $A_{t-}$ be the value of assets at time $t$, just before new contributions are added or benefit payments are made at time $t$.
\end{itemize}
We now define $\mathcal{L}_{t-}(\aitch, \vartheta)$
to be the discounted value of the accrued benefits at time $t$, just before new contributions are added or benefit payments are made, where the real indexation rate used to project the accrued pensions from time $t-1$ is $\aitch$, and the bonus level applied at time $t$ is $\vartheta$. Thus
\begin{equation}
\mathcal{L}_{t-}(\aitch, \vartheta) := \vartheta \sum_{\xi=0}^{M-1} N^\xi_t \, B^{\xi,\cum}_{t-1} \, \sum_{k=0}^\infty p^{\xi}_{t,k} \, \nu^{\xi}_{t,k} \, I_{t,k}(\aitch) \, {\mathbf 1}^{\xi, \Ret}_{t+k}.
\label{eq:actuarialvaln}
\end{equation}

Equating the asset value with the discounted benefits yields a function of the unknown real indexation rate $\aitch$ and the unknown bonus level $\vartheta$, i.e.
\begin{equation}
A_{t-} = \mathcal{L}_{t-}(\aitch, \vartheta)
\label{eq:asset_equal_liabilities}
\end{equation}
A solution for the pair $(\aitch,\vartheta)$ is first sought by setting $\vartheta:=1$ in (\ref{eq:actuarialvaln}) and solving for the constant $\aitch$.  There are three possible outcomes at each time $t$.
\begin{itemize}
    \item If $\aitch \in [-\DCPIt, h^{\upper}_t]$ then $h_t := \aitch$ is the real indexation rate declared at time $t$, and no bonus is awarded, i.e. $\theta_t:=1$.  
    \item If $\aitch < -\DCPIt$ then the real indexation rate declared at time $t$ is $h_t := -\DCPIt$.  The expression (\ref{eq:actuarialvaln}) is solved for $\vartheta$ with the free variable $\aitch:=-\DCPIt$.  The solution is the declared bonus, i.e. $\theta_t := \vartheta$, and it corresponds to a nominal benefit cut, i.e. $\theta_t<1$.  
    \item Similarly, if $\aitch > h^{\upper}$ then $h_t := h^{\upper}$ and expression (\ref{eq:actuarialvaln}) is solved for $\vartheta$ with the free variable $\aitch:=h^{\upper}$.  The solution is the declared bonus, that is, $\theta_t := \vartheta$, corresponding to a uniform increase in the accrued benefits, i.e. $\theta_t>1$.
\end{itemize}

\begin{figure}[!htbp]
\begin{centering}
\includegraphics[width=0.6\linewidth]{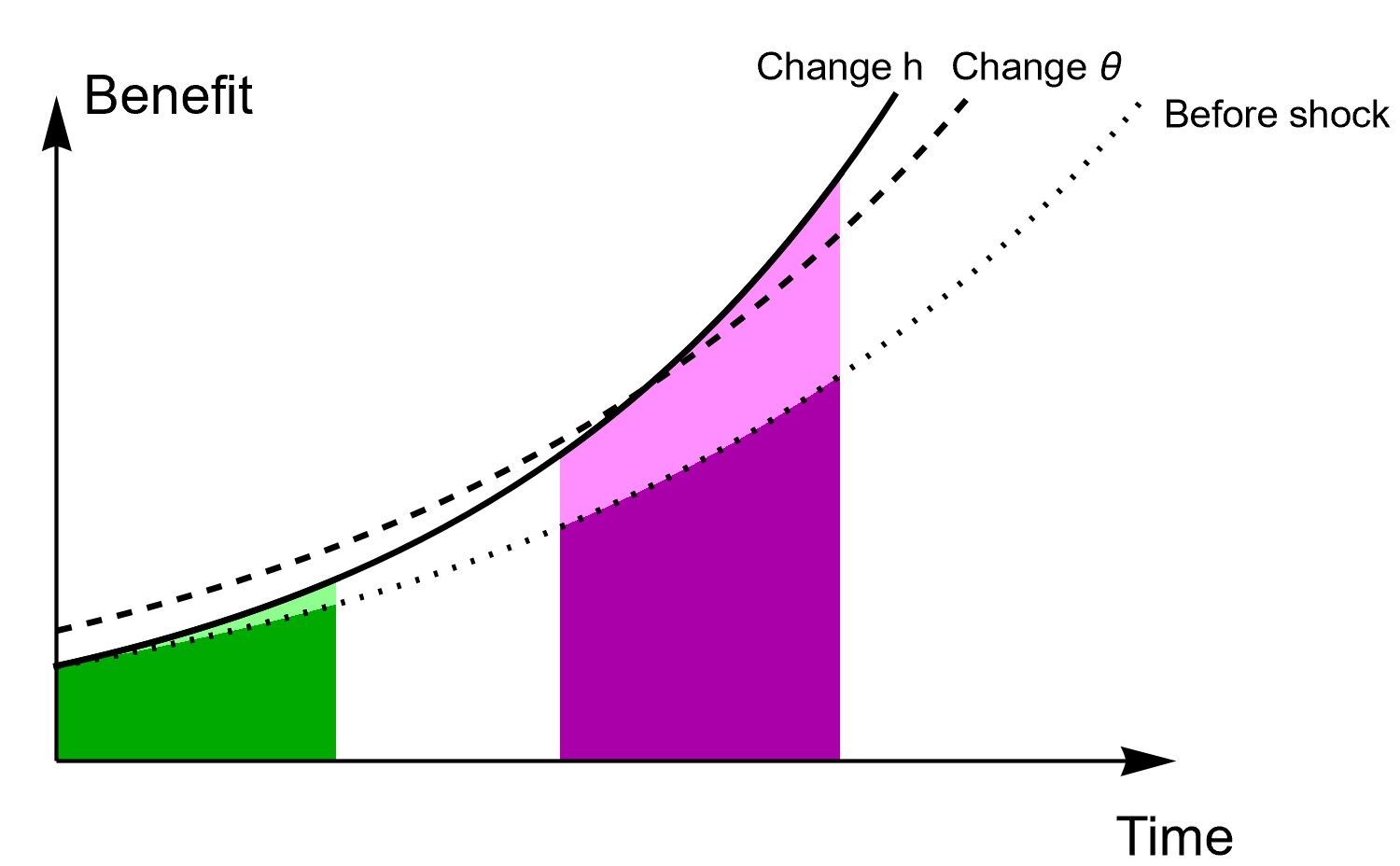}
\caption{
A stylised diagram of how increased benefits might be distributed among different age groups when the market outperforms expectations.
}
\label{fig:stylised}
\end{centering}
\end{figure}

In Figure \ref{fig:stylised}, we show a stylised picture of how the projected benefits change under two different pension increases. The dotted line depicts the projected benefits over time just before a shock occurs. In the example shown, markets have performed better than expected. Suppose that the concomitant increase to benefits can be distributed either as a uniform one-off increase by changing the one-off bonus level $\vartheta$, shown as a dashed line, or by changing the nominal indexation rate, shown as a solid line.

The shaded regions illustrate the projected retirement benefits that are expected by two different generations. The left-hand coloured block in Figure \ref{fig:stylised} shows the income received during retirement for a generation that has already retired.  Its area represents the total income.  The relative increase in their benefits when the nominal indexation rate $h$ increases, depicted with lighter shading, is small.

Similarly, the right-hand coloured block indicates the retirement income for a generation that is yet to retire.  The paler shaded area is much larger, showing that their retirement benefits are much more sensitive to changes in $h$. The increase in their annual pension in retirement is higher than that of the retired generation, compared to before the shock.  If the shock results in a reduction in benefits, then the young would have a larger relative decrease in their pension at retirement, compared to the old.

Thus, using annual indexation rates results in lower volatility in the annual benefit amount paid to older generations.  In exchange, there is a greater fluctuation in projected benefits for the younger generations.  In other words, there is a risk transfer from the older to the younger generations. This argument is heuristic and should be seen as motivational; a more precise argument is made in Section \ref{sec:smoothing}.

In contrast, using only a one-off bonus level means that annual benefit amounts change by the same percentage.  No pension smoothing would take place.  

However, in our shared indexation schemes, the actual adjustment to benefits is a combination of an indexation rate (which is compounded annually to project future benefits) and a one-off bonus (which is applied only at its calculation time and is not re-applied in each projected year).  The combination of using both indexation rates and one-off bonuses means that there are limits on the transfer of risk from the old to the young. 

In RMCPP, nominal benefit cuts (corresponding to $\theta_t < 0.95$) are spread over two or three years with the goal of trying to keep reductions to at most 5\% each year.  Positive bonuses (i.e. $\theta_t > 1$) can occur if they are used to mitigate against these multi-year reductions.  In our flat-accrual scheme design, benefit cuts and bonuses are applied in full immediately, for simplicity.

%The calculation of benefits and contributions is the key difference between flat- and dynamic-accrual schemes and we discuss how these calculations are performed later in a separate section for each scheme.

\subsubsection{Asset and discounted benefit value dynamics}

The scheme's asset value evolves as
\[
A_{t-}=\begin{cases}
0 & \textrm{if $t = 0$}, \\
(1 + R_t) A_{t-1} & \textrm{if $t=1,2,\ldots$}.
\end{cases}
\]
Here, $R_t$ denotes the return realised by the fund on its investments over the time period $(t-1,t]$.

The assets of the fund at time $t$, after new payments have been received and annual pension payments have been made, is
\begin{equation}
A_{t} := A_{t-} + \sum_{\xi=0}^{M-1} N^\xi_t (C^\xi_t - B^{\xi,\cum}_t {\mathbf 1}^{R,\xi}_t).
\label{eq:increments}
\end{equation}
After new payments have been received and annual pension payments have been made at time $t$, the discounted benefit value is
\begin{equation}
\mathcal{L}_{t} := \sum_{\xi=0}^{M-1} N^\xi_t \, B^{\xi,\cum}_{t} \, \sum_{k=1}^\infty p^{\xi}_{t,k} \, \nu^{\xi}_{t,k} \, I_{t,k}(h_t) \, {\mathbf 1}^{\xi, \Ret}_{t+k}.
\label{eq:liability}
\end{equation}
Note that the benefits paid out at time $t$ are excluded from the calculation of $\mathcal{L}_{t}$.  Additionally, as $\theta_t$ is already incorporated into $B^{\xi,\cum}_{t}$, the bonus level is neutral (i.e. $\theta=1$) in this calculation. For brevity, we will sometimes call the quantity ${\mathcal L}_t$ the liability at time $t$, although, since the fund does not
promise any particular pension payment, it is not strictly speaking a liability.

It is straightforward to show that
\[
\begin{split}
\mathcal{L}_{t} &=  \mathcal{L}_{t-}(h_t,\theta_t)\\
&\hphantom{=:} - \underbrace{\sum_{\xi=0}^{M-1} N^\xi_t \, B^{\xi,\cum}_{t} \, {\mathbf 1}^{\xi, \Ret}_{t}}_{\text{Benefits paid out at time $t$}} + \underbrace{\sum_{\xi=0}^{M-1} N^\xi_t \, B^{\xi}_{t} \, \sum_{k=1}^\infty p^{\xi}_{t,k} \, \nu^{\xi}_{t,k} \, I_{t,k}(h_t) \, {\mathbf 1}^{\xi, \Ret}_{t+k}}_{\text{Discounted new benefits accrued at time $t$}}.
\end{split}
\]
Since $A_{t-} = \mathcal{L}_{t-}$, substitution from \eqref{eq:increments} leads to
\begin{align}
\underbrace{A_t - \mathcal{L}_{t}}_{\substack{\text{Asset value less} \\ \text{discounted benefits} \\ \text{at time $t$}}} &= \underbrace{\sum_{\xi=0}^{M-1} N^\xi_t \, C^\xi_t}_\text{Contributions paid at time $t$} \nonumber \\
&\hphantom{=:}
- \underbrace{\sum_{\xi=0}^{M-1} N^\xi_t \, B^{\xi}_{t} \, \sum_{k=1}^\infty p^{\xi}_{t,k} \, \nu^{\xi}_{t,k} \, I_{t,k}(h_t) \, {\mathbf 1}^{\xi, \Ret}_{t+k}}_{\text{Discounted new benefits accrued at time $t$}}.
\label{eq:newbenvalnjump}
\end{align}

The last expression shows that if new contributions do not equal the value of the newly-accrued benefits, then there will be an immediate scheme gain or loss at time $t$.  This is in spite of the asset value equalling the discounted benefits immediately before the contributions are paid at time $t$, i.e.  $A_{t-} = \mathcal{L}_{t-}(h_t,\theta_t)$.  

For this reason, the contribution rate is chosen so that, when a flat-accrual CDC scheme has a stable population, the right-hand side of equation \eqref{eq:newbenvalnjump} is zero.  As the same contribution rate is used in the dynamic-accrual scheme for our numerical illustrations, there are different changes in the scheme's funding position immediately after new contributions are made, depending on whether the scheme is a flat- or dynamic-accrual scheme, as discussed in Sections \ref{SUBSECflatdesign} and \ref{SUBSECdynamicdesign}.

\subsection{Flat-accrual scheme design} \label{SUBSECflatdesign}

A flat-accrual CDC scheme has a design similar to a typical UK DB scheme. Like DB schemes, income benefits in flat-accrual CDC schemes are accrued at a constant proportion of salaries in exchange for a fixed rate of salaries being paid as contributions.  This means that a member who is approaching retirement accrues the same amount of annual benefit as a member at the start of their career, assuming they pay the same contribution amount.

This leads to a difference in the value of the benefits accrued to the value of the contributions made in respect of those accrued benefits.  Younger active members contribute more than the value of the benefits they are accruing, whereas older active members contribute less.

Let $S^\xi_t$ denote the salary of individuals of type $\xi$, at time $t$. We write $\mathbf{1}^{C,\xi}_t$ for the indicator function taking the value $1$ if individuals of type $\xi$ make a contribution in year $t$, and 0 otherwise.

Let the constant $1/\beta$ be the accrual rate, that is the proportion of salaries accrued as an annual benefit at time $t$ by all currently contributing members.  Then the benefit amount accrued at time $t$, by each individual of type $\xi$, due to the contributions made at time $t$ in the flat-accrual model is 
\begin{equation}
B^\xi_t=\frac{1}{\beta} \, S^\xi_t \, \mathbf{1}^{C,\xi}_t,
\label{eq:defBeta}
\end{equation}
which has a discounted value at time $t$ of
\begin{equation}
\frac{1}{\beta} \, S^\xi_t \, \mathbf{1}^{C,\xi}_t \, \sum_{k=1}^\infty p^{\xi}_{t,k} \, \nu^{\xi}_{t,k} \, I_{t,k}(h_t) \,   \, {\mathbf 1}^{\xi, \Ret}_{t+k}.
\label{eq:PVnewbenflat}
\end{equation}
Let the constant $\alpha>0$, be the proportion of salaries which is paid at time $t$ in respect of all members who make a contribution at time $t$.  Then the amount of contribution made at time $t$, in respect of each individual of type $\xi$, is
\begin{equation}
C^\xi_t = \alpha \, S^\xi_t \, \mathbf{1}^{C,\xi}_t.
\label{eq:defAlpha}
\end{equation}
It is assumed here that neither the contribution rate $\alpha$, nor the accrual rate $1/\beta$, change over time.  In reality, both may change as a result of management decisions, but this is anticipated to happen infrequently.

The contribution rate $\alpha$ is chosen so that the right-hand side of equation \eqref{eq:newbenvalnjump} is zero, when the scheme has a stable population.  This approach implies that there is collective financial fairness at the time of each contribution payment.  However, there is not individual financial fairness at each time. 
The contribution $C^\xi_t$ is, in general, not equal to the discounted value of the newly  accrued benefit, which is seen by comparing \eqref{eq:PVnewbenflat} to \eqref{eq:defAlpha}. Moreover, the total contributions may not match the total additional discounted benefits.
Consequently, new benefit accrual at time $t$ causes a sudden change in the scheme's overall funding position, which can be calculated using \eqref{eq:newbenvalnjump}.
This movement away from a 100\% funding position is only rectified at the next year's valuation.  The gain or loss due to the new contributions are then shared among all the scheme members through the pension increase factor. As a result, the level of indexation will typically vary from year to year, even if one assumes a constant economic model.

If we assume a deterministic model, then this gives rise to dynamical equations for $h_t$. Suppose we allow the number of types of individual, $M$, and the time, $t$, to tend to infinity. Assume that the demographics of the fund and the economic factors all converge to long-term limits.  In that event, we expect that $h_t$ will also converge to a long-term limit $h^\infty$ given by solving the equation
\begin{equation}
\lim_{t\to \infty}
\frac
{\sum_{\xi=0}^{\infty} \frac{1}{\beta} \, S^\xi_t \, \mathbf{1}^{C,\xi}_t \, \sum_{k=1}^\infty p^{\xi}_{t,k} \, \nu^{\xi}_{t,k} \, I_{t,k}(h^\infty) \,   \, {\mathbf 1}^{\xi, \Ret}_{t+k}}
{\sum_{\xi=0}^{\infty} \alpha \, S^\xi_t \, \mathbf{1}^{C,\xi}_t} = 1.
\label{eq:steadyState}
\end{equation}
The numerator in this equation is determined by equation \eqref{eq:PVnewbenflat} and represents the total discounted benefits at time $t$. The denominator is determined by equation \eqref{eq:defAlpha} and represents the total contributions made at time $t$. If the demographics of the contributors to the scheme stabilise from some time $\tau$ onwards, the fraction on the left-hand side will become constant for $t\geq \tau$, so instead of the limit, one can simply work with values of $t$ that are chosen to be sufficiently large. In the demographic model we use in our simulations, one may take any $t\geq 0$.

In a constant economic model, equation \eqref{eq:steadyState} can be used to determine the ratio between the contribution rate $\alpha$ and the benefit accrual rate $\frac{1}{\beta}$, which gives a desired target level of indexation $h^\infty=h^\target$. More generally, in a stochastic, but stationary, economic model, one can compute central estimates for all terms and use equation \eqref{eq:steadyState} to estimate the ratio of contributions to benefits that will achieve a target long-term median indexation. Conversely, however one chooses the ratio of contributions and benefits, this will determine the long-term median indexation of the fund.

In a flat-accrual scheme, the mismatch in a given year between a member's discounted benefit and their contributions may be substantial. It is important to consider whether this is fair to members.  In practice, the contribution calculated in equation \eqref{eq:defAlpha} is really the sum of the member contribution and the sponsoring employer's contribution made on behalf of a particular member (as is the case in \eqref{eq:defVhat}, too).  However, formally, the employer's contribution is not allocated to any specific member, only to the fund as a whole. This enables money to be distributed unevenly across generations, which, as we discuss in detail later, is an essential feature of this type of ``DB-lite'' scheme.

If a member is in the scheme for their entire working life, one might imagine that lifetime benefits will match lifetime contributions. However, in general, this will not be the case. This is because the earliest generations receive disproportionately high benefits and later generations must, in effect, pay interest on the debt this creates. As a result, later generations receive less than they pay in, an effect we call drag. It is only in the long-term steady state that all scheme members are treated equally, and in this steady state no member will receive as much benefit as they pay in. We quantify this effect in Theorem \ref{thm:analyticFormula} below.

The use of flat-accrual means that if a member leaves the scheme early, they can expect to have over-paid for the value of what they can transfer out of the scheme.  This is considering the contributions paid by both member and employer.   Similarly, it is not fair for existing members if someone close to retirement joins and their contributions are not sufficient to cover the value of their accrued benefits.  This perception of unfairness may exist even if the employer's contribution is not attributed to each individual member, but rather to the scheme as a whole.  Existing members would suffer a decrease, albeit likely small, to make good this individual deficit.  Lastly, the calculation of the annual pension increases on accrued benefits can lead to significant changes in accrued pensions, depending on the scheme maturity and asset value.  We discuss this further in Section \ref{sec:crossSubsidy}.

\subsection{Standard dynamic-accrual scheme design} \label{SUBSECdynamicdesign}

The term dynamic-accrual, is intended to suggest that accrual rates vary with market conditions. We define a {\em standard dynamic-accrual CDC scheme} to be one where the value of the new benefit accrued by a contribution matches the amount of that contribution. 

A standard dynamic-accrual CDC scheme is intended as an alternative to a DC fund and could be well-suited to multi-employer schemes. In a multi-employer scheme, it is important to minimize intergenerational cross-subsidies as different employers may have different demographics of their workforce and would presumably not wish to provide an overall subsidy to another employer (see \cite{mcinally}).

In our studied dynamic-accrual scheme, the new benefit accrued at time $t$ by the corresponding contribution $C^\xi_t$ made at time $t$, is given
by defining the price of a unit of nominal benefit by
\begin{equation}
\label{eq:defVhat}
\hat{V}^\xi_t := \frac{1}{I_{t,0}(h_t)} \, \sum_{k=1}^\infty I_{t,k}(h_t) \, \nu^\xi_{t,k} \, p^\xi_{t,k} \, {\mathbf 1}^{R,\xi}_{t+k},
\end{equation}
and then choosing $B^{\xi}_t$ to solve
\begin{equation}
C^\xi_t = B^{\xi}_t \hat{V}^{\xi}_t.
\end{equation}
By equation \eqref{eq:newbenvalnjump}, this ensures that
\begin{equation}
A_t - \mathcal{L}_{t} = 0.
\label{eq:after}
\end{equation}

The basis used in \eqref{eq:defVhat} is the same as that used for the calculation of the pension increase awarded at time $t$. Assuming all interest rates are positive, the ratio of benefits to contributions will decrease with age, reflecting the time-value of money. 

Using the valuation basis to calculate the new accrued benefits means that there is no change in the scheme's overall funding position due to new benefit accrual.  Since the new contributions made at time $t$ are equal to the value of the new accrued benefits, the right-hand side of \eqref{eq:newbenvalnjump} equals zero.  Consequently, if the valuation basis is borne out in practice, then the asset value will automatically equal the discounted benefit value at the next and subsequent time periods, for the selection $(\aitch, \vartheta):=(h_t,1)$, i.e. equation \eqref{eq:asset_equal_liabilities} holds at the next time.%  This is unlike the flat-accrual scheme, where new benefit accrual can often result in the total asset value being different to the discounted benefits even if model assumptions are borne out.

For the contributions paid into the scheme at time 0, an assumption is needed about the level of future pension increases in order to calculate the first benefits accrued.  We assume that the real indexation rate is a constant, $h^{\target}$, at time 0 and hence set $h_0 := h^{\target}$ in \eqref{eq:defVhat}.  For contributions made at time $t=1,2,\ldots$, the real indexation rate $h_t$ calculated at time $t_{-}$ is used in \eqref{eq:defVhat}.

We have seen when designing a flat-accrual fund and in a mean-reverting economic model, it is possible to choose a target level of indexation $h^\target$ and set the contribution and benefit accrual rates to ensure $h_t$ tends to revert to this level in the long-term. There is no equivalent mechanism to set long-term indexation in a standard dynamic-accrual fund. Setting $h_0=h^\target$ will have a negligible influence on the values of $h_t$ for large $t$. Instead, if one wishes to target a long-term median level for $h_t$, one must choose the bounds on $h_t$ to limit its fluctuations and use benefit cuts and bonuses to constrain $h_t$ within reasonable limits. We will see a numerical example of this in Section \ref{sec:cap2}.

As nominal indexation rates, discount rates and expectations of future lifetimes will vary over time, the amount of benefit calculated using \eqref{eq:defVhat} will also vary continuously for a fixed contribution amount.  For occupational pension schemes in the UK, it is usual to have infrequently changing conversion factors for members.  This is unlike life insurance companies, who may update daily their life annuity conversion rates.  Although the trustees of a CDC pension scheme may prefer to have a rarely changing table of values for members, we do not explore the implications here as we focus on dynamically-changing benefit accrual only. 

In a flat-accrual scheme, $\nu^\xi_{t,k}$ varies with $\xi$ and reflects their lifestyle strategy. Some of the pension providers we spoke to were considering developing dynamic-accrual CDC funds where $\nu$ depends upon $\xi$ in the same way. However, as all funds are pooled and subject to the same risks, if one allows $\nu$ to vary with $\xi$, this will result in different
generations being charged different amounts for the same future cashflows. This runs against the design goal of minimizing intergenerational cross-subsidies, as can be confirmed numerically using the techniques of Section \ref{sec:crossSubsidy}.
To avoid this issue, we selected an overall strategy for the dynamic-accrual fund that was designed to reflect the cumulative effect of each generations lifestyle strategy and chose the discount rate $\nu^\xi_{t,k}$ to match this overall strategy, and hence to be independent of $\xi$. See Section \ref{sec:dynamicStrategy}.

\section{Modelling details}
\label{sec:modelling}

\subsection{Demographic assumptions}

We will assume that the following modelling assumptions hold in all our simulations except where we explicitly state otherwise.

The earliest entry age to the scheme is age $x_0 := 25$.  An individual receives their first pension payment at age $\NRA := 65$ and pension payments are made annually in advance.  Every individual is assumed to be dead at age $\omega := 120$, the limiting age.  The fund is closed to new contributions after $100$ years.

In our simulations, the type $\xi \in \{0,1,2,\ldots,M-1\}$, represents the generation number of the individual.  At time 0, there are $\NRA-x_{0}+1 \leq M$ generations in the scheme, with generation $\xi=0$ of age $\NRA-1$, generation $\xi=1$ of age $\NRA-2$, and so on, up to generation $\xi=\NRA-x_{0}+1$ who are age $x_0$ at time 0. At time 1, generation $\xi=\NRA-x_{0}+2$ joins the scheme at age $x_0$, and the new entrants continue to flow steadily into the scheme at each integer time. Thus we have the following equations:
\begin{equation}
\begin{split}
\age(\xi,t) &:= \NRA - \xi + t - 1, \\
\mathbf{1}^{R,\xi}_t &:= \mathbf{1}_{\age(\xi,t) \geq \NRA},  \\
\mathbf{1}^{C,\xi}_t &:= \mathbf{1}_{\age(\xi,t) \in [x_0,\NRA)}.
\end{split}
\label{eq:ageAndXi}
\end{equation}
for $\xi = 0,1,2,.\ldots, M-1$.

There is the same number of individuals in each generation when they join the scheme, so that the number of survivors satisfy $N^0_0=N^1_0=\cdots=N^{\NRA-x_0+1}_0=N^{\NRA-x_0+2}_1=N^{\NRA-x_0+3}_2=\cdots$. Additionally, before reaching age $\NRA$, each individual in generation $\xi$ who makes a contribution at time $t$ is paid the annual rate of salary $S^{\xi}_t := S_t$ at that time.  Therefore, there are no promotional salary increases.  Salaries increase annually at time $t$ at the rate $g$ per annum, such that $S_t = S_0 (1+g)^t$.

All scheme members survive to age $\NRA$.  Thereafter, an individual who is alive at age $x$ has the same probability of dying over the next year, regardless of their generation.  In our simulations, the survival probabilities from age $\NRA$ are computed using the S1PMA tables produced by the Institute and Faculty of Actuaries' Continuous Mortality Investigation.

Let us write $p(x,n)$ for the probability of an individual surviving $n \geq 0$ years from age $x$, given that the individual is alive at age $x$. Then
\begin{equation}
%p^\xi(t,n)=:p(\age(\xi,t), n).
p^\xi_{t,n}=: p(\age(\xi,t), n).
\label{eq:xiAndP}
\end{equation}

We assume throughout that there are sufficient members so that individual longevity risk can be perfectly hedged at all ages.  As a consequence, we use the proportion of survivors, rather than the number of survivors, in our calculations.

\subsection{Economic assumptions}

We use a number of economic models with varying levels of sophistication in this paper, but all the models share the same values for the long-term median of various risk factors as shown in Table \ref{table:riskFactors}.
Where applicable, values are chosen to match the UK's Office for Budget Responsibility figures for September 2024 \cite{obr24}. The stock volatility in all stochastic models is $15.3\%$ per annum to match the model of \cite{armstrongMaffraPennanen}.

\begin{table}[h!tbp]
\begin{center}
\begin{tabular}{ll}
\hline
Economic variable & Long-term median \\
\hline
Stock growth & 7.73\% p.a. \\
Wage growth & 3.83\% p.a.\\
CPI growth & 2.00\% p.a.\\
Index-linked bond growth & 4.36\% p.a. \\
\hline
\end{tabular}
\end{center}
\caption{Long-term medians in our economic models.}
\label{table:riskFactors}
\end{table}

\subsection{Scheme design assumptions}

\subsubsection{Bounds on the nominal indexation rate}

For both CDC schemes, the real indexation rate at time $t \in \mathbb{N}_0$ is constrained to lie in the range $h_{t} \in [-\DCPIt, 0.05]$, so that $h^{\upper}_t:=0.05$.  Any further adjustments to benefits required to equate the discounted benefits with the asset value, are dealt with by changing the bonus level $\theta_t$ to a value other than one.

Consequently, real indexation rates granted on benefits are capped at $5\%$ for both types of CDC fund, so nominal rates are  capped at approximately $q_t + 5\%$. The floor for nominal rates is set at $0\%$.  Our industry consultation suggested that a benefit cap of $\DCPIt+2\%$ would be considered a more reasonable level for a standard dynamic-accrual scheme, but as we explain in more detail later, choosing the lower cap results in a decreasing median pension in retirement. The RMCPP does not have an upper bound on $h_t$, but we have included one to aid comparison between the fund types.

\subsubsection{Benefit accrual rate}

For the flat-accrual CDC scheme, the benefit accrual rate is set to $1/\beta=1/80$. This matches the value for the RMCPP. For the dynamic-accrual scheme, the benefit accrued depends on the amount of contribution, and is determined using equation \eqref{eq:defVhat}.

\subsubsection{Setting the contribution rate}
\label{sec:steadyStateComputation}

To aid comparability between scheme types, the same contribution rate $\alpha$, is used for both the flat-accrual and dynamic-accrual CDC schemes.

Within the flat-accrual CDC scheme, $\alpha$ is computed as the rate required to deliver the desired benefits.  To do this, it is assumed that the membership is a stable population and that, within the chosen economic model, the economic variables have attained their long-term medians, which are displayed in Table \ref{table:riskFactors}.  We refer to these central estimates as \emph{steady-state rates}. 

We now assume the fund is targetting a long-term indexation rate $h^\infty$, that matches the long-term CPI growth. Using equation \eqref{eq:steadyState} for the resulting constant economic model we may compute $\alpha$. The resulting contribution rate is $\alpha=4.84\%$. This is then used as the contribution rate for all the other fund types we simulate.  Sensitivity testing shows that the contribution rate is most sensitive to the CPI and stock growth rate assumptions.  For example, jointly increasing CPI by 1\% per annum and lowering the stock growth rate by 1\% per annum gives a contribution rate of just over 8\%.  Further lowering the stock growth rate by another 1\% results in a contribution rate of around 11\%.

Since the argument used to derive the contribution rate assumes a deterministic model and does not take into account benefit cuts or bonuses, one cannot expect that a future nominal indexation rate of CPI is in any precise sense the ``average'' rate of increase that will occur in simulations.

The contribution level in our simulations is much lower than in the RMCPP. The Trust Deed and Rules \cite{royalMailTrustDeeds} specify a total contribution rate of 14.9\% of pensionable pay (10.9\% from the employer, 4\% from the member) but does not state how these contribution rates were derived. Thus, no explicit target rate of indexation is set in the Trust Deed and Rules. Since we are not considering issues such as spousal benefits, it would not be appropriate to directly use the RMCPP contribution levels in our simulations. Nevertheless, since this contribution rate is much larger than the value of $\alpha$ we use, one would expect that the indexation achieved in the RMCPP will, on average, be above $\DCPI$. The previous Royal Mail defined benefit scheme targeted the Retail Price Index, which can be approximated as $\DCPI+1\%$.  

\subsection{Investment strategy: Flat-accrual scheme}

The investment objective of RMCPP is to give scheme members exposure to the returns on return-seeking assets (which are defined in RMCPP's Trust Deed \& Rules  \cite[Schedule 1, C.3]{royalMailTrustDeeds}).  Specifically, the objective is to invest in line with a discounted benefit-weighted investment strategy, after a cash investment that is sufficient to meet the scheme's liquidity requirements is met.  The discounted benefit-weighted investment strategy is an interpolation between 100\% low-risk assets, defined as bonds and other assets which have an investment-grade credit rating, and 100\% return-seeking assets.  Return-seeking assets are assets which have an expected return in line with that on the FTSE All World Index, 50\% hedged to sterling and weighted according to market capitalisation, with an expected volatility of absolute return in the range $[65\%, 100\%]$ of the same index.

The consequent RMCPP investment strategy is similar to a typical default investment strategy of a defined contribution pension scheme, but with a higher risk strategy that extends well into retirement. Members aged 67 years or younger are assumed to invest entirely in return-seeking assets with respect to their discounted benefit value. Those aged 90 years or older invest entirely in low-risk assets.  Linear interpolation between these two extremes is done for members between age 67 and 90 years.  The aggregate scheme investment into return-seeking assets is the average of the percentage across the membership, weighted by each member's discounted benefit.  The discounted benefit values at the last scheme valuation are used in the calculation.

We assume that our shared indexation CDC schemes have a similar objective.  However, we ignore cash liquidity requirements and assume that the schemes invest in a combination of risky stocks and index-linked bonds.

The flat-accrual CDC scheme uses a life-styling strategy for each member of $100\%$ in the risky asset up to age $\NRA$ years, tapering linearly to $0\%$ at age $\NRA+20$.  These percentages are based on the discounted benefits of each member at the last valuation date.  Aggregating the percentages over the entire scheme membership gives the scheme's investment in the risky asset.  The remaining assets are invested in index-linked bonds.

\subsubsection{Investment strategy: dynamic-accrual scheme}
\label{sec:dynamicStrategy}

The dynamic-accrual CDC fund applies a deterministic investment strategy. It is calculated as the median investment strategy of the flat-accrual CDC scheme.   This is both for reasons of simplicity and to have broadly the same level of investment risk in both scheme types. 

\begin{figure}[!htbp]
\begin{centering}
\begin{tabular}{cc}
\includegraphics[width=0.45\linewidth]{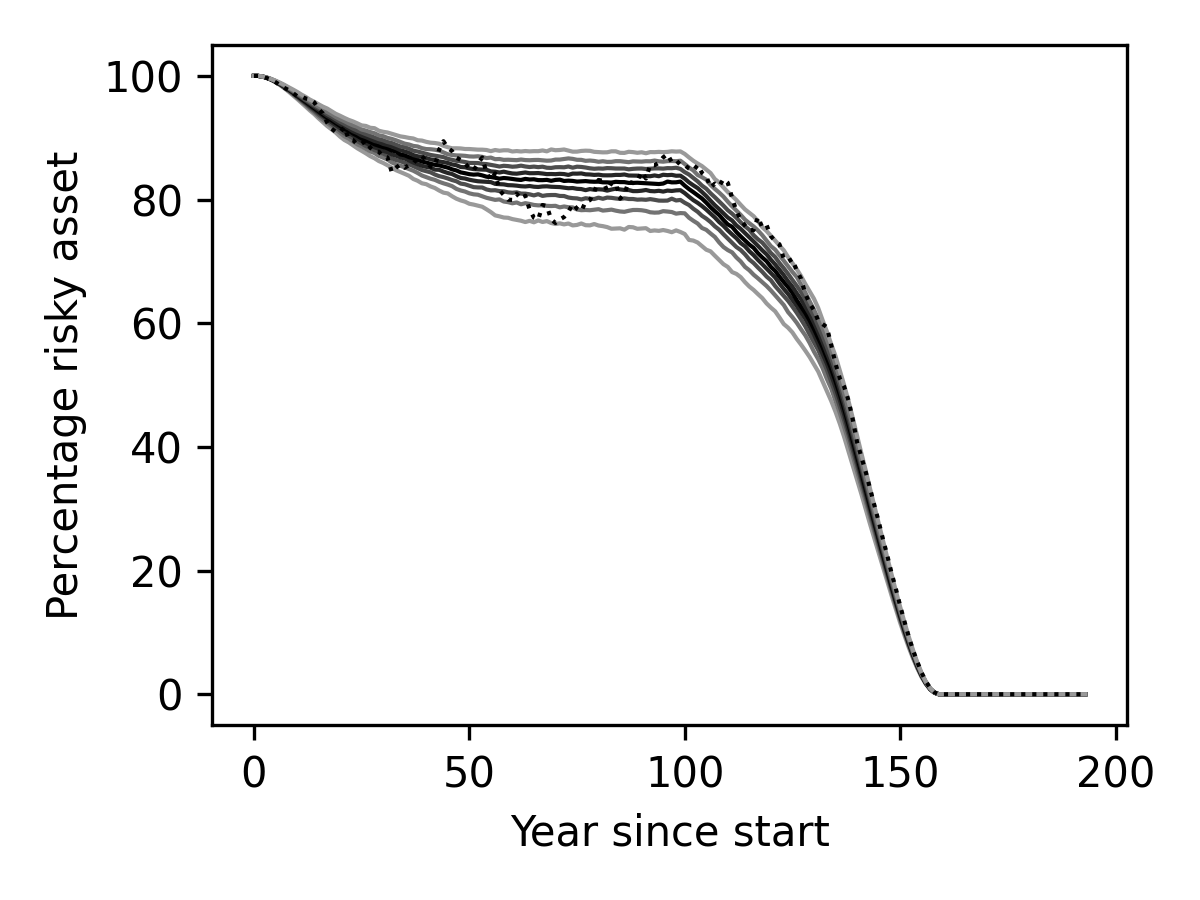} & 
\includegraphics[width=0.45\linewidth]{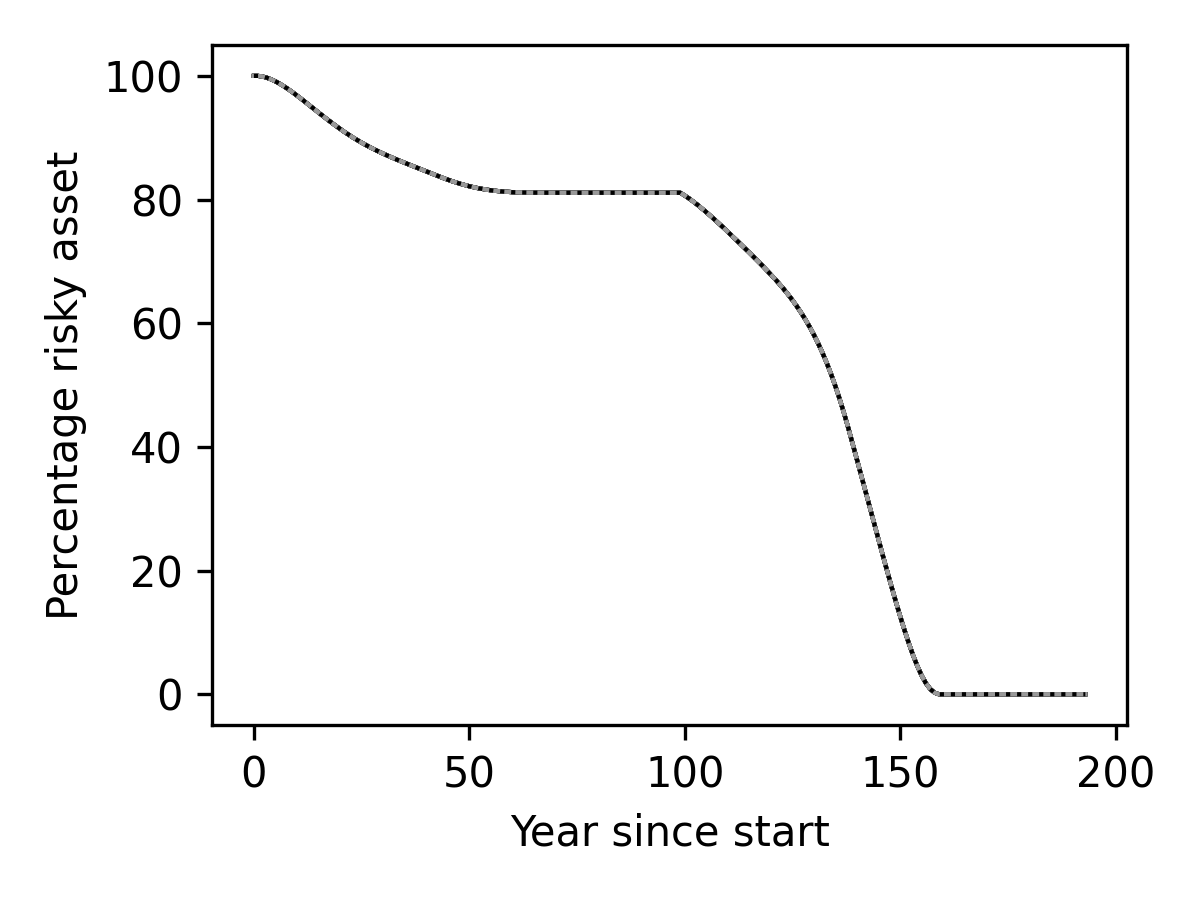}
\end{tabular}
\caption{(Left) fan diagram showing the deciles of the proportion
invested in risky assets in the flat-accrual CDC fund each year. (Right) the deterministic proportion invested in risky asset each year in the standard dynamic-accrual scheme. The economic model used for this simulation is as described in Section \ref{sec:esgResults}}
\label{fig:investmentStrategy}
\end{centering}
\end{figure}

To choose the strategy for the standard dynamic-accrual fund, we simulate the flat-accrual fund using a constant economic model where all risk-factors grow at the median values. From this, we can read off the overall proportion invested in risky assets each year, and then use this to determine the investment-strategy of the standard dynamic-accrual fund. As shown in Figure \ref{fig:investmentStrategy}, the result is that our standard dynamic-accrual fund follows a similar strategy to the median strategy adopted by the flat-accrual fund.

Alongside all our computations we perform various
tests to validate our models. See Appendix \ref{sec:testing} for details of some of the key sanity checks.

\section{Cross-subsidies}
\label{sec:crossSubsidy}
\subsection{Deterministic modelling for flat-accrual CDC}
\label{sec:infiniteHorizon}

On an intuitive level, the decision to provide a fixed proportion of salary as benefit
for a given level of contribution can be expected to benefit investors who are approaching retirement. This is because younger investors have longer to invest their assets than older investors, so, to be financially fair, one would expect younger investors to receive a higher nominal benefit entitlement for a given contribution. Similar existing intergenerational cross-subsidies occur within DB schemes and are well understood. The Equality Act 2010 contains specific provisions to ensure that DB schemes are lawful despite the age-discrimination inherent in such cross-subsidies \cite{equalityActPensions}.

At each time, the amount paid by each investor can be compared to the present value of the additional cashflows they will receive. We will call the difference the {\em instantaneous profit or loss}. Notice that this is a profit and loss relative to the full contribution consisting of both the employee and employer's contribution. Since the employer's contribution is not specifically allocated to any individual, one could phrase this effect in terms of different age groups receiving different effective employer contributions, rather than in terms of a profit or loss. We use the terminology of profit and loss because this seems more natural when discussing dynamic-accrual schemes and we wish to use a consistent approach to all schemes.

If we assume a deterministic economic model, we can quantify the instantaneous profit or loss analytically.   For interest rates $i$ and $j$ and positive integer $n>0$, define
\[
\ax{\angl{n}}^{i,j} := \sum_{k=1}^{n} \left( \frac{1+j}{1+i} \right)^{k} \quad \textrm{and} \quad \ax**{\angl{n}}^{i,j} := \sum_{k=0}^{n-1} \left( \frac{1+j}{1+i} \right)^{k}.
\]

\begin{theorem}
\label{thm:analyticFormula}
Consider a flat-accrual CDC scheme which has a stable population.  Before retirement, each individual contributes the amount $\alpha S_t$ at time $t$, where $\alpha$ is the contribution rate as a proportion of salary.   Contributions are made in advance for an integer $n$ years by each member, before retirement.  The new annual benefits accrued at time $t$ is $1/\beta \times S_t$ per member, where $1/\beta$ is the accrual rate.  

Assume a constant economic and demographic model, in which $S_t =(1+g)^t$ and accrued benefits receive annual pension increases at the constant effective rate $\constantindexrate$ p.a. both before and after retirement.  Asset returns are a constant effective rate $R$ per annum.  It is assumed that the scheme is funded 100\% at all times under this model.  For the latter assumption to hold, the contribution rate $\alpha$ is calculated by equating the contributions paid at time $t$ with the new benefits accrued at time $t$, so that the right-hand side of equation (\ref{eq:newbenvalnjump}) is zero.

Under this contribution rate, the instantaneous percentage gain made by a member who is exactly $k$ years from retirement is the additional value of the new benefits accrued by the member over the contribution made by that member, i.e. 
\begin{equation}
%\left( \frac{n \netinterest^{k}(\netinterest-1)}{\netinterest^n-1} - 1 \right) \times 100\%, \quad \text{where }\netinterest:=\frac{1+R}{1+\constantindexrate}.
\left( \frac{n}{\ax{\angln}^{R,\constantindexrate}} \, \left(\frac{1+\constantindexrate}{1+R}\right)^{k} - 1 \right) \times 100\%.
\label{eq:dbPNL}
\end{equation}
A negative value indicates a loss. As $R \to \constantindexrate$, the instantaneous percentage gain tends to zero.

Suppose the member could alternatively have invested their contributions in an individual DC fund and then used the accumulated DC funds at retirement to purchase a single-life annuity that increases at the rate $\constantindexrate$ p.a.  Assume that there is no additional fee charged when purchasing the annuity to cover systematic longevity risk.

The \emph{benefit ratio \textrm{BR}} at retirement, is the ratio of the annual benefits from the CDC scheme to the equivalent from the individual DC fund at the first time of retirement.  It is given by
\begin{equation}
\textrm{BR} = \left( \frac{1+g}{1+R} \right)^{n} \, \frac{n \, \ax{\angl{n}}^{g,\constantindexrate}}{\ax{\angl{n}}^{R,\constantindexrate} \, \ax**{\angl{n}}^{R,g}}.
%\frac{(\netinterest - 1) (\constantindexrate+1) n \netinterest ^n (R-g) \left((\constantindexrate+1)^n-(g+1)^n\right)}{(R+1) (\constantindexrate-g) \left(\netinterest^n-1\right) \left((R+1)^n-(g+1)^n\right)}.
\label{eq:dbRatio}
\end{equation}
Finally, as $R \to g$ and as $R \to \constantindexrate$, $\textrm{BR} \to 1$.
\end{theorem}

See Appendix \ref{appendix:proofs} for a proof. Note that since this result
only refers to a single asset, we can use a constant economic model to compute the value of benefits unambiguously without worrying about the possibility of arbitrage. As we explain in Section \ref{sec:dynamicAccrual}, one must be cautious when using a constant economic model to evaluate benefits in a CDC fund. Since this theorem is predicated on the assumption that indexation is constant, it can only be expected to give reasonable estimates for a CDC fund when indexation is maintained at its long-term average level. Thus, the result estimates the average behaviour of a CDC fund, but the instantaneous profit may be somewhat different from these estimates in any particular year.

We see from equation \eqref{eq:dbPNL} that the level
of intergenerational cross-subsidy will typically increase with the annual interest rate $R$.
If we assume $R$ is given by the nominal growth of index-linked bonds,
using the values in Table \ref{table:riskFactors}, we find from equation \eqref{eq:dbPNL} that the instantaneous profit for the youngest generation would be $-39.4\%$ each year, and for the oldest
generation it is $53.0\%$. We may read from this that, in a DB scheme, the benefit entitlement received each year by the oldest generation is worth approximately $2.5$ times as much as the benefit entitlement received by the youngest generation.
If we assume $R$ is given by equity growth and repeat the calculation,
we can estimate that, in a flat-accrual CDC scheme, the benefit entitlement of the oldest generation generation is worth almost $9$ times as much as that of the youngest generation.

\begin{figure}[h!tbp]
\begin{center}
\begin{tabular}{cc}
\includegraphics[width=2in]{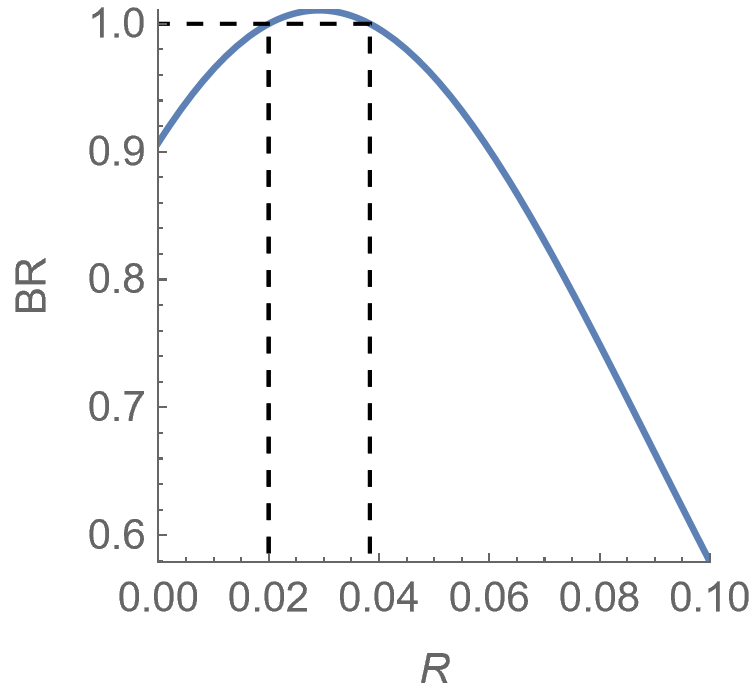} &
\includegraphics[width=2in]{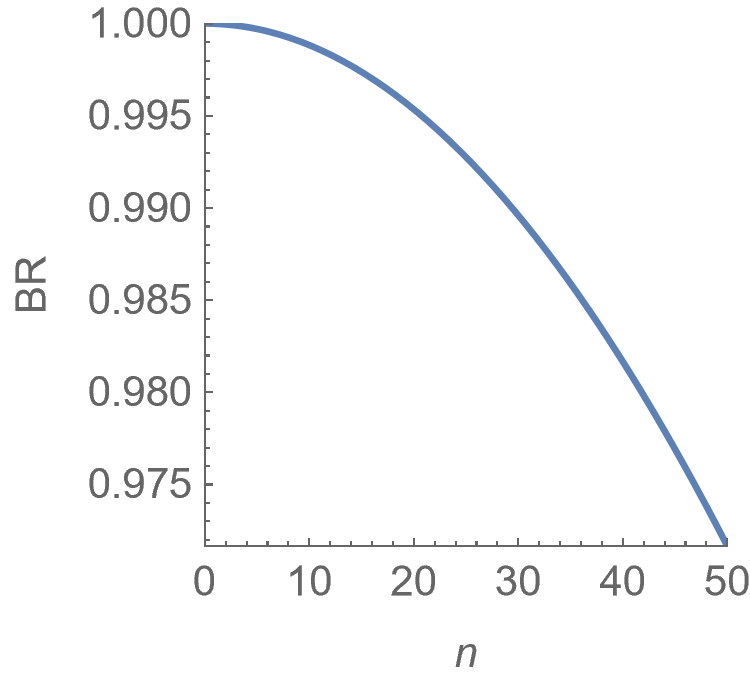} \\
(a) & (b)
\end{tabular}
\end{center}
\caption{The ratio $\textrm{BR}$ of the benefits from a DB fund and a DC fund investing in the risk-free asset, (a) as a function of the interest rate $R$ used for discounting, and (b) as a function of the number of years $n$ from joining the scheme until retirement. All non-varying parameters are as in Table \ref{table:riskFactors}. In (a), n=40. In (b), $R$ is equal to the stock growth rate given in Table \ref{table:riskFactors}. The dotted lines in (a) indicate that the ratio is equal to $1$ when the risk-free rate is equal to wage inflation or price inflation.}
\label{fig:benefitRatio}
\end{figure}

Equation \eqref{eq:dbRatio} quantifies what we call {\em infinite-horizon effects} that can occur in both flat-accrual CDC schemes and DB schemes.
It shows that when a flat-accrual CDC scheme is in its steady state, it may not be actuarially fair.

To see this, in Figure \ref{fig:benefitRatio} (a), we plot
the ratio of the CDC and DC annual pension payments when all rates match the values in Table \ref{table:riskFactors}, except for the risk-free rate which we allow to vary. The rate of indexation in this figure is equal to price inflation. This ratio is also equal to the ratio of the value of fixed-accrual CDC benefits to payments in a constant economic model.
When the interest rate $R$ is less than wage inflation but higher than the rate of indexation in the steady state of the scheme, each generation receives a pension that is worth slightly more than the value of their accumulation contributions. This is possible because they are able to consume some of the savings of subsequent generations, who then take some of the savings of the next generation, and so on \emph{ad infinitum}. Because wage inflation dominates indexation, this results in each generation being better off in the steady state. This is, of course, a form of Ponzi scheme. If subsequent generations decided they no longer wanted to participate in the scheme resulting in its closure, the fund would leave its steady state and the remaining members would receive an income worth less than they have paid in. We call this an infinite-horizon effect because it is only sustainable if one assumes the fund operates over an infinite time period.

As $R$ increases above wage inflation, the level of intergenerational unfairness increases. As early generations have been overpaid, the fund begins to accumulate a debt and the payments that can be sustained in the long term are reduced by the interest that must be paid on this debt. 
If the fund invests primarily in risky assets, resulting in a large value for $R$, the value of the lifetime benefit received in a flat-accrual scheme will be less than the accumulated value of the contributions made over their working lifetime.

If one wants to ensure that the scheme is actuarially fair over each investor's lifetime and that the funding position is unchanged after new contributions are paid, one must have either that $R$ is equal to the level of indexation or that $R$ is equal to wage inflation. Neither of these options is desirable in a flat-accrual CDC scheme.

Theorem \ref{thm:analyticFormula} shows that infinite-horizon effects
also exist in DB schemes, but will be small if the risk-free rate is close to wage inflation. This is the case for the OBR assumptions shown in Table \ref{table:riskFactors}. This explains why infinite-horizon effects are an important issue to consider for flat-accrual CDC funds, but have not attracted much attention in the study of DB funds.

In Figure \ref{fig:benefitRatio} (b), we plot the ratio of the DB and DC payoffs
as a function of $n$, the number of years from joining the scheme to retirement. This is equal to the ratio of the CDC and DC payoff in the steady state of the schemes. In this plot, we have taken all rates to match the OBR assumptions and with $R$ equal to stock growth. The plot shows the longer the time over which contributions are made, $n$, the worse the value of the flat-accrual CDC scheme. This explains why the report \cite{upton} found that flat-accrual CDC sometimes barely outperforms a DC scheme followed by annuitization. In that report, it was assumed that individuals join the scheme at age $20$ and retire at age $67$, giving a value of $n=47$, substantially higher than the value $n=40$ used in this paper.

Our results also explain the wide variation in the estimates of the potential benefits of CDC that have been found in different studies. The report \cite{wtw} gives a particularly high estimate for the benefit of CDC schemes, but its methodology tacitly assumes actuarial fairness over each member's lifetime and so it cannot
be applied to flat-accrual schemes.

\subsection{Stochastic modelling of cross-subsidies in flat-accrual schemes}

In order to get a complete picture of intergenerational transfers of wealth, we must use a pricing methodology which takes into account the fact that $h$ will vary over time.  The liability given by formula \eqref{eq:liability} is an actuarial estimate of the value of the benefits that are expected to be received by the members of a CDC scheme. However, this formula is only an approximation as it does not consider the fact that the nominal indexation rate $h$ varies with time.  Instead, it assumes that $h$ continues at its current level. 

To get a complete picture, we simulate the different types of CDC scheme in a Black-Scholes model with constant rates of price inflation and wage inflation.  We assume that the scheme membership is an infinitely large, stable population and that there is no systematic longevity risk, so that mortality can be treated deterministically.    The continuously-compounded annual return on the riskless asset is a constant $r$.  The price process of the risky asset follows geometric Brownian motion with constant drift $\mu$ and constant volatility $\sigma$.  When simulating the dynamics of CDC funds in this model, the ``central-estimates'' and ``projections'' required are computed using the mean values predicted by this Black-Scholes model.

In the Black--Scholes model, the pension that a generation receives can be viewed as a path-dependent derivative contract in the stock price.  This is because once one knows the full history of the stock price, one can unambiguously compute the payments received in retirement. Here, we assume that we know what others will contribute to the fund.  In effect, we are assuming that employer contributions are sufficiently high to ensure that there is never any incentive for members not to contribute and, similarly, that there is no incentive to leave the scheme. Subject to this assumption, the theory of derivative pricing in the Black--Scholes model then allows one to compute, at any time, an unambiguous market value for these payments. We can then directly compare the value of the benefits received to the contributions paid in respect of each scheme member.

The central point is simply that there is a single, undisputed pricing methodology in the Black-Scholes model and it does not correspond exactly to the ``central-estimate'' approach to pricing benefits implied in equation \eqref{eq:liability}, even when one uses the same discount rate.  In effect, applying risk-neutral pricing allows us to calculate the monetary value of cross-subsidies experienced by each generation of members.  It also allows us to estimate the accuracy of equation \eqref{eq:liability} as an estimate of future benefits.

\begin{figure}[!htbp]
\begin{centering}
\includegraphics[width=0.9\linewidth]{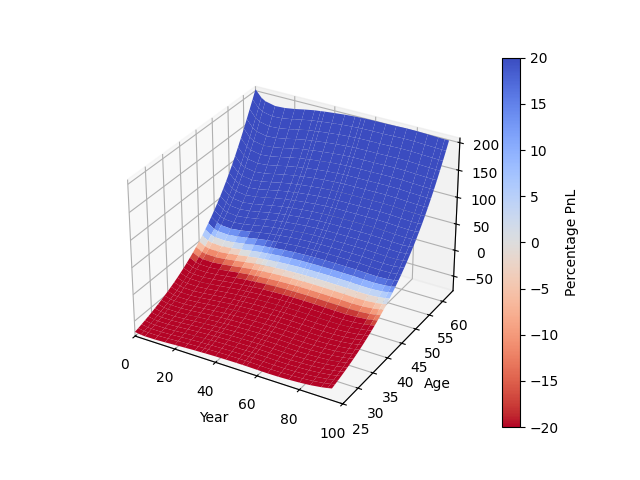}
\caption{Expected instantaneous profit for investors in a flat-accrual CDC fund, by age and year of operation, evaluated using $20,000$ Monte Carlo scenarios.}
\label{fig:pnlSingleByAge}
\end{centering}
\end{figure}

Fix a time $t$ and consider the cashflows that arise from time $t$ for all members of generation $\xi$, due to the contributions they pay at time $t$.  Let $B^{t,\xi}_s$ denote the total annual amount of nominal benefit at time $s \geq t$ that was accrued by generation $\xi$ at time $t$, by their contributions paid at time $t$, i.e.
\[
B^{t,\xi}_s:=\begin{cases}
B^\xi_t & \text{if $s = t$}; \\
\theta_s (1+h^\nom_s) B^\xi_{s-1} & \text{if $s>t$}.
\end{cases} 
\]
Let $X^{t,\xi}_s$ denote the total random cashflow paid or received by all members of generation $\xi$ at time $s \geq t$, as a result of their contributions paid at time $t$. We
have
\[
X^{t,\xi}_s:=\begin{cases}
-C^\xi_t N^\xi_t & \text{if $s=t$}; \\
B^{t,\xi}_s N^\xi_s \mathbf{1}^{R,\xi}_s & \text{if $s>t$}. \\
\end{cases}
\]
%Let $r$ be the annualised continuously-compounded return on the risk-free asset.  
Let $E^{\mathbb Q}_t$ denote the expectation taken in the risk-neutral
measure ${\mathbb Q}$, conditional on the information available at time $t$.  Write ${\mathbb P}$ for the physical measure.  The instantaneous profit at time $t$ for generation $\xi$, is given by the random variable
\[
E^{\mathbb Q}_t\left( \sum_{s=t}^{\infty} e^{-r s} X^{t,\xi}_s \right),
\]
with a negative value indicating a loss.

We define the {\em expected instantaneous profit at time $t$} to be the $\mathbb P$-measure expectation of this, namely
\[
E^{\mathbb P} \left( E^{\mathbb Q}_t\left( \sum_{s=t}^{\infty} e^{-r s} X^{t,\xi}_s \right) \right).
\]
To compute this value, we simulate economic scenarios
using the stock drift under the physical measure, $\mathbb P$, up to the time of retirement. We then perform the remaining simulation using the stock drift for the pricing measure, $\mathbb Q$. This means that the expected instantaneous profit at a particular time $t$ can be computed using a single Monte Carlo simulation, but a separate simulation is needed for each time.

Figure \ref{fig:pnlSingleByAge} shows the expected instantaneous profit made by each generation in each year when the fund is open to contributions. As one would expect, older investors benefit considerably from the flat-accrual CDC structure.
This figure indicates that the average benefit entitlement of the oldest generation is worth approximately $10$ times more than the average benefit entitlement of the youngest generation, which is in line with the estimate of $9$ times using Theorem \ref{thm:analyticFormula}. We conclude that despite using a constant economic model, Theorem \ref{thm:analyticFormula} is able to account for most of the intergenerational cross-subsidy in the flat-accrual scheme.

\begin{figure}[!htbp]
\begin{centering}
\includegraphics[width=0.5\linewidth]{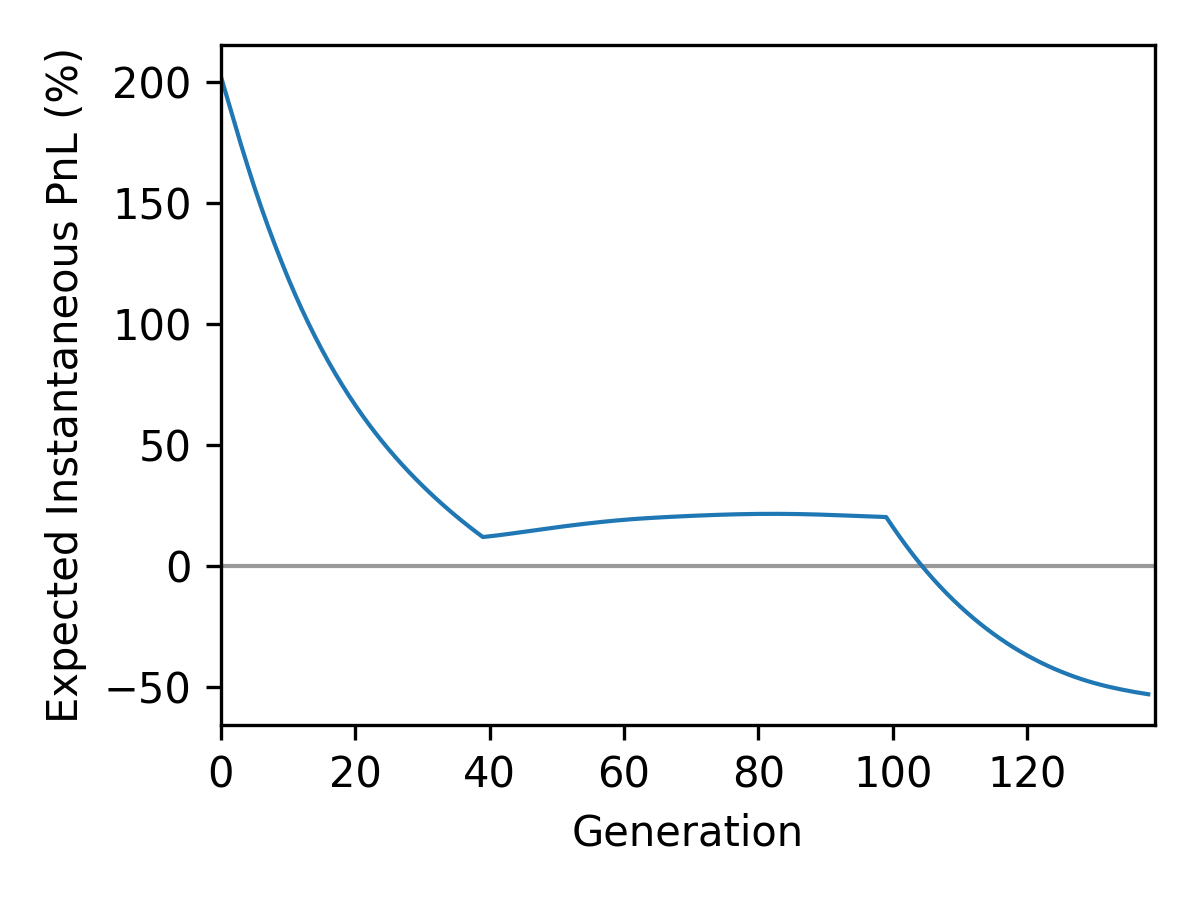}
\caption{Total lifetime expected instantaneous profit 
for investors in a flat-accrual CDC fund by generation,
evaluated using the data points in Figure \ref{fig:pnlSingleByAge},
 with linear interpolation used for missing years.}
\label{fig:pnlSingle}
\end{centering}
\end{figure}

\begin{figure}[!htbp]
\begin{centering}
\includegraphics[width=0.5\linewidth]{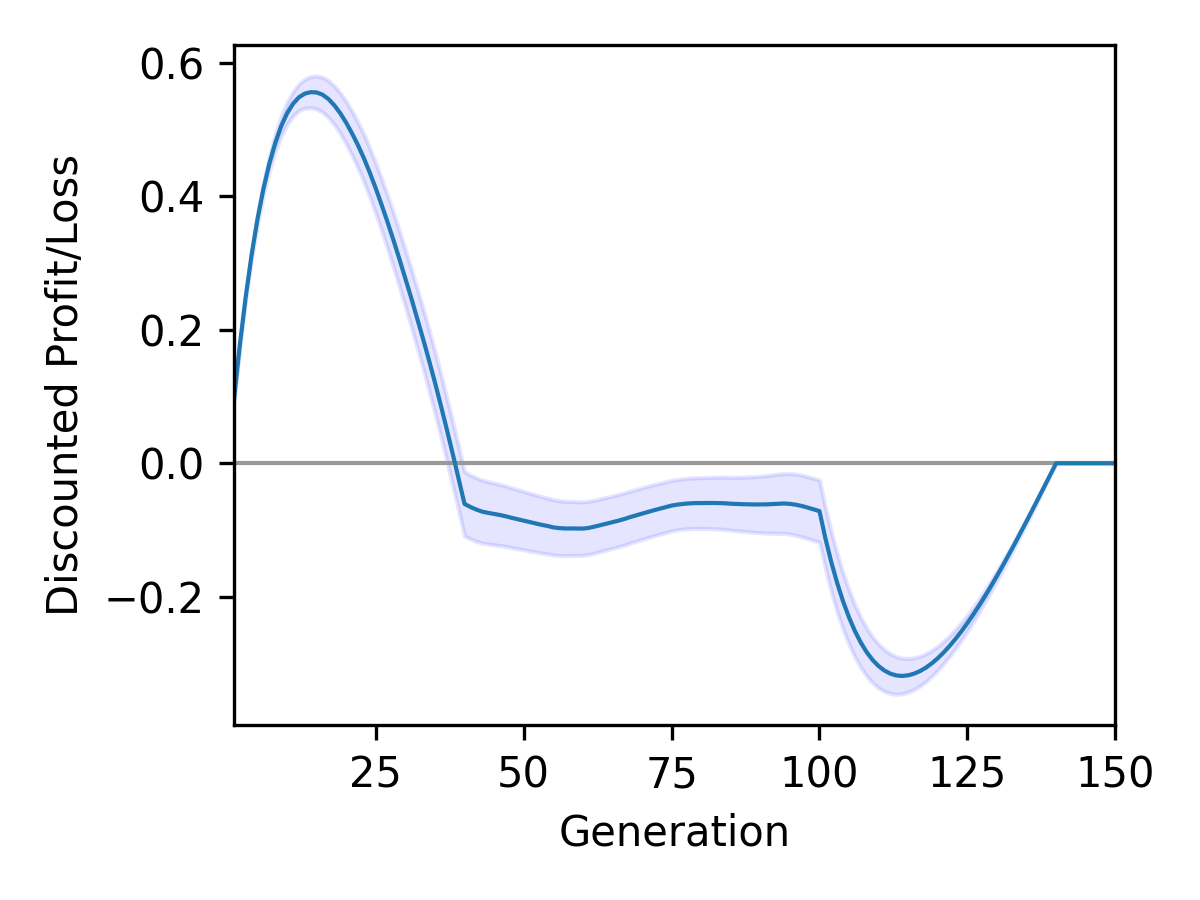}
\caption{Total lifetime discounted value of net cashflows at time 0 for each generation as a proportion of one year's salary, calculated with $100,000$ Monte Carlo simulations. A 95\%-confidence interval is shown (but barely perceptible).}
\label{fig:pnlSingleQMeasure}
\end{centering}
\end{figure}

In Figure \ref{fig:pnlSingle}, we show the aggregate effect for each generation,
\[
\sum_{t=0}^{\infty} E^{\mathbb P} \left( E^{\mathbb Q}_t\left( \sum_{s=t}^{\infty} e^{-r s} X^{t,\xi}_s \right) \right).
\]
Under the latter measure, there are two groups who have the most extreme outcomes.  The first group consists of the older members who joined when the scheme started; they have a large total expected instantaneous profit.  These members joined when they were close to the retirement age of $65$ years.  For example, generation $\xi=0$ is age 64 years and generation $\xi=1$ is age 63 years at time 0.  Their contributions would not, in isolation, have funded all their retirement benefits.

However, as the generation number increases, the contribution history also increases and the total expected instantaneous profit declines. The nadir is reached by generation $\xi=39$, who are the first generation to make contributions over the entirety of their working life.

We emphasize that it is a deliberate design feature to give older members at the start of the scheme's life a pension payment that is not entirely funded by their \emph{de facto} individually attributable contributions.  The employer's contribution is to the scheme as a whole, and not owned by individual members.  This mirrors what happens in UK DB pension schemes, which the flat-accrual CDC scheme is designed to replace.

The second group with extreme outcomes comprises the members who joined the scheme shortly before it closed, when they were quite young.  Their contributions were higher than the value of their accrued benefits, and thus they have a considerable expected loss.

We compute the total risk-neutral value of the cashflows received by each generation, evaluated at time $0$, that is
\[
E^{\mathbb Q} \left( \sum_{t=0}^\infty \sum_{s=t}^{\infty} e^{-r s} X^{t,\xi}_s \right),
\]
with the results shown in Figure \ref{fig:pnlSingleQMeasure}.  We see that when the scheme membership is fully mature (which we identify as the middle phase when all members have a full contribution history), the risk-neutral value at time $0$ of each generation's pension cashflows is net negative.  In contrast, Figure \ref{fig:pnlSingle} shows that the expected instantaneous profit is net positive. One can interpret these results as the earlier generations speculating on stock growth on behalf of subsequent generations, and also using some of the subsequent generations' capital to increase their own pensions.

The total area under the graph in Figure \ref{fig:pnlSingleQMeasure} is equal to zero. However, as we have remarked, when the scheme is fully mature, members receive an amount with a negative risk-neutral value at time 0.  In other words, their contributions are higher than the value of the benefits they receive.  This is a result of the infinite-horizon effects discussed above. 

\medskip

We conclude that, for flat-accrual schemes, the use of a stochastic pricing
methodology reproduces the results found using a constant economic model. One practical consequence of our observations is that
employer contributions must be high in a flat-accrual CDC scheme in
order to guarantee that it is always in an employee's interests to contribute to the scheme. 
In addition, a younger, potential member of a flat-accrual CDC scheme should ask what, if any, recompense they will receive for subsidizing older generations if they change employer. Similarly, a younger member
should ask what will be done if a scheme is closed, to ensure that they receive the intergenerational cross-subsidy they expected. These are equally valid questions for members of DB schemes to ask, but they are far more pressing for potential members of a flat-accrual CDC scheme.

\subsubsection{Age-related accrual}
\label{sec:ageBased}

To reduce the cross-subsidies of a flat-accrual scheme, one might
consider an age-related accrual scheme where the amount of annual benefit accrued decreases with age.   For an individual of age $x$ with annual salary rate $S_t$, suppose at time $t$ their additional annual benefit is of amount
\[
B_t = \frac{1}{\beta} (1 + d)^{\NRA-1-x} S_t,
\]
for some constant discounting level $d$. In Figure \ref{fig:surfaceAge} in the appendix, we show the expected instantaneous profit cross-subsidies that occur across different generations when $d=3.5\%$. This level was chosen as it is approximately equal to the difference between stock growth and bond growth rates, so can be expected to reduce the cross-subsidies to a similar level to that seen in a DB scheme invested in gilts (as can be seen by applying Theorem \ref{thm:analyticFormula}). We will also see later that this scheme reduces the drag effect seen in the flat-accrual scheme.

\subsection{Dynamic-accrual schemes}
\label{sec:dynamicAccrual}

We now compute the instantaneous profit for standard dynamic-accrual schemes using a stochastic economic model. If our formula for liabilities 
priced the value of benefits in a CDC fund perfectly accurately,  we would expect the instantaneous profit to be zero. Thus, our results evaluate the discrepancy between the mathematically rigorous methodology of risk-neutral pricing and the heuristic central-estimate approach implicit in equation \eqref{eq:liability}.

As described in Section \ref{sec:dynamicStrategy}, we chose the investment strategy for the standard dynamic-accrual CDC fund to match the observed time-dependent strategy of a flat-accrual scheme in a deterministic version of the economic model. This ensures that we are using the same discount rate for all members in equation \eqref{eq:liability}.

\begin{figure}[!htbp]
\begin{centering}
\includegraphics[width=0.9\linewidth]{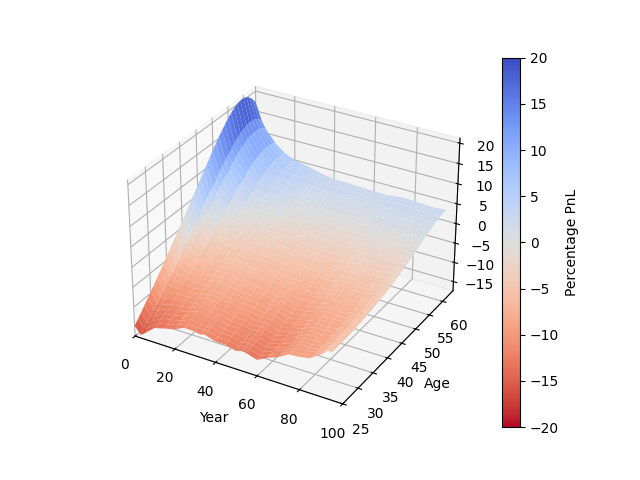}
\caption{Expected instantaneous profit and loss for members in a standard dynamic-accrual CDC fund, by age and time, evaluated using $20,000$ Monte Carlo scenarios.}
\label{fig:pnlMultiByAge}
\end{centering}
\end{figure}

\begin{figure}[!htbp]
\begin{centering}
\includegraphics[width=0.45\linewidth]{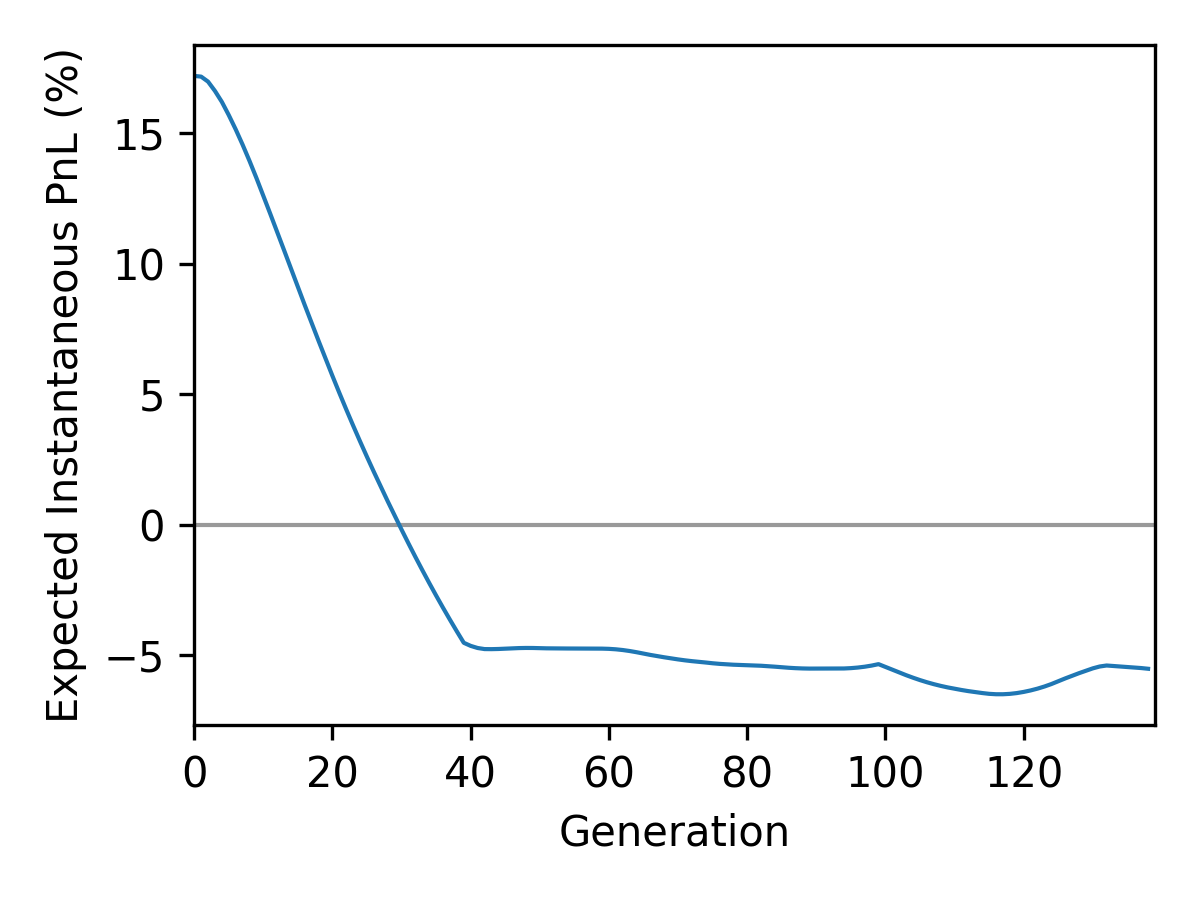}
\includegraphics[width=0.45\linewidth]{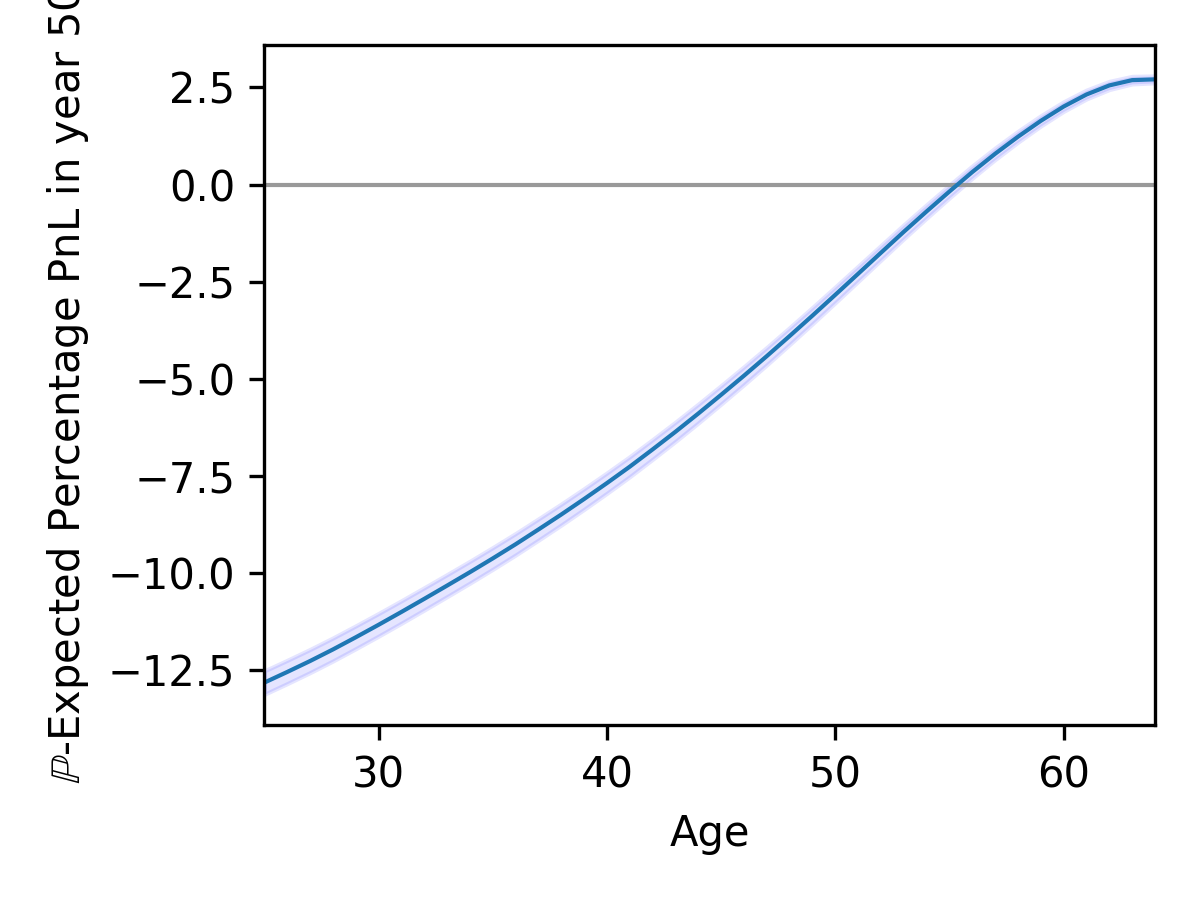}
\caption{Expected instantaneous profit for members of a standard dynamic-accrual CDC fund, by generation (left plot) and by age at time 50 (right plot). The left plot was evaluated using the points in Figure \ref{fig:pnlMultiByAge}, with linear interpolation between years. The right plot was evaluated with $100,000$ Monte Carlo scenarios and the $95\%$ confidence interval is shown (but barely perceptible).}
\label{fig:pnlMulti}
\end{centering}
\end{figure}

Figures \ref{fig:pnlMultiByAge} and \ref{fig:pnlMulti} show the
expected instantaneous profit for the standard dynamic-accrual fund.
Note, that the right-hand chart in Figure \ref{fig:pnlMulti}, shows a slice
of the 3D plot in Figure \ref{fig:pnlMultiByAge} at time 50.

The intergenerational disparities are, as one would expect, much smaller
than those seen in a flat-accrual fund (note the difference in the vertical scale between  Figures \ref{fig:pnlSingleByAge} and \ref{fig:pnlMultiByAge}) but are still non-trivial.
Of the first members of the fund, the oldest generation does particularly well and the youngest generation does particularly badly.  

\begin{figure}[!htbp]
\begin{centering}
\includegraphics[width=\linewidth]{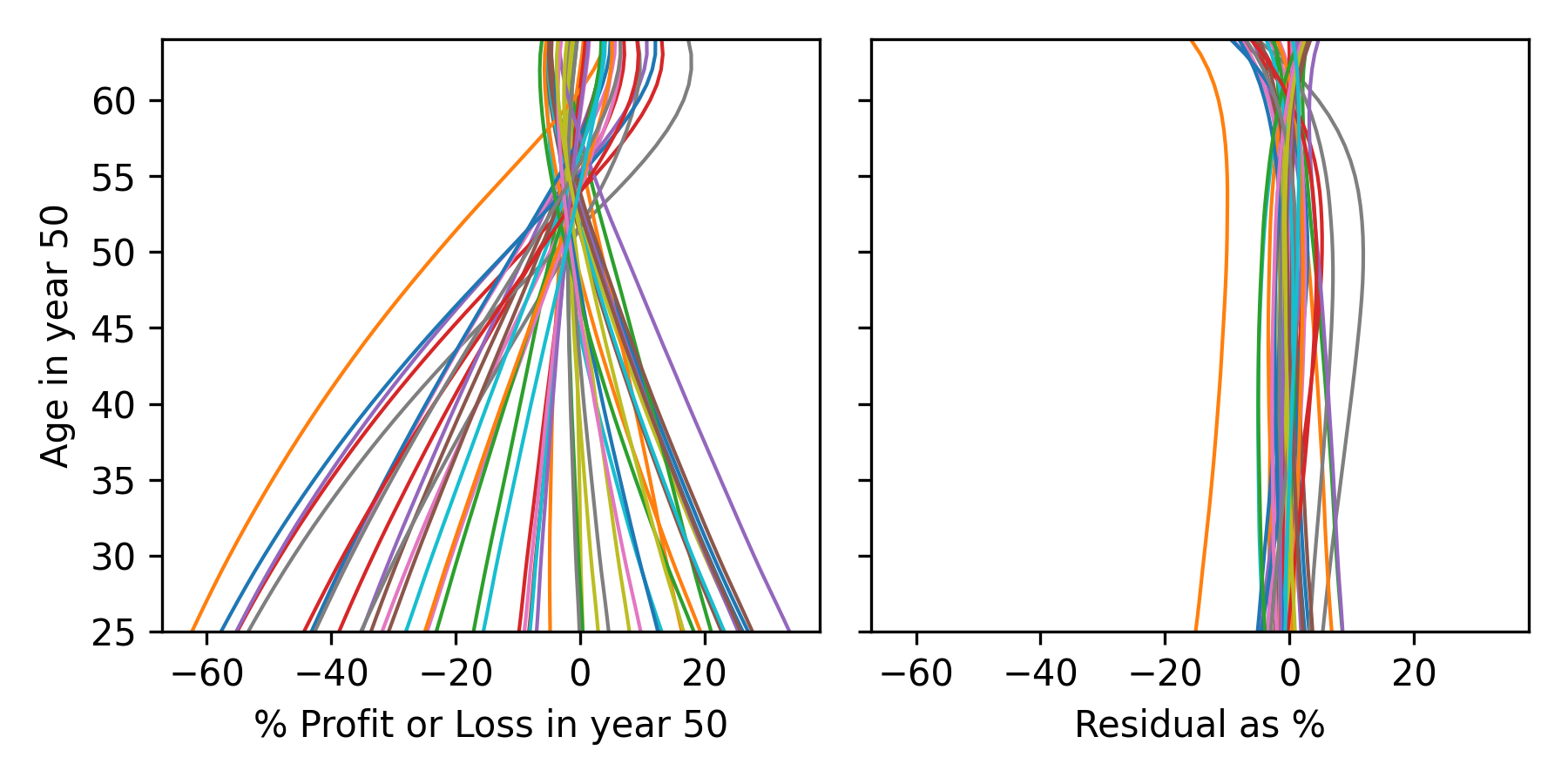}
\caption{Left plot: 50 random scenarios simulated in the physical measure showing the instantaneous profit made by each member at time $t=50$, according to their age. Each curve represents a different scenario.  Interpolation is used between integer ages to obtain the curves. The profit in each scenario is computed using a nested Monte Carlo simulation in the pricing measure with $50,000$ samples. Right plot: the residual of the model defined by \eqref{eq:linearModel}, expressed as a percentage of $V^\xi_t$, the Monte Carlo estimate of \eqref{eq:defVhat}.}
\label{fig:year50}
\end{centering}
\end{figure}

The instantaneous profit is a random variable, so we should consider more than simply its expected value. In Figure \ref{fig:year50}, we show 50 different scenarios for the
instantaneous profit of members by their age at time $t=50$.
This plot is generated using nested Monte Carlo simulations. We use the physical
measure ${\mathbb P}$ to generate 50 different stock price paths up to time $t=50$, and then perform
a Monte Carlo pricing calculation using the pricing measure ${\mathbb Q}$ to calculate the profit made at time $t=50$.

To understand why equation \eqref{eq:defVhat} is an inaccurate pricing formula,
we need to consider how this heuristic could be justified in the first place.
If one begins by assuming a Black-Scholes model and then computes the value of the benefits, one could then take the limit of that value as the stock volatility tends to 0.
This will give what we call a {\em small-noise estimate} for the value of the benefits. We believe that this
methodology could be developed to allow one to rigorously price derivatives using a constant economic model without introducing arbitrage, and, in particular, would provide
a rigorous basis for the heuristic pricing formula
given in equation \eqref{eq:defVhat}.

If the stock volatility is sufficiently small that the upper and lower barriers for $h$ are never hit, the small-noise approximation might give reasonable
results. However, in practice, the barriers are very likely to be hit and so the long-term average value of $h$ is not determined by the current value of $h$, but by the choice of barriers. Since \eqref{eq:defVhat} depends upon values of $h$ projected a considerable time into the future, we cannot expect the small-noise approximation
to be accurate. As a result, one expects that when $h_t$ is high, equation \eqref{eq:liability} will overprice benefits.  When $h_t$ is low, it will underprice benefits.

To test this explanation, and to also obtain a more accurate pricing model, first recall the value
of a unit of nominal benefit at time $t$, $\hat{V}^{\xi}_t$, defined in \eqref{eq:defVhat}.  Let us write $V^{\xi}_t$ for the value computed using our Monte Carlo calculation.
We then perform a linear regression to find a model of the form
\begin{equation}
\log(V^{\xi}_t) \sim c_0 + c_x\, \age(\xi,t) + c_{h} \, h_t + c_{x,h} \, \age(\xi,t) \, h_t,
\label{eq:linearModel}
\end{equation}
using the same 50 scenarios used to obtain Figure \ref{fig:year50}. This gives a total of $2000=50 \times 40$ data
points for the regression.
All coefficients are significant with $p<0.1\%$, and their values are summarized in Table \ref{table:coefficients}.
For comparison, the coefficients for a similar model for
$\log(\hat{V}^\xi_t)$ (defined in equation \eqref{eq:defVhat}), are also shown.
Comparing the coefficient $c_h$, confirms our prediction that when $h_t$ is large, benefits will be overpriced by $\hat{V}^\xi_t$.

\begin{table}
\begin{center}
\begin{tabular}{rrr}
                 & $\log(V^\xi_t)$ & $\log(\hat{V}^\xi_t)$ \\ \hline
$c_0$       &  $-1.952$      & $-1.412$ \\
$c_x$       &  $0.06808$      & $0.05855$  \\
$c_h$       &  $26.06$       & $72.76$  \\
$c_{x,h}$   &  $-0.1597$     & $-1.002$ 
\end{tabular}
\end{center}
\caption{Estimated coefficients of the linear model \eqref{eq:linearModel}
computed for both the empirical value of 1 unit
of nominal benefit,
$\log(V^\xi_t)$, and the discounting-formula estimate, $\log(\hat{V}^\xi_t)$.}
\label{table:coefficients}
\end{table}

The $R^2$-statistic for the model for $\log(V^\xi_t)$ is $0.9984$. A plot of the residuals as a percentage of the values of $V^\xi_t$ is shown in Figure \ref{fig:year50}. We conclude that estimating $V^{\xi}_t$ using equation \eqref{eq:linearModel} will be considerably more accurate than using the estimate $\hat{V}^\xi_t$. The 
$R^2$-statistic for the model for $\log(\hat{V}^\xi_t)$ is $1.0000$.

Treating age as a categorical variable did not significantly reduce the maximum value of the residuals, so we conclude that the benefits of using a more complex model will be limited. A precise calculation of the value of the benefits would be path dependent and so there is a limit on what can be achieved using only the age and $h_t$ as input.

\subsubsection{Statistically-calibrated accrual}

It is natural to consider pricing benefits using
the statistical model given by the coefficients in Table \ref{table:coefficients}, instead of using the liability value in \eqref{eq:liability}. One cannot simply charge the amount
given by $V^\xi_t$ in equation \eqref{eq:linearModel}, as then equation \eqref{eq:after}
will be violated. Instead, one can charge individuals an amount $c\, V^\xi_t$  
where the constant $c$ is chosen to ensure that equation \eqref{eq:after} holds.
We refer to this method of pricing benefits as {\em statistically-calibrated} dynamic-accrual.

As one might expect, this succeeds in reducing the average intergenerational cross-subsidies, as shown in Figure \ref{fig:surfaceStatistical} in the Appendix.
It also reduces the random pricing errors to a size similar to the residuals of our statistical model.

One potential downfall with this approach is that
in any incomplete market model, there may not be universal agreement on the price of benefits.
However, it seems reasonable to expect that the prices
obtained by using a Black--Scholes approximation to an incomplete market model will yield fairer outcomes
than using the liability formula \eqref{eq:liability}. We see in the sequel
that using the model with coefficients given by Table \ref{table:coefficients}, yields slightly better outcomes in a steady state when we perform the simulation in a more realistic market model; see Figure \ref{fig:medianLifetimeMeans}. 
One could, if desired, attempt to obtain a yet more accurate pricing model by calculating the value of benefits in this new scheme and then recalibrating the linear model.

\section{Simulations in a richer economic model}
\label{sec:esgResults}

\subsection{The economic model}

So far we have used a constant economic model and a Black--Scholes model
for parsimony. In particular, when computing cross-subsidies, the Black-Scholes model allowed us to value benefits unambiguously. In this section, we will evaluate the overall
performance of CDC schemes without considering the value of benefits and this allows us to use a richer economic model.

For the simulations detailed in this section, we simulate
all risk factors using a minor variation of the economic scenario generator (ESG) described in \cite{armstrongMaffraPennanen} (see Appendix \ref{sec:esg} for further details).
This scenario generator
allows one to input views about long-term median values for
different risk-factors and these are chosen as shown in Table
\ref{table:riskFactors}. Although this scenario generator allows one to simulate 
population mortality, we do not use this feature and continue to use
the S1PMA tables produced by the Continuous Mortality Investigation.

To compute projected values of CPI to apply in equation 
\eqref{eq:actuarialvaln}, we use median values as these are
simple to project in our ESG. Our ESG is designed to output index-linked
long-term bond yields and we use these combined with the long-term
median estimate for CPI to determine the discount rate for riskless assets in equation \eqref{eq:actuarialvaln}. We also assumed
that the bond-portfolio is index-linked and chosen to match the liabilities exactly. The returns on the bond portfolio can then be computed by pricing this portfolio using the index-linked long-term bond yields supplied by the ESG. The net result of these assumptions is that if the fund is invested 100\% in bonds, then $h$ will remain constant unless the lower bound on indexation is hit due to changes in CPI.

In our ESG, the stock follows geometric Brownian motion, so we can compute
the predicted mean returns and use this to compute the appropriate
discount rate for risky assets in equation \eqref{eq:actuarialvaln}.

All plots in this section were performed using 100,000 Monte Carlo simulations.

\subsection{Annual increments}

\begin{figure}[!htbp]
\begin{centering}
\begin{tabular}{cc}
\includegraphics[width=0.45\linewidth]{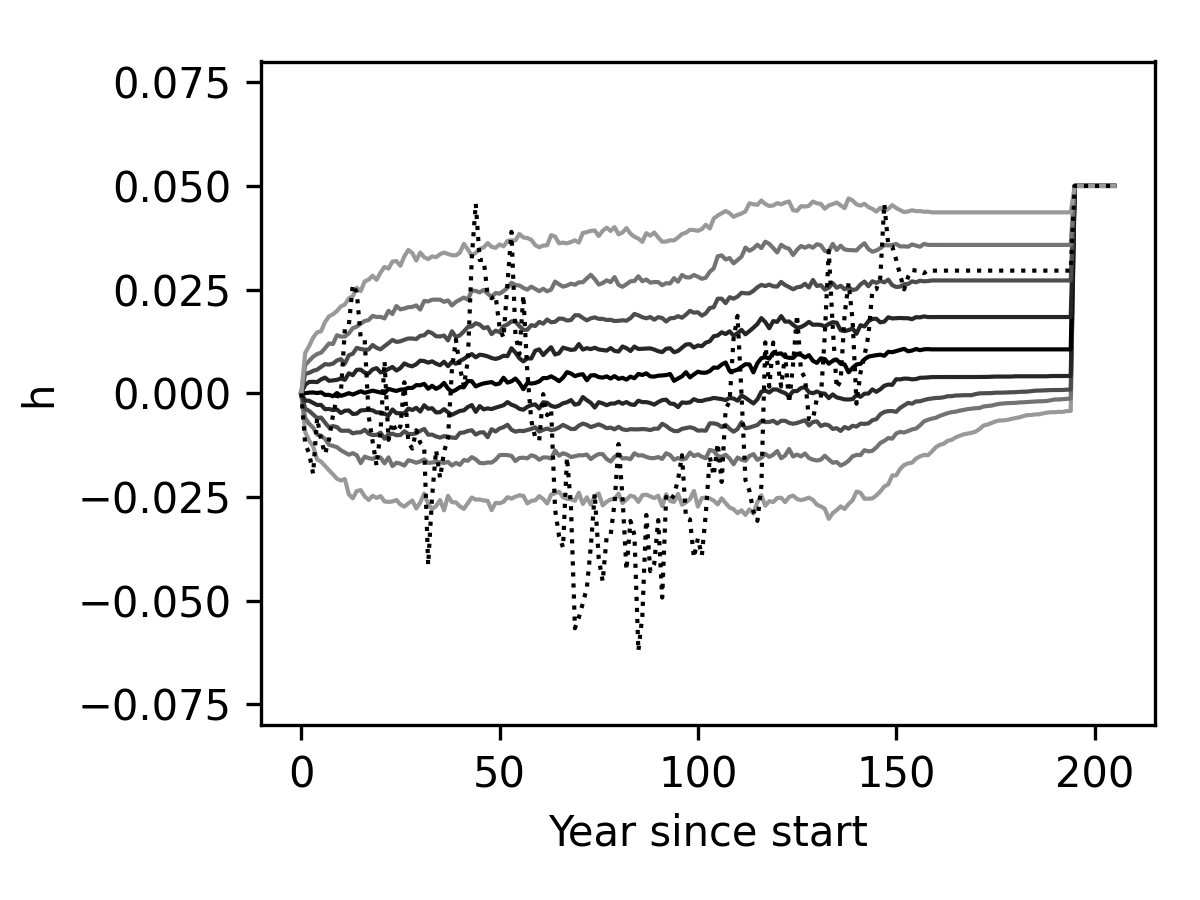} & 
\includegraphics[width=0.45\linewidth]{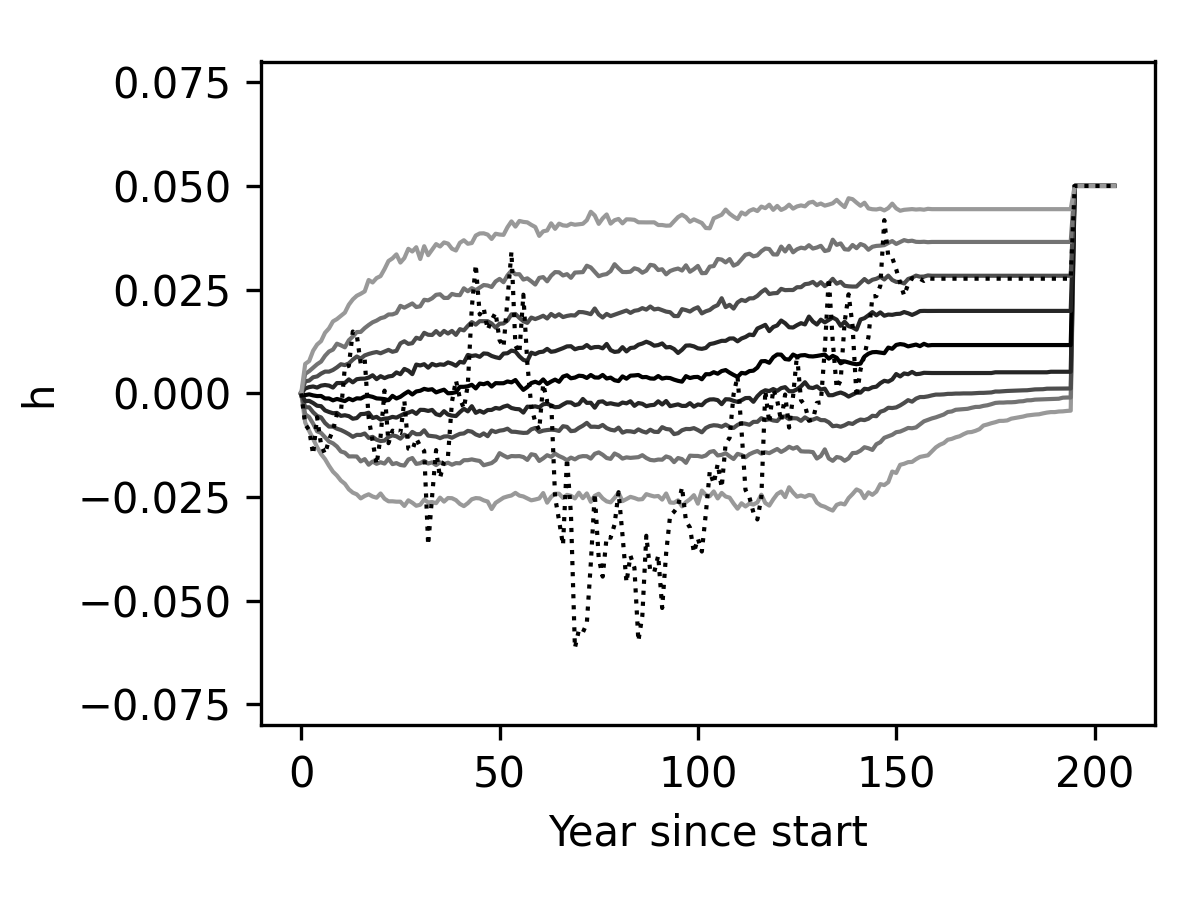} \\
Flat-accrual & Dynamic-accrual
\end{tabular}
\caption{Fan diagrams of the real indexation rate, $h_t$, in each year of the simulation.}
\label{fig:hAgainstTime}
\end{centering}
\end{figure}

In Figure \ref{fig:hAgainstTime}, we plot a fan-diagram of
the real indexation rate $h$, in different years of our simulation. In all fan-diagrams in this paper,
we plot all nine deciles together
with an example scenario to indicate the volatility.

The median value of $h$
for the flat-accrual scheme is approximately zero until the closure
of the scheme.  At the end of the scheme's lifetime, it is invested entirely in risk-free assets, so $h$ will remain constant unless a benefit cut occurs due to changes in inflation pushing the total $h+\DCPI$ below zero. A bonus is not possible because these are determined by the value of $h$, rather than $h+\DCPI$. This is why, at the closure of the scheme, the lower percentiles in the distribution of $h$ begin to increase. The constancy of $h$, excluding benefit cuts, is a consequence of our decision to model a fund with an infinitely large membership. For finite populations, $h$ becomes increasingly volatile at scheme closure as liabilities cannot be perfectly hedged and due to the short duration of these liabilities; this requires comparatively large adjustments in $h$.

The dynamic-accrual scheme behaves similarly. As the 95th percentiles are very close to the cap and floor levels, it seems that bonuses and benefit cuts each occur on the order of once every twenty years, excluding the period after scheme closure.

\begin{figure}[!htbp]
\begin{centering}
\begin{tabular}{cc}
\includegraphics[width=0.45\linewidth]{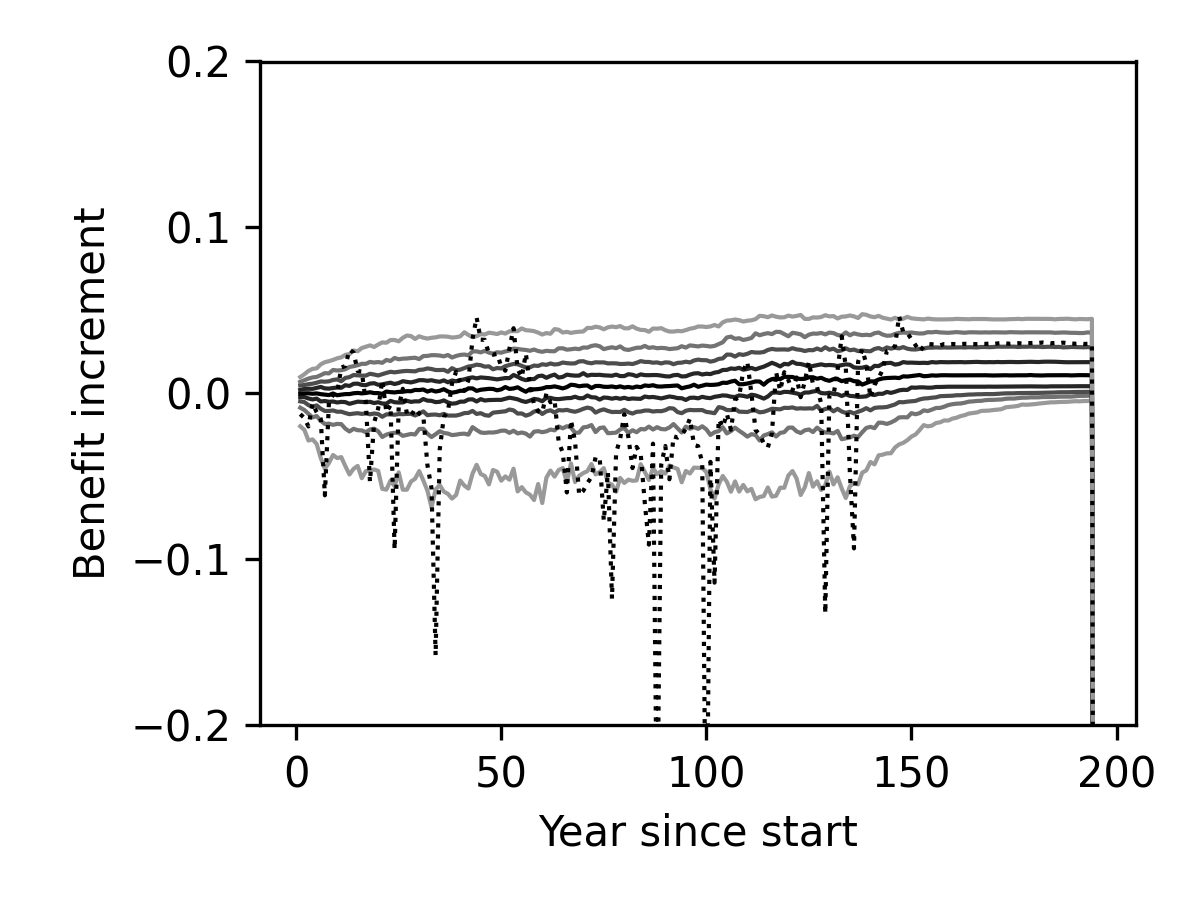} & 
\includegraphics[width=0.45\linewidth]{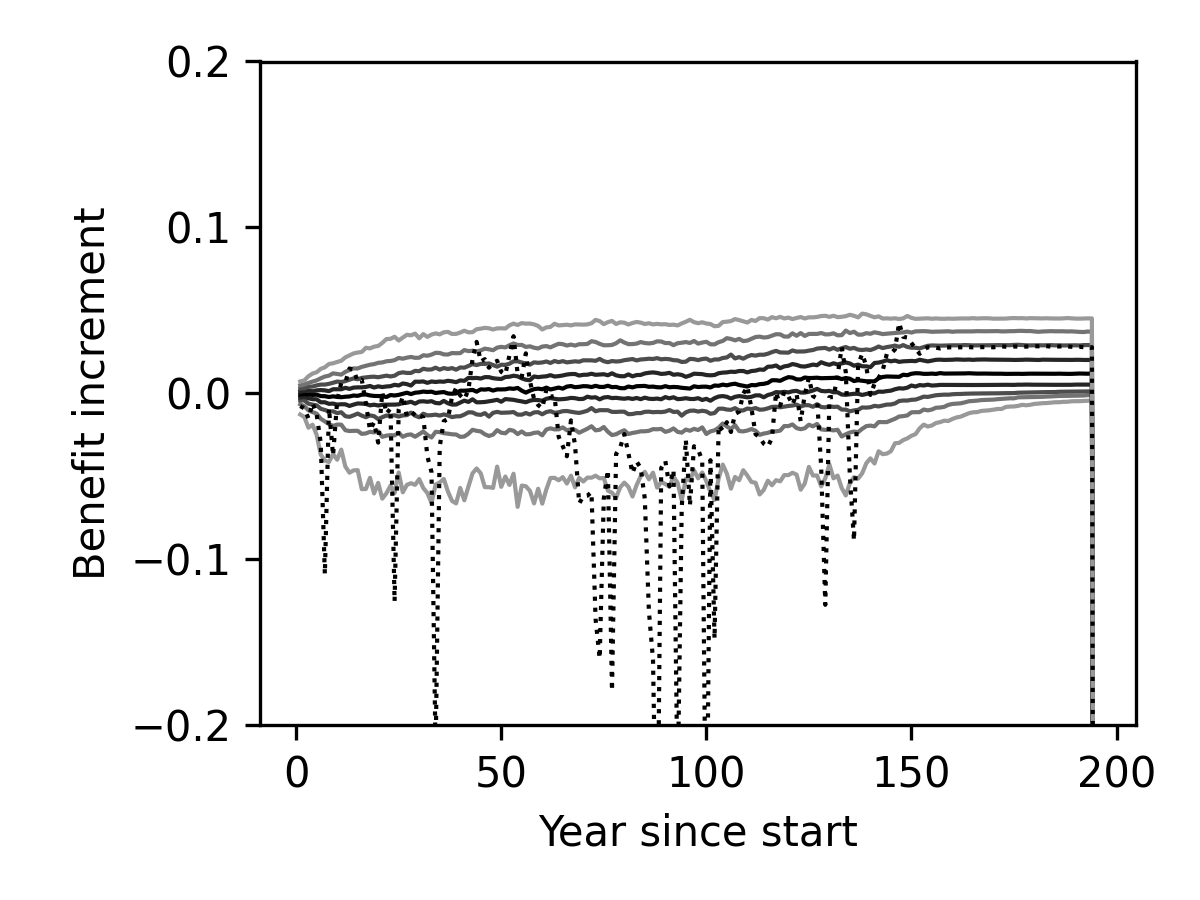} \\
Flat-accrual & Dynamic-accrual
\end{tabular}
\caption{Fan diagrams of the benefit increase/decrease over inflation 
including bonuses/cuts in each year of the simulation.}
\label{fig:incrementAgainstTime}
\end{centering}
\end{figure}

The indexation level $h$, does not include information about cuts or bonuses. In Figure \ref{fig:incrementAgainstTime}, we plot the
actual increase or decrease to the nominal benefit amount each year
including cuts and bonuses. One sees that the impact of cuts and bonuses
are significant.

\subsection{Pensions for a typical generation}

\begin{figure}[h!tbp]
\begin{centering}
\begin{tabular}{cc}
\includegraphics[width=0.45\linewidth]{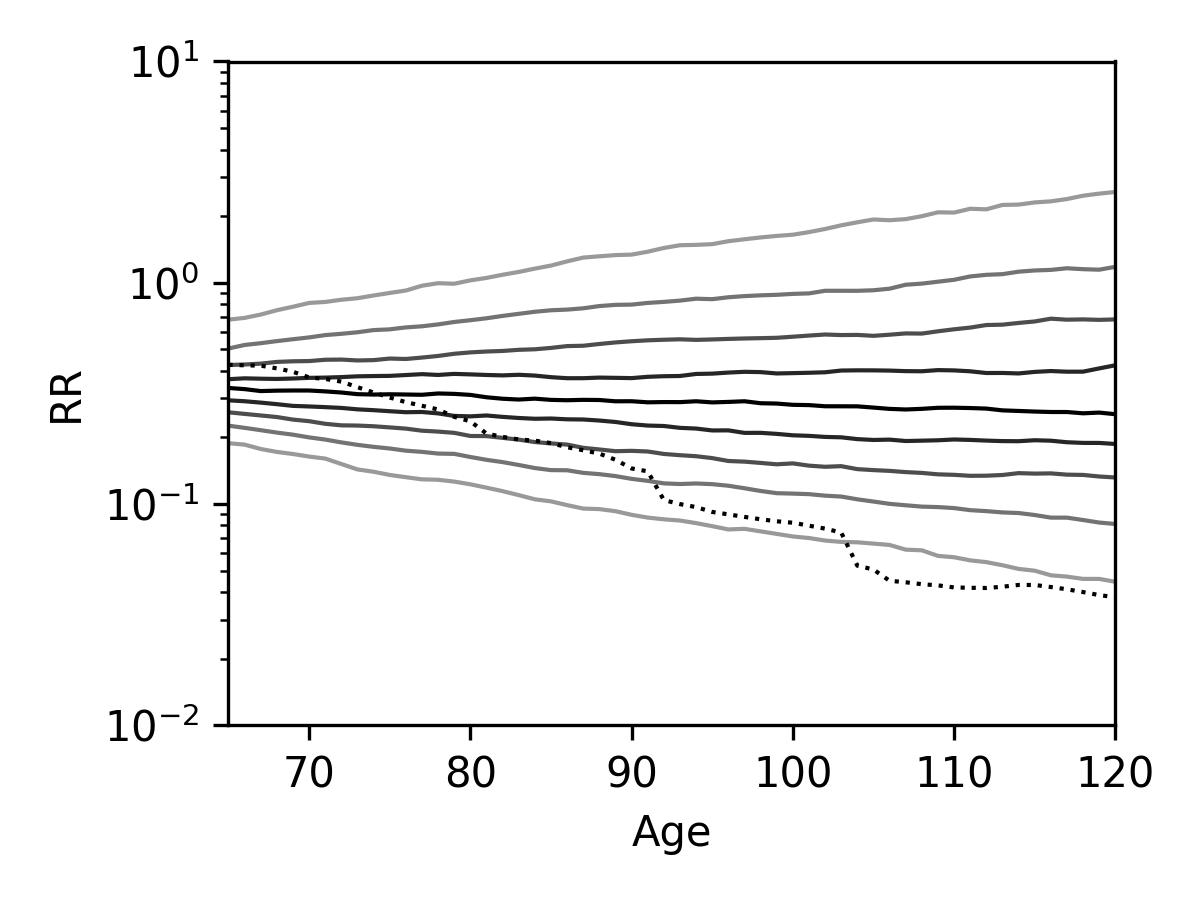} & 
\includegraphics[width=0.45\linewidth]{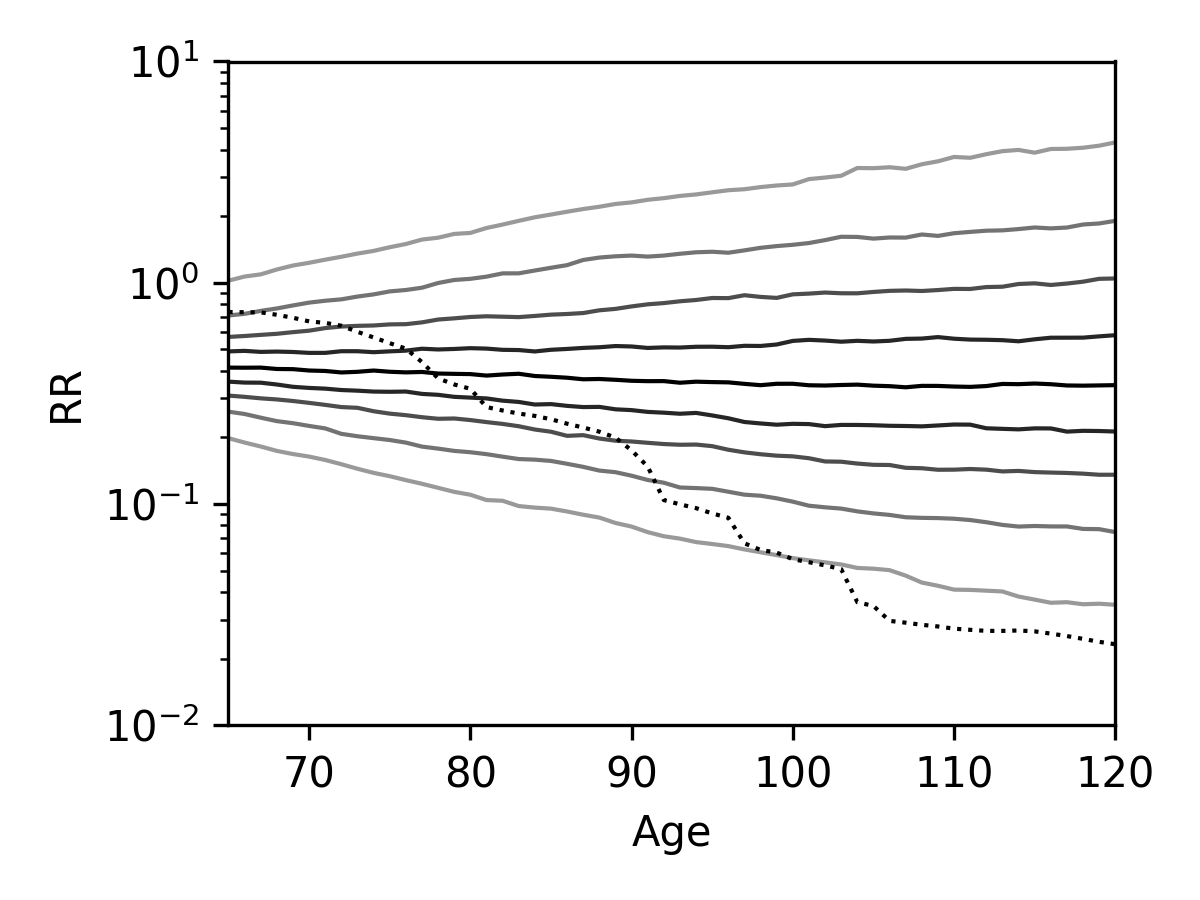} \\
Flat-accrual & Dynamic-accrual \\
\includegraphics[width=0.45\linewidth]{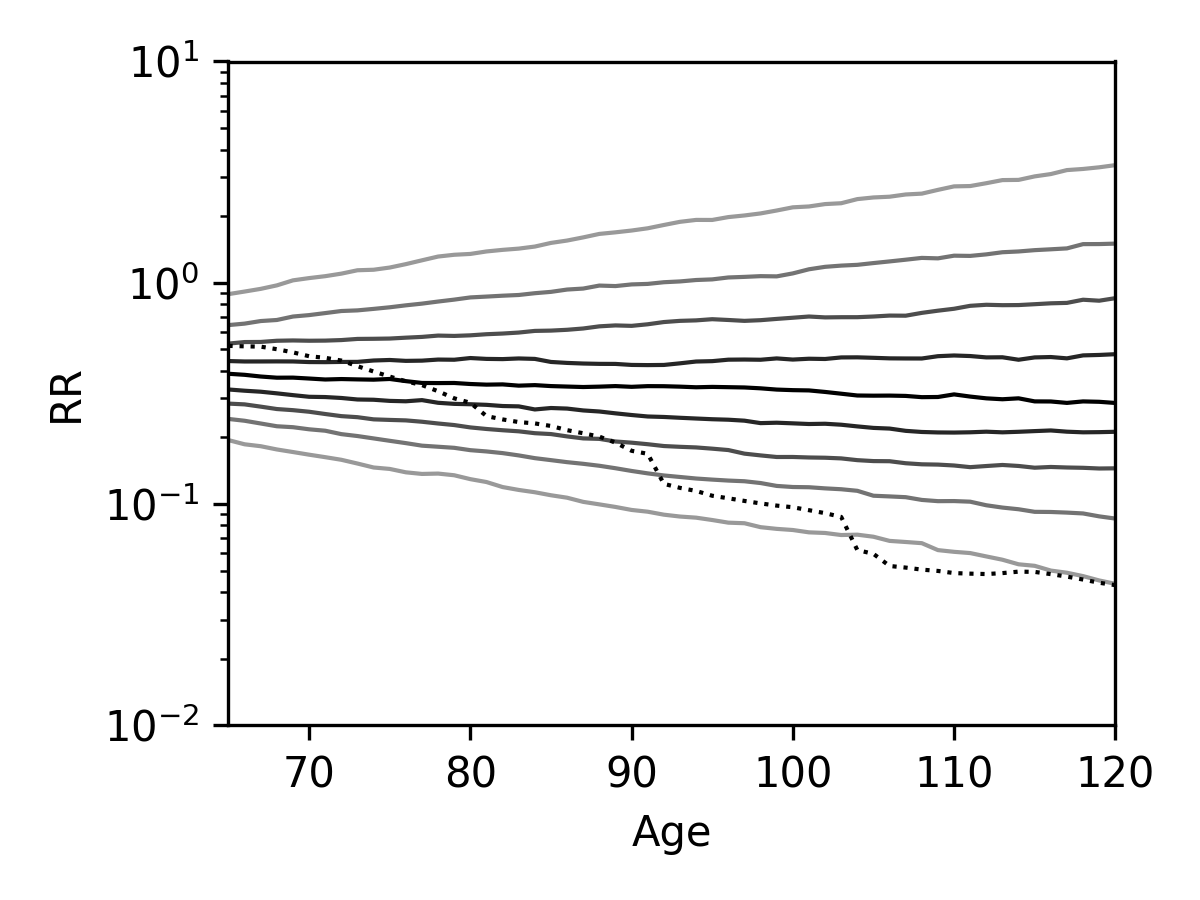} & 
\includegraphics[width=0.45\linewidth]{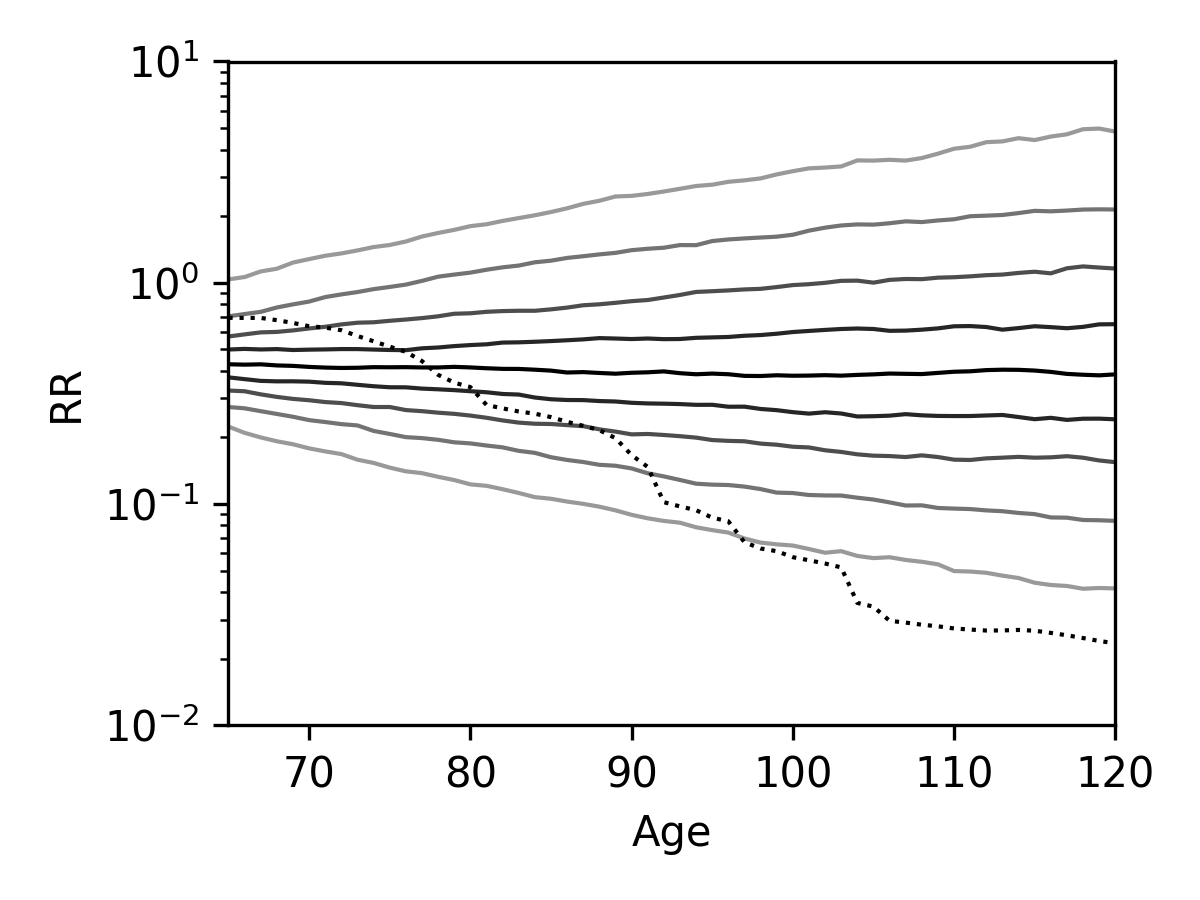} \\
Age-based accrual & Statistically calibrated accrual \\
\includegraphics[width=0.45\linewidth]{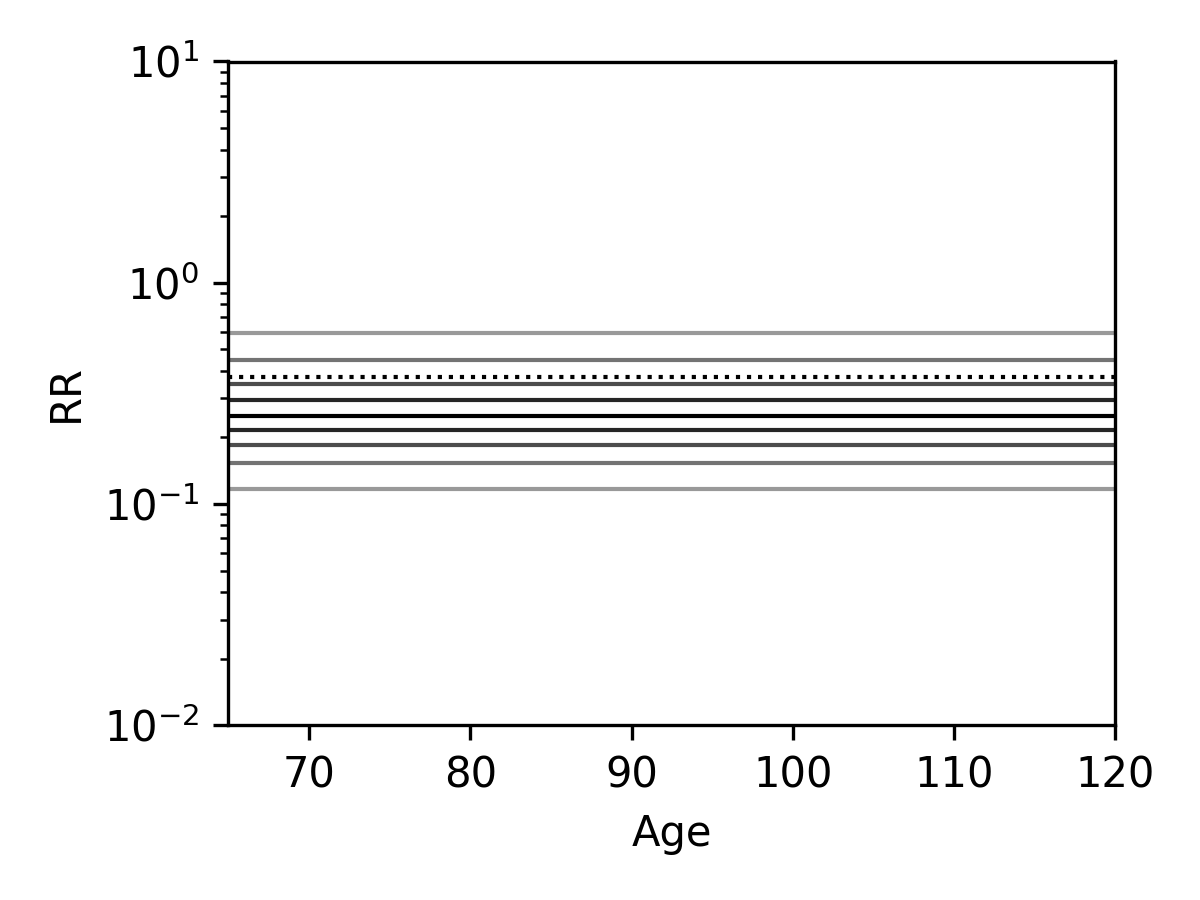} & 
\includegraphics[width=0.45\linewidth]{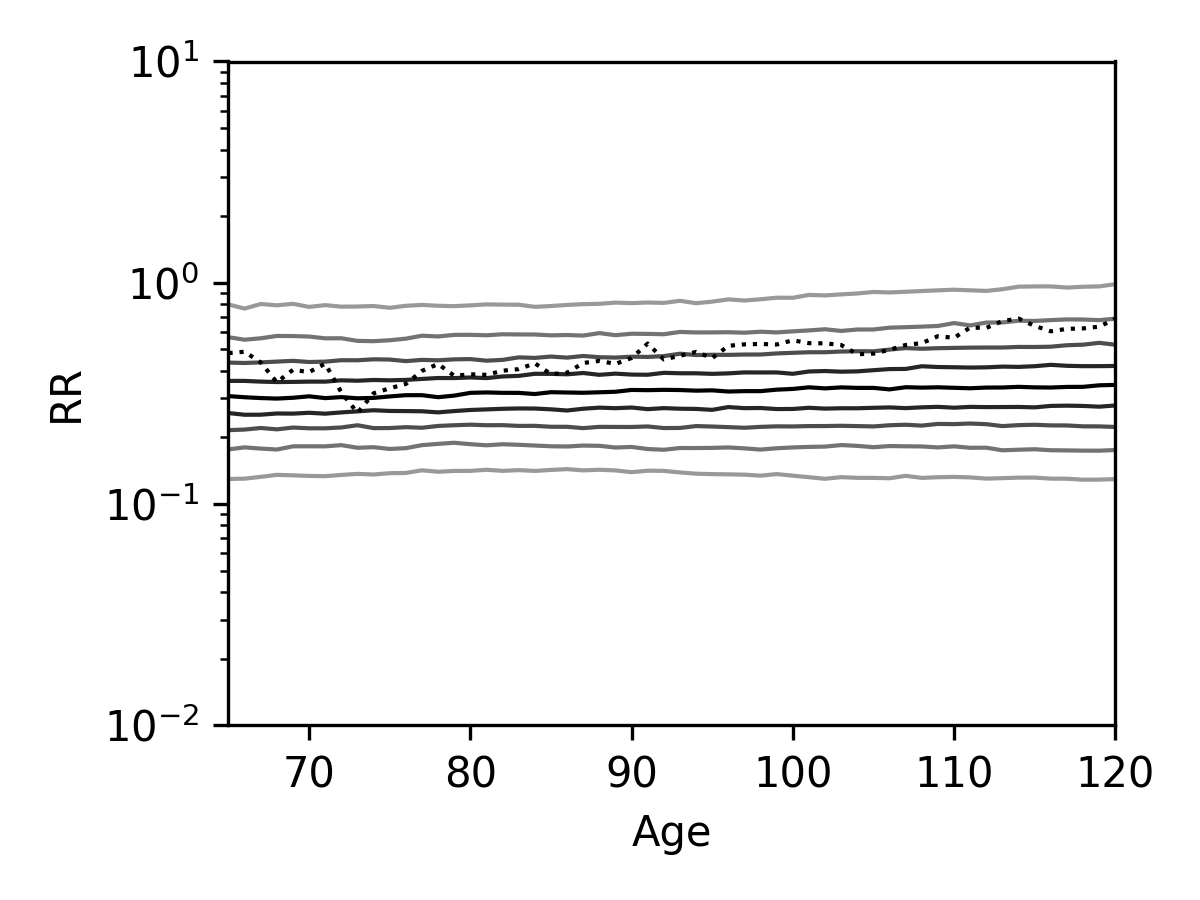} \\
DC + Annuity & Pooled annuity fund \\
\end{tabular}
\caption{Fan diagrams of the log replacement ratio by age for the 60th generation.}
\label{fig:generation60}
\end{centering}
\end{figure}

\begin{figure}[h!tbp]
\begin{centering}
\includegraphics[width=0.45\linewidth]{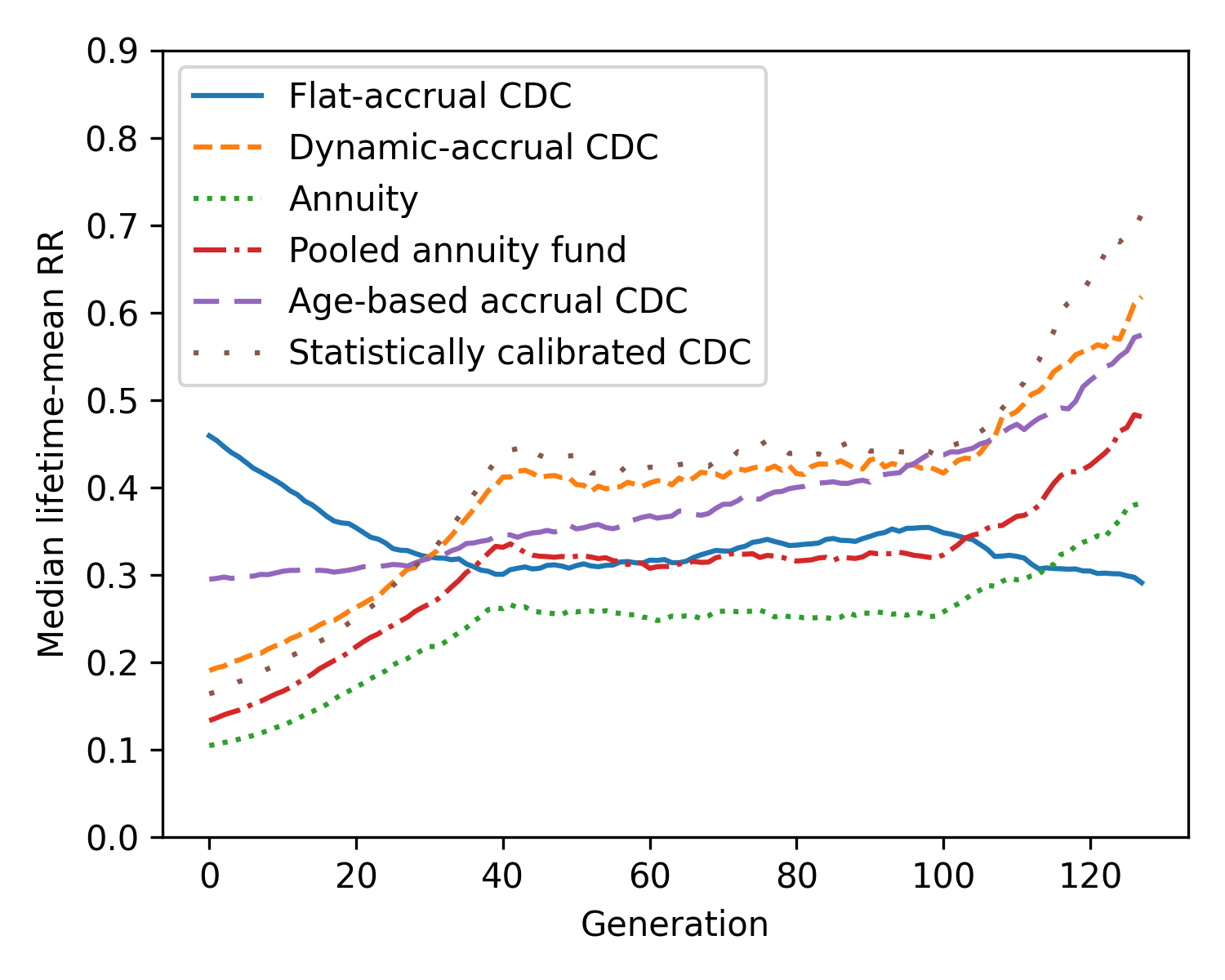}
\includegraphics[width=0.45\linewidth]{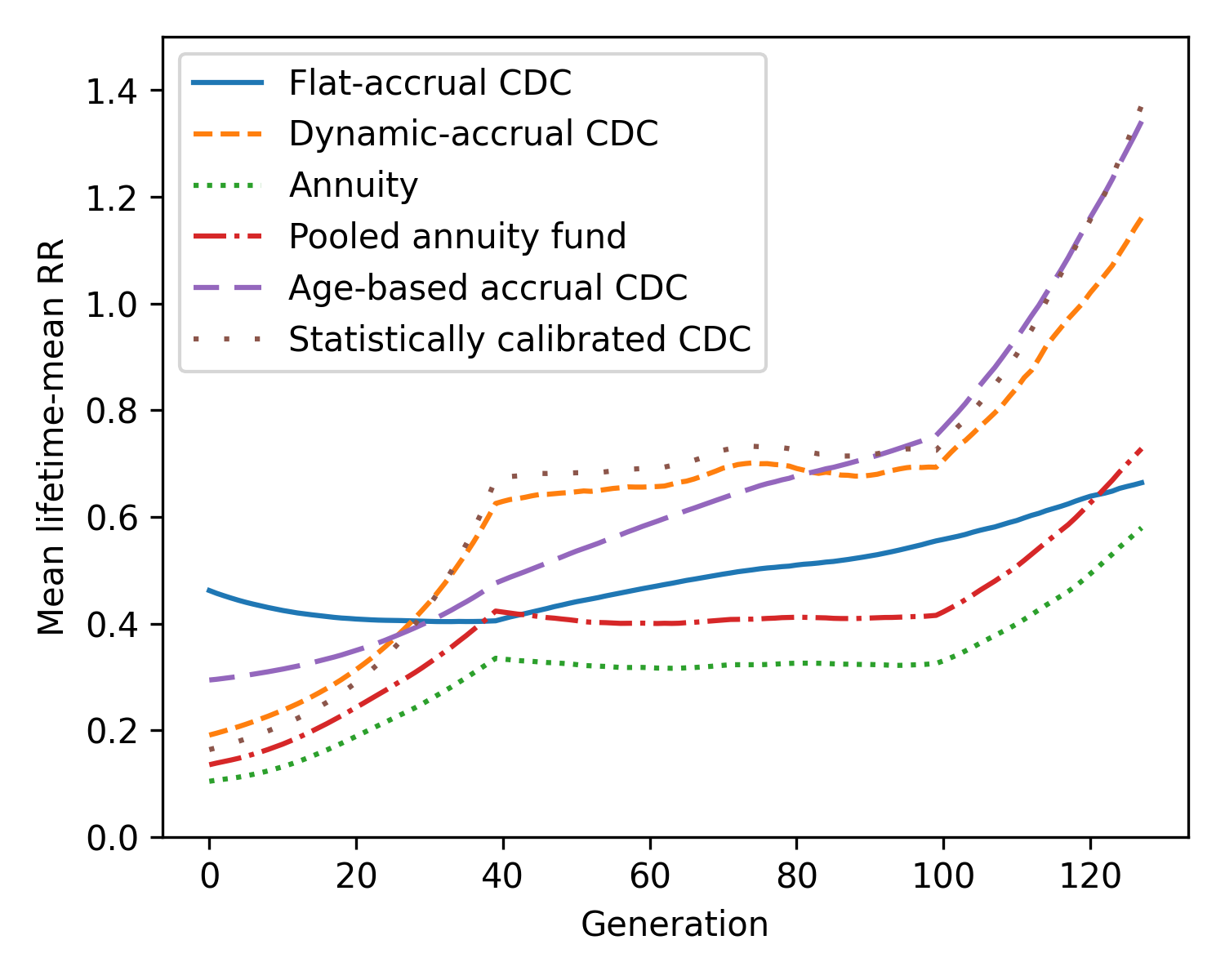}
\caption{
Left plot: Median lifetime-mean replacement ratio. Right plot: Mean lifetime-mean replacement ratio.  Both plots are generated using our economic scenario generator and a target of $\DCPI+0\%$ for the CDC funds. 
}
\label{fig:medianLifetimeMeans}
\end{centering}
\end{figure}

We define the {\em replacement ratio} to be the ratio between
the annual benefits at a given time during retirement and the annual salary rate exactly one year before retirement, indexing the salary with CPI to the given time.
Figure \ref{fig:generation60} shows fan-diagrams of the logarithm of the replacement ratio for the 60th generation in a number of different schemes. The 60th generation is chosen as they have
retired by the time the fund closes, but the fund has had sufficient time to settle into an approximate ``steady state''. In fact, as can be seen from Figure \ref{fig:medianLifetimeMeans}, the CDC funds never truly enter a steady-state, but nevertheless broadly similar outcomes are experienced by generations 60 to 80.

We see that in a flat-accrual CDC scheme targeting $\DCPI+0\%$, the median replacement ratio is close to constant in retirement, but the width of the fan-diagram continues to increase throughout retirement. Since
the median income is close to constant in real terms, the CDC scheme can be viewed as on average achieving the target.

In the dynamic-accrual CDC scheme, median real-terms pensions remain constant over time and are somewhat higher than for the flat-accrual scheme. The dynamic-accrual scheme appears to straightforwardly outperform the flat-accrual scheme in the sense that for all the plotted percentiles, the dynamic-accrual scheme always gives a better outcome.

The age-based accrual scheme introduced in Section \ref{sec:ageBased} performs somewhere between flat- and dynamic-accrual schemes. Although it is not clear from this picture, we will see later that the statistically-calibrated scheme slightly outperforms the standard dynamic-accrual scheme.

For comparison, we show the income in retirement of a DC-plus-annuity scheme. In this scheme, individuals in a DC scheme invest 100\% of their assets in the risky asset until age 55, and nothing in the riskless asset.  They then taper linearly to 0\% in the risky asset at age 65, when they purchase an index-linked single life annuity. The price of the annuity is calculated by using the discount rate for long-term bonds together with the SP1MA mortality distribution used in the simulation. An additional charge of $5\%$ is then applied to represent the charges that would need to be levied by the annuity provider to account for systematic longevity risk.

As another comparison, we show the income in retirement of a pooled annuity fund. In this scheme, the fund invests at 100\% in the risky asset until age 55, tapering down to 33\% at age 65. Each year in retirement, the assets of anyone who dies are shared with the survivors. To compute the consumption rate, suppose that the expected return on assets in a given year is $R$. Compute the cost of a level single-life annuity which pays one unit at the start of each year using the mortality distribution used in the simulation and the discount rate $R$. The pension paid out from the pooled annuity fund is equal to the value of the assets in the fund divided by the life annuity cost. As Figure
\ref{fig:generation60} shows, this scheme results in an approximately
constant real terms income in retirement, but is able to offer a slightly better income than a DC-plus-annuity scheme. The pooled annuity fund benefits
from lower costs (as there is no need to charge for systematic longevity risk) and higher returns (as there is some investment in the risky asset), but this does result in some annual fluctuations in pension income.
The pooled annuity fund and the DC fund with annuitisation both show lower median outcomes than the flat- and dynamic-accrual schemes, but do have less risk later in retirement.

Let $\zeta^\xi_t$ denote the replacement ratio of the
pension received by individuals of type $\xi$ at time $t$.
We define the lifetime-mean
replacement ratio for individuals of type $\xi$ as
\[
\left(\frac{\sum_t {\mathbf 1}^{C,\xi}_t}
{\NRA-x} \right)
\frac{\sum_t N^{\xi}_t \, {\zeta}^\xi_t \, {\mathbf 1}^{R,\xi}_t}
{\sum_t N^{\xi}_t \, {\mathbf 1}^{R,\xi}_t}.
\]
The first factor scales the value according to the number of years for which a generation has contributed in order to facilitate
comparison between generations.

A plot of the median lifetime-mean replacement ratio for each generation
is given in Figure \ref{fig:medianLifetimeMeans}, together with a plot of the mean lifetime-mean replacement ratio.  Assuming that investors are risk-averse, the median lifetime-mean replacement ratio seems a better measure of the quality of a pension.

Using the median lifetime-mean as a metric in the flat-accrual scheme, the earliest generations receive a proportionately better pension than those when the scheme is fully mature, and the final generations receive a worse pension. The dynamic-accrual scheme works in a manner which is closer to being actuarially fair: the earliest generations receive a proportionately lower pension as their investments have not had time to grow; the latest generations receive a proportionately larger pension as all their payments were made early in their employment before the scheme was closed in year 100.
One can see the behaviour of the dynamic-accrual scheme is similar to that of the pooled annuity fund and the DC+annuity scheme, both of which are actuarially fair.

When the scheme is fully mature, the dynamic-accrual CDC scheme gives a higher median lifetime-mean replacement ratio than the flat-accrual scheme. This can be attributed to the drag effect discussed in Section \ref{sec:infiniteHorizon}.

The age-based accrual scheme is intended to approximately replicate the level of cross-subsidies seen in a DB scheme. The first generation receive a lower median lifetime-mean replacement ratio than those in the steady state of the scheme, but this will be compensated for by the fact that their pension income is less risky.
As a result of reducing the cross-subsidies, the drag effect is reduced. This confirms the intuition that it is overpaying the earliest generations that causes comparative underperformance of flat-accrual CDC.

The statistically-calibrated scheme contains slightly less intergenerational cross-subsidy than the standard dynamic-accrual scheme and this results in slightly better performance in the steady state of the scheme.

As a simple numerical comparison, we give the median lifetime-mean replacement ratio for the 60th generation. This is 32.6\% for the flat-accrual CDC scheme, 42.6\% for the dynamic-accrual CDC scheme, 26.3\% for the DC-plus-annuity scheme, and 33.0\% for the pooled annuity fund. By this metric, the flat-accrual scheme outperformed DC and annuity by a factor of  23\%.

\section{Smoothing}
\label{sec:smoothing}

CDC funds transfer the fluctuations in predicted pension incomes from older generations to younger generations. The idea is that the fund can then invest a greater proportion in risky assets without this resulting in large fluctuations in pension income for those who have already retired.

Let us call the partial derivative of the log of the current market value of an individual's benefit with respect to $h$, the {\em $h$-duration} of the benefit.

\[\text{$h$-duration}:=\frac{\partial \log(V^\xi_t)}{\partial h}\] 
This definition
mirrors that of the modified duration of a bond.
We can calculate the $h$-duration of a member's benefit using the coefficients in Table 
\ref{table:coefficients}, and we see that $h$-duration increases with time to retirement. Intuitively this occurs for the same reason that a bond's duration increases with time to maturity.

In a CDC fund in years with no benefit cuts, fluctuations in asset values are accounted for by changing $h$. As a result, the higher the $h$-duration of a member's benefits, the more sensitive it will be to changes in the asset price. This means that individuals who are close to retirement will experience less fluctuations in their pension income than individuals who are a long way from retirement. This is a key design goal of shared-indexation CDC as it results in smoother pensions in retirement by shifting the risk of asset volatility from older to younger members.

Notice that the coefficient $c_{x,h}$ in our linear models, is much lower in the model for $\log(V^\xi_t)$
than it is in the model for $\log(\hat{V}^\xi_t)$. This means that if one attempts to estimate how much smoothing will occur by looking at the sensitivity of liabilities to $h$, one will significantly over-estimate the amount of smoothing that occurs.

\begin{figure}[htbp!]
\begin{center}
\begin{tabular}{cc}
\includegraphics[width=0.45\linewidth]{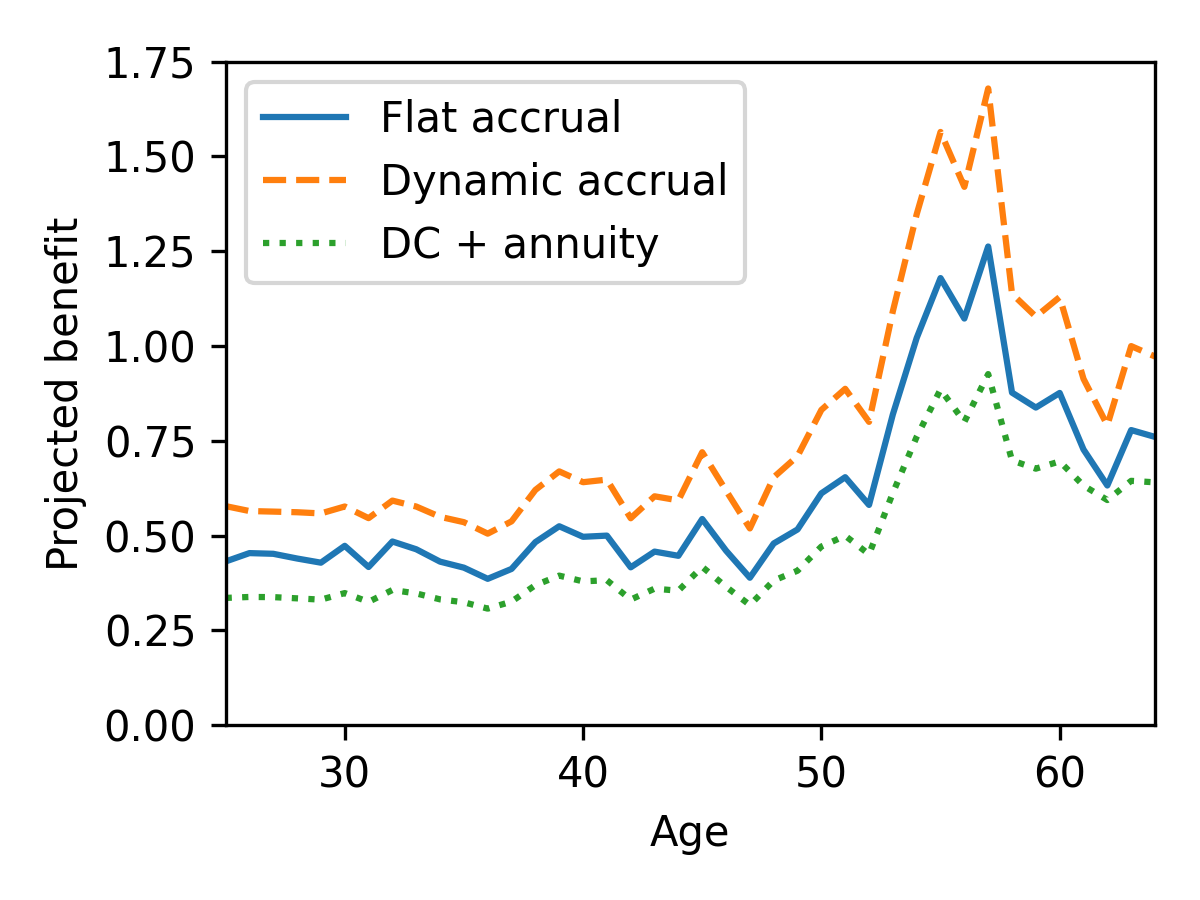} & 
\includegraphics[width=0.45\linewidth]{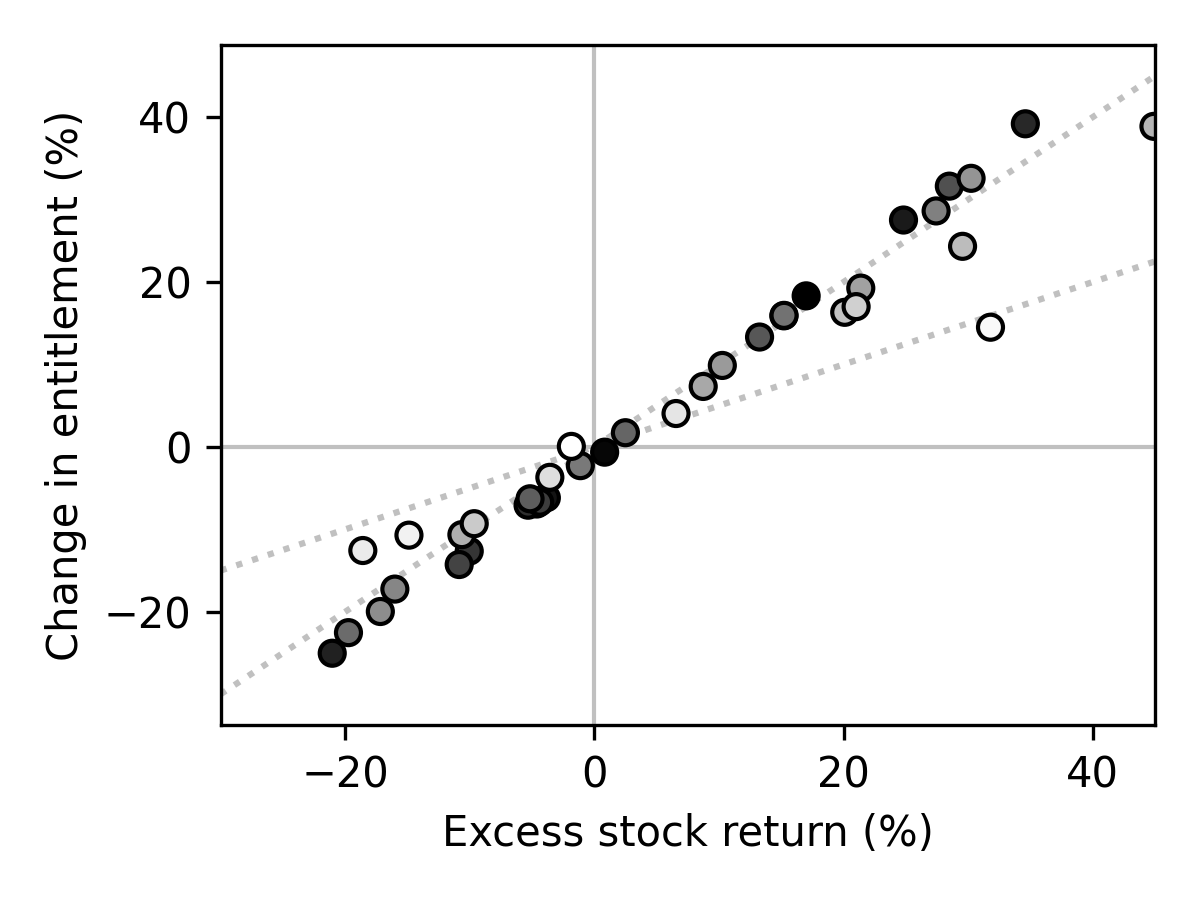} \\
(a) & (b)
\end{tabular}
\end{center}
\caption{(a) Projected mean lifetime-mean replacement ratio for one typical cohort against age. (b)
Annual fluctuation in projected mean benefit entitlement at retirement for the same cohort due to investment returns, plotted against the corresponding excess stock returns. Age is indicated by the shading of the points, with darker points used when the cohort was young. The lines through the origin with gradient $1$ and $\tfrac{1}{2}$ are shown for reference.}
\label{fig:smoothing}
\end{figure}

Figure \ref{fig:smoothing} (a) illustrates an example scenario which shows how the projected mean lifetime-mean replacement ratio varies for one cohort over their lifetime. The roughness of the two shared-indexation designs is very similar to that of a DC+annuity scheme. The latter is somewhat smoother in the final years before retirement due to the lifestyling strategy. Figure \ref{fig:smoothing} (b) considers the same scenarios in the dynamic-accrual scheme and shows how the projected accrued benefit entitlement at retirement fluctuates with the stock price. We see that for younger cohorts, their projected benefit entitlement fluctuates slightly more than excess stock returns.  For the oldest cohort, the projected benefit entitlement fluctuates approximately half as much as excess stock returns. We conclude that shared-indexation schemes do provide some smoothing of pension incomes before retirement, but it is a relatively modest effect compared to some estimates that have been made in the literature. For example, in \cite{catherineDonnelly},
smoothing is explicitly estimated using
such sensitivities with prices estimated
using \eqref{eq:liability}. As a result, this estimates much higher smoothing than is found through simulation.

Aon's report \cite{aon} contains a chart (Chart 5, p.\ 31)
which suggests that in a CDC design similar to our flat-accrual scheme, projected benefits are almost entirely smooth in the run up to retirement. This can be
explained by their study computing projected benefits using a deterministic model rather than a simulation. Our results suggest that this deterministic calculation is likely to be inaccurate. 

Smoothing of benefits may also be over-estimated due to a confusion between nominal benefit entitlements and projected benefit entitlements. Nominal benefit entitlements vary much more smoothly than projected benefit entitlements. This is because the change in projected benefits is a function of both the current nominal benefit and the level of indexation $h_t$. If one only looks at changes in nominal benefits each year, one will miss the changes in $h_t$. As an example of the potential for confusion, \cite{governmentActuaries} uses
the phrase ``projected benefits'' to describe what we call nominal benefits.

Examining the example trajectories plotted in the fan diagrams in Figure \ref{fig:generation60}, we see
that similar levels of smoothing of pension outcomes are achieved across the CDC designs. While the CDC trajectories are smoother than that for the pooled annuity fund, this is a short-term smoothing rather than a long-term decrease in risk. As indicated by the funnel shapes of the CDC diagrams, risk increases significantly throughout retirement in our CDC funds.
The reports \cite{popat} and \cite{owadally}
both seek to quantify the risk of CDC investments,
but only measure the standard deviation of incomes
at the time of retirement. As a result, these studies will have tended to understate the risks of CDC schemes.

\section{Changing the target level}
\label{sec:cap2}

So far we have considered the flat-accrual scheme with a target indexation level of $\DCPI+0\%$. This aids
comparisons with other schemes that target a constant income in retirement. However, to avoid
frequent benefit cuts, one can target a higher indexation level, though this requires increasing
contributions.

\begin{figure}[!htbp]
\begin{centering}
\begin{tabular}{cc}
\includegraphics[width=0.45\linewidth]{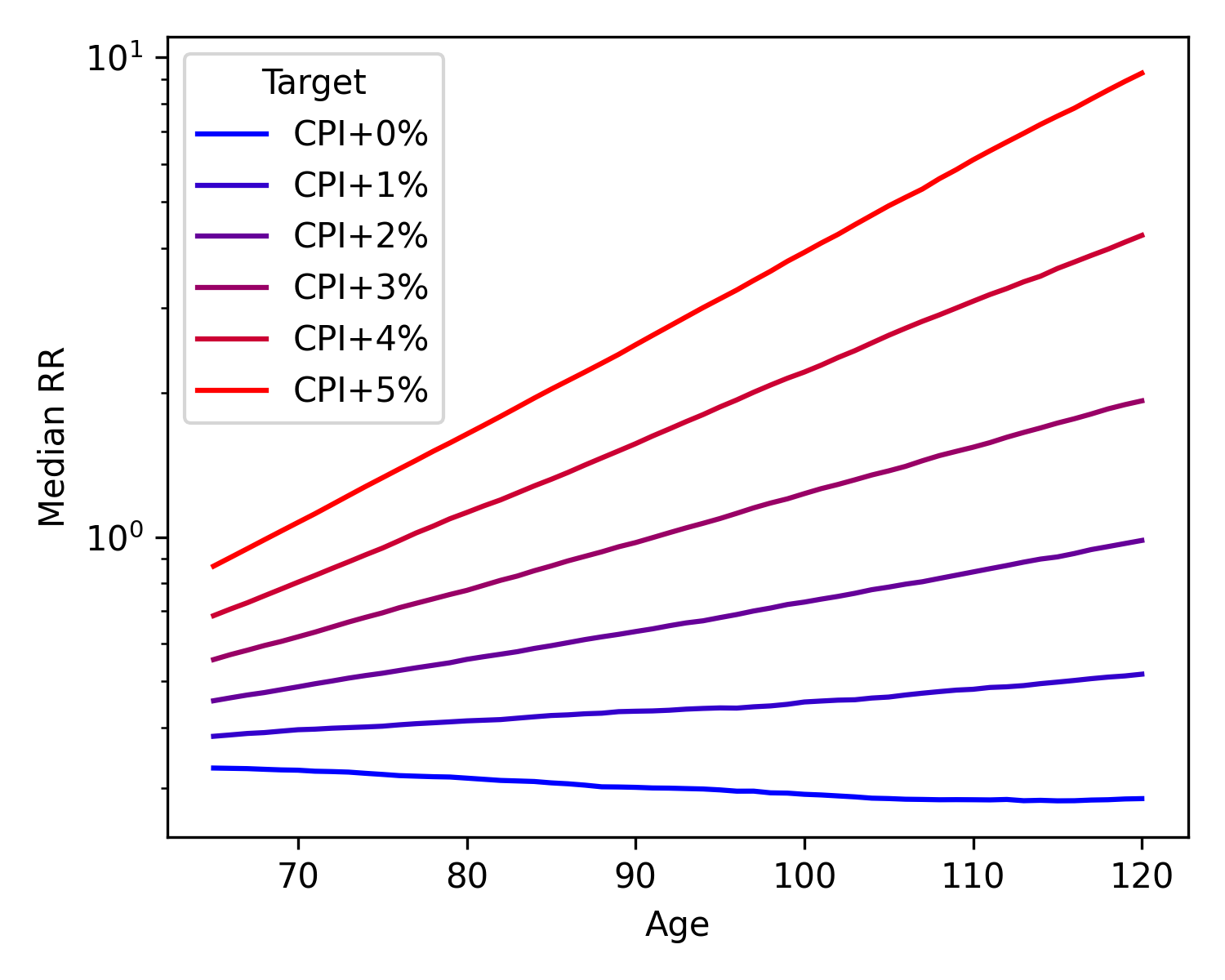} & 
\includegraphics[width=0.45\linewidth]{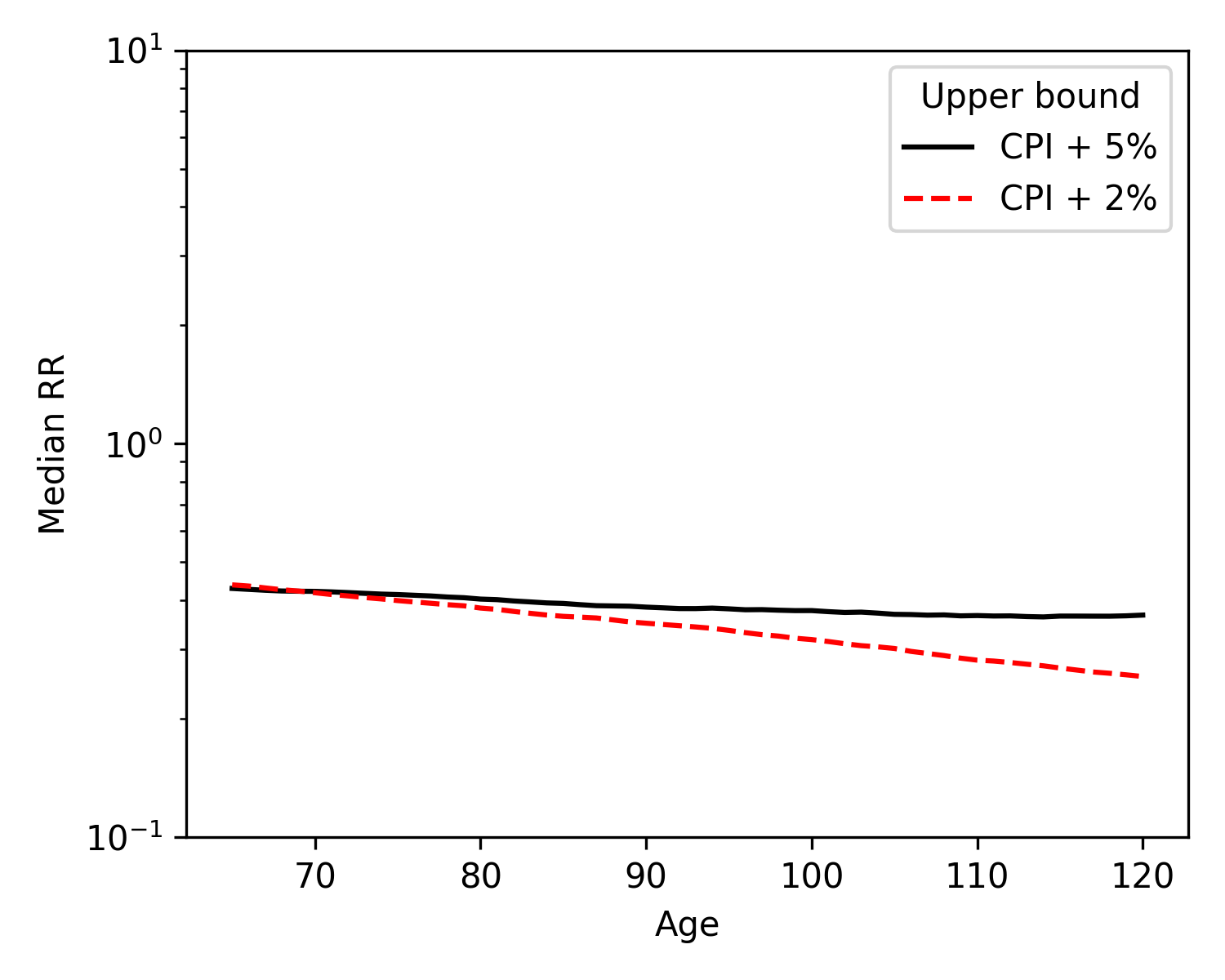}
\end{tabular}
\caption{Plots of median income (log scale) against age for generation 60. Left plot: for a flat-accrual scheme with the target level of indexation being varied. Right plot: for a dynamic-accrual scheme with the level at which bonuses are made being varied.}
\label{fig:changingtargets}
\end{centering}
\end{figure}

We plot the results of simulations targeting different levels of real indexation rate in a flat-accrual scheme (left plot of Figure \ref{fig:changingtargets}).
As we increase the target level, we also increase the level at which bonuses are awarded by the same amount. When targetting $\DCPI+0\%$ the median retirement income decreases throughout retirement. If the median income in retirement were to increase at a rate of $1\%$ per year, it should increase by a total of about 73\% from age 65 to 120. Thus, we see that the target levels do not correspond to the achieved rate of increase of median incomes in retirement. We note, however, that the method we are using to target a particular level of indexation is just a heuristic. 

These results show that it would be better to use a line search over simulations to identify the required contribution rate. Since one can achieve different levels of indexation by varying the ratio of benefits to contributions, one should not be overly concerned that the heuristic is imperfect.

As the dynamic-accrual scheme does not incorporate a target level, only an initial level $h_0$, we cannot adjust
the pension increments in a dynamic-accrual scheme in the same way.
 If one wishes to choose a long-term target for indexation, one can attempt to do this by adjusting the boundaries for cuts and bonuses or by adjusting the proportion invested in risky assets. The former option may be unattractive as a fund currently targeting $\DCPI+5\%$ will appear very expensive to investors. The latter may be unattractive as changing this will simultaneously change the risk profile of the fund.
We have taken the approach of varying the boundary for bonuses in our simulations. The right plot of Figure \ref{fig:changingtargets} shows
the median income in retirement for a dynamic-accrual scheme when indexation is capped at $\DCPI + 2\%$ and when it is capped at $\DCPI+5\%$, plotted against age. As this plot shows,
even if we have an upper bound of $\DCPI+5\%$ on indexation, the median income in retirement in the dynamic-accrual scheme does not quite manage to keep up with inflation.

\section{Scheme behaviour when investment returns are not as expected}
\label{sec:modelRisk}

\begin{figure}
\begin{centering}
\includegraphics[width=0.45\linewidth]{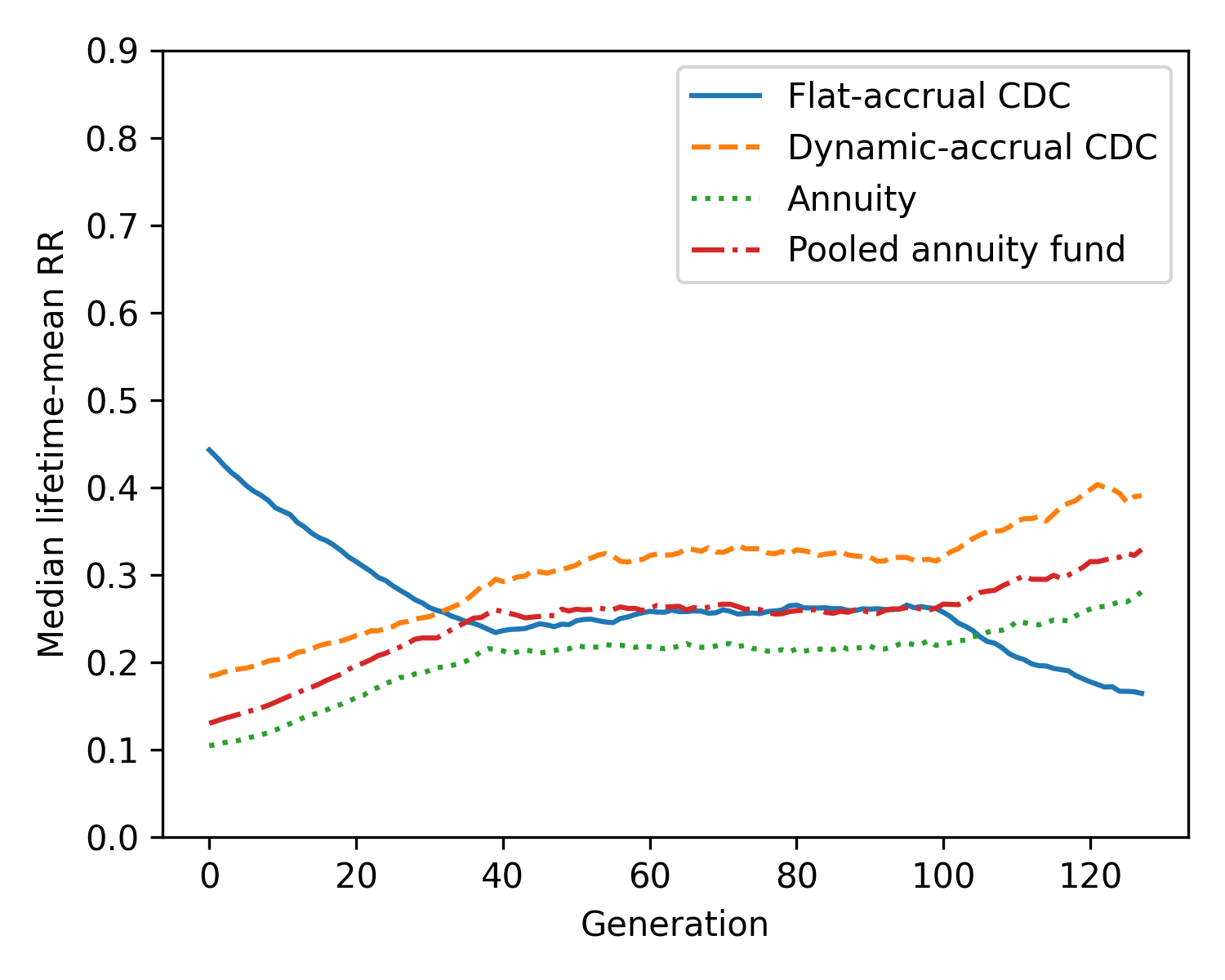}
\includegraphics[width=0.45\linewidth]{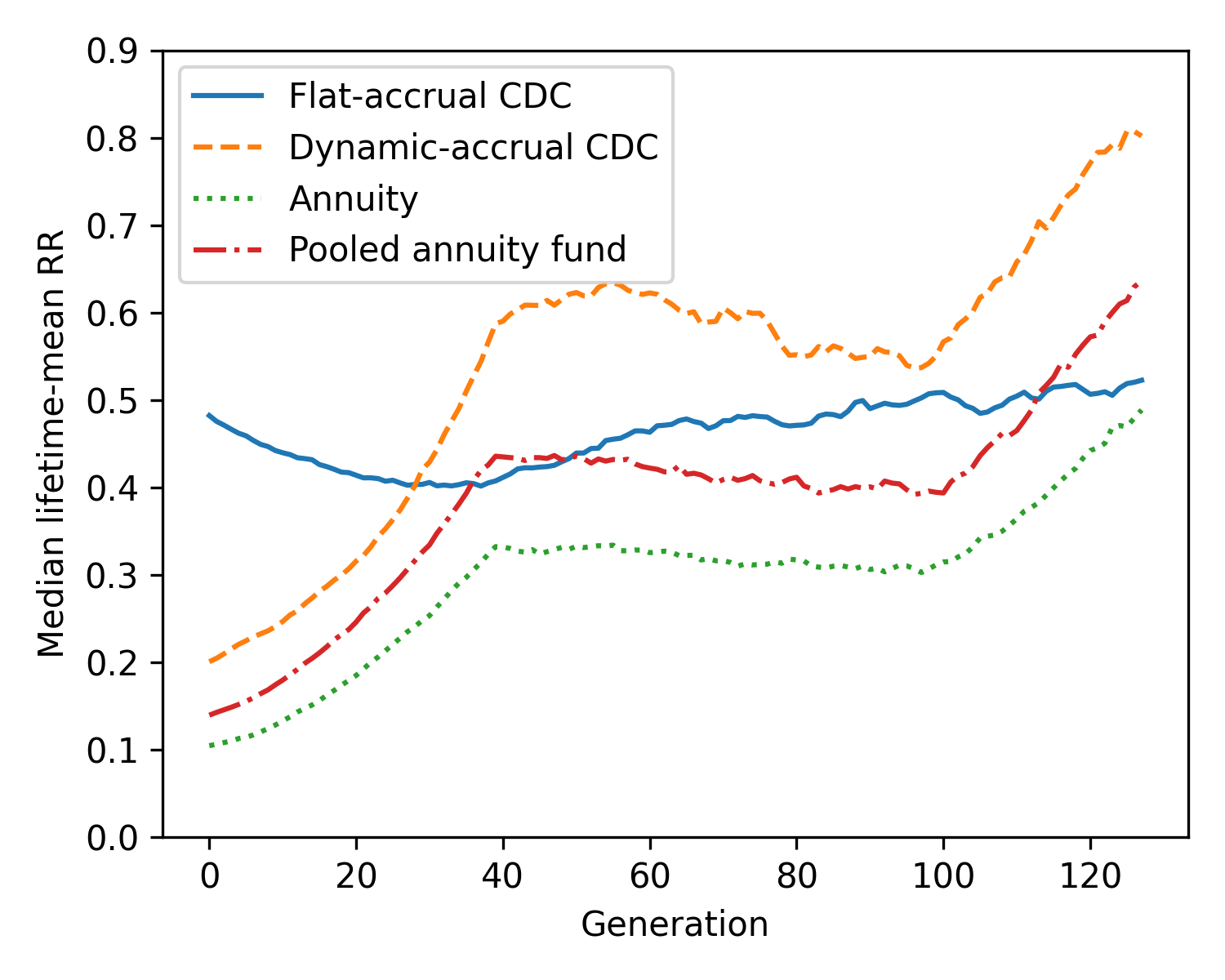}
\caption{Median lifetime-mean replacement ratio by generation when investment returns are not as expected. The chart on the left shows the case when stock returns are 1\% lower than expected, the chart on the right shows the case when stock returns are 1\% greater than expected.}
\label{fig:investmentReturnsShifted}
\end{centering}
\end{figure}

In Figure \ref{fig:investmentReturnsShifted}, we show the effect
of running our simulations when the scenario generator is adjusted to generate data with stock returns on average 1\% greater or less than expected, but where the fund is managed as before.
In this case, one sees that if stock returns are greater than expected, later generations in the flat-accrual scheme benefit more than earlier generations. Similarly, if stock returns are lower than expected, later generations experience a greater penalty. The size of these effects are enough to significantly change which generation benefits the most from the scheme.

In the dynamic-accrual scheme, the fund responds in a way that appears broadly consistent with actuarial fairness: the longer an individual invests their funds, the more they benefit from unexpectedly high stock prices. Generations in the middle of the scheme all receive very similar pensions even when the investment returns are not as expected.

\section{Conclusions}

The flat-accrual CDC fund seeks to mirror the high levels of intergenerational cross-subsidy seen in DB schemes, but we find this subsidy is approximately four times greater in the flat-accrual CDC scheme. The flat-accrual scheme smooths income in retirement. It can lead to a higher average income in retirement than DC+annuity, but this will not necessarily occur due to infinite-horizon effects. While such effects also exist in DB schemes, they will be small.
In order to reproduce similar levels of cross-subsidy to the DB schemes they replace, a better alternative is to offer varying benefit entitlements by age. This reduces the infinite-horizon issues experienced in these funds and result in higher average retirement incomes.

The standard dynamic-accrual fund does not target a specific benefit level in the same way as a flat-accrual scheme, and this makes it harder to control factors such as the frequency of benefit cuts or the median annual rate of pension increase. The dynamic-accrual CDC design results in much lower intergenerational cross-subsidies than a flat-accrual scheme, though they are still present. Much of the level of cross-subsidies in dynamic-accrual schemes can be predicted using a linear model in $h$ and age, with a path-dependent residual.
Using this linear model to match benefits and
contributions reduces the pricing errors. Because the dynamic-accrual scheme has much lower intergenerational cross-subsidies than a flat-accrual scheme, the infinite-horizon effects are small. This explains why the average retirement income is higher in dynamic-accrual schemes.

For both types of scheme it is important to distinguish between nominal benefits and projected benefits.
Nominal benefits should not be presented to members as though they were projected benefits.

This paper only considers shared-indexation CDC schemes. The pooled annuity fund gives an example of a collective scheme which provides a much stronger fairness guarantees than a shared-indexation fund: for all generations, the value of the cashflows received is equal to the value of the cashflows invested. In future work we will investigate the extent to which the benefits of CDC funds can be reproduced while maintaining similarly strong guarantees of fairness.

\section*{Acknowledgments}

We would like to thank the PPI and the members of our project advisory board for their feedback on earlier drafts of this work. This research is funded by Nuffield grant FR-000024058. High Performance Computing was completed using \cite{create}.

\medskip

\bibliography{cdc}

\begin{thebibliography}{10}

\bibitem{armstrongMaffraPennanen}
S.~Alvares~Maffra, J.~Armstrong, and T.~Pennanen.
\newblock Stochastic modeling of assets and liabilities with mortality risk.
\newblock {\em Scandinavian Actuarial Journal}, 2021(8):695--725, 2021.

\bibitem{aon}
AON.
\newblock The case for collective {DC}, 2013.

\bibitem{irri}
D.~Blake.
\newblock Independent review of retirement income: Report, 2016.

\bibitem{bonenkamp}
J.~Bonenkamp and E.~Westerhout.
\newblock Intergenerational risk sharing and endogenous labour supply within funded pension schemes.
\newblock {\em Economica}, 81(323):566--592, 2014.

\bibitem{create}
CREATE.
\newblock King's {C}omputational {R}esearch, {E}ngineering and {T}echnology {E}nvironment ({CREATE}), 2024.

\bibitem{cui}
J.~Cui, F.~De~Jong, and E.~Ponds.
\newblock Intergenerational risk sharing within funded pension schemes.
\newblock {\em Journal of Pension Economics \& Finance}, 10(1):1--29, 2011.

\bibitem{governmentActuaries}
The Government~Actuary’s Department.
\newblock Modelling collective defined contribution schemes.

\bibitem{catherineDonnelly}
C.~Donnelly.
\newblock Inter-generational crosssubsidies in the {UK}’s first {CDC} pension scheme.

\bibitem{obr24}
Office for Budget~Responsibility.
\newblock Fiscal risks and sustainability report – {September} 2024 – charts and tables: supplementary tables, 2024.

\bibitem{gollier}
C.~Gollier.
\newblock Intergenerational risk-sharing and risk-taking of a pension fund.
\newblock {\em Journal of Public Economics}, 92(5-6):1463--1485, 2008.

\bibitem{cdcRegulations}
UK~Government.
\newblock {The Occupational Pension Schemes (Collective Money Purchase Schemes) Regulations 2022, UK Statutory Instruments} 2022 no. 255, 2022.

\bibitem{haan}
J.~Haan, Z.~Lekniute, and E.H.M. Ponds.
\newblock Pension contracts and risk sharing--a level playing field comparison.
\newblock {\em Available at SSRN 2741542}, 2015.

\bibitem{equalityActPensions}
UK~Statutory Instrument.
\newblock {The Equality Act (Age Exceptions for Pension Schemes) Order 2010 (SI 2010/2133)}, 2010.

\bibitem{mcinally}
K.~McInally.
\newblock Multi-employer {CDC} schemes: can they be fair?

\bibitem{owadally}
I.~Owadally, R.~Ram, and L.~Regis.
\newblock An analysis of the {Dutch-style} pension plans proposed by {UK} policy-makers.
\newblock {\em Journal of Social Policy}, 51(2):325--345, 2022.

\bibitem{royalMailHandbook}
Royal Mail Collective~Pension Plan.
\newblock {The Collective Plan's Handbook}, 2024.

\bibitem{popat}
S.~Popat, C.~Curry, T.~Pike, and C.~Ellis.
\newblock Modelling collective defined contribution schemes, 2015.

\bibitem{tprWhatIsCDC}
The~Pension Regulator.
\newblock Collective defined contribution ({CDC}).

\bibitem{royalMailTrustDeeds}
{Slaughter and May}.
\newblock {Royal Mail Collective Pension Plan, Definitive Trust Deed and Rules}, 2024.

\bibitem{abi}
S.~Taylor and R.~Ward.
\newblock Collective {DC} modelling (prepared for the {ABI}), 2023.

\bibitem{upton}
J.~Upton.
\newblock Quantifying multi-employer and single employer {CDC} outcomes, 2024.

\bibitem{wtw}
Willis~Towers Watson.
\newblock How {CDC} pension levels compare with other types of schemes, 2020.

\bibitem{wilkinson}
Lauren Wilkinson.
\newblock How might {CDC} develop in the {UK}?, 2008.

\end{thebibliography}

\appendix

\section{Proofs}
\label{appendix:proofs}

\begin{proof}[Proof of Theorem \ref{thm:analyticFormula}]
Write $\ddot{a}_{\NRA}$ for the expected value at age $\NRA$ of paying in advance a single-life annuity, increasing at the annual effective rate $\constantindexrate$, from age $\NRA$ until death and discounted at the annual effective rate $R$, i.e.
\[
\ddot{a}_{\NRA} = \sum_{k=0}^{\infty} p(\NRA,k) \left(\frac{1+\constantindexrate}{1+R} \right)^{k}.
\]
Consider the flat-accrual CDC scheme.  As the scheme has a stable population, without loss of generality, we can do our calculations as if there is only one member per generation before retirement and fractional members thereafter.

First calculate the accrual rate $\alpha$.  The additional liability of the fund at time $t$, due to the new benefits accrued by $n$ generations of contributing members, is
\[
\beta^{-1} (1+g)^t \, \ax{\angl{n}}^{R,\constantindexrate} \, \ddot{a}_{\NRA}.
\]
As the additional liability should match the additional contribution $n \alpha S_t$ paid by these $n$ contributing generations, we compute the contribution rate to be
\begin{equation}
\alpha = (n\beta)^{-1} \ax{\angl{n}}^{R,\constantindexrate} \, \ddot{a}_{\NRA},
\label{eq:dbSteadyState}
\end{equation}
independent of time $t$.  The instantaneous percentage gain of an individual who is $k$ years from retirement is
\[
%\left(\frac{A B \netinterest^k }{C}-1\right) \times 100\%.
\left( \frac{\beta^{-1} S_t \left( \frac{1+\constantindexrate}{1+R} \right)^{k} \, \ddot{a}_{\NRA}}{\alpha S_t} - 1 \right) \times 100\%.
\]
Substituting in the last expression for $\alpha$ from \eqref{eq:dbSteadyState} gives equation \eqref{eq:dbPNL}.

Now turn to calculating the Benefit Ratio.  Consider an individual who joins the flat-accrual CDC scheme at integer time $t_0$.  The total annual benefit accrued by them at the start of retirement is
\[
\sum_{k=0}^{n-1} \beta^{-1} S_{t_0+k} (1+\constantindexrate)^{n-k} = \beta^{-1} (1+g)^{t_0} \, (1+\constantindexrate)^{n} \,\sum_{k=0}^{n-1} \left( \frac{1+g}{1+\constantindexrate} \right)^{k}.
\]
Substituting for $\beta^{-1}$ using equation \eqref{eq:dbSteadyState}, we find that their first annual CDC pension payment is
\begin{equation}
B^{\cum,\CDC}:=
\frac{\alpha \, (1+g)^{t_0+n}}{\ddot{a}_{\NRA}}  \, \frac{n \, \ax{\angl{n}}^{g,\constantindexrate}}{\ax{\angl{n}}^{R,\constantindexrate}}.
\label{eq:dbFirstPensionPayment}
\end{equation}

Suppose instead the individual invests their contributions into a individual DC scheme.  Then the accumulation of their contributions at retirement is
\[
\sum_{k=0}^{n-1} \alpha S_{t_0+k} (1+R)^{n-k} = \alpha \, (1+g)^{t_0} \, (1+R)^{n} \, \sum_{k=0}^{n-1} \left( \frac{1+g}{1+R} \right)^{k}.
\]

Purchasing an index-linked annuity at retirement means that the individual DC member's first annual pension payment is 
\[
B^{\cum,\DC}:= \frac{\alpha \, (1+g)^{t_0} \, (1+R)^{n} \, \ax**{\angl{n}}^{R,g}}{\ddot{a}_{\NRA}}.
\]
As $\textrm{BR} = B^{\cum,\CDC} / B^{\cum,\DC}$, we obtain equation \eqref{eq:dbRatio}.

Finally, it follows from $\lim_{R \to g} \ax**{\angl{n}}^{R,g} = n$, $\lim_{R \to \constantindexrate} \ax{\angl{n}}^{R,\constantindexrate} = n$ and $\lim_{g \to \constantindexrate} \ax{\angl{n}}^{g,\constantindexrate} = n$, that $\textrm{BR} \to 1$ as $R \to \constantindexrate$ and $R \to g$. 
\end{proof}

\section{Economic scenario generator}
\label{sec:esg}

To generate economic scenarios we initially used the scenario generator described in \cite{armstrongMaffraPennanen}. However, we then found that
there was a surprisingly high probability of benefit cuts
at start-up of a CDC scheme. This could be traced to the handling of CPI inflation in the model of \cite{armstrongMaffraPennanen}, which allows deflation to occur with a relatively high probability. See Figure 8 of \cite{armstrongMaffraPennanen}.

To address this issue we performed an additional transformation of the variable used to model CPI, similar to the transformations used for the yields to maturity in the original model. The original paper defined a variable
\[
I_t = \log\left( \frac{\CPI_t}{\CPI_{t-1}} \right)
\]
and used this as one of the vector components in a vector auto-regressive model. We defined a transformed variable
\[
\tilde{I}_t = \log\left( I_t + \mu \right)
\]
where $\mu=0.01$ and used this as a vector component of the model in place of $I_t$.

We then calibrated the model using the same data set as \cite{armstrongMaffraPennanen} and applied the views described in Table \ref{table:riskFactors} to determine the long-term medians of the risk factors.

\section{Testing}
\label{sec:testing}

The results in this paper depend heavily upon simulations whose results are hard to predict. In particular, Figure \ref{fig:year50} is difficult to validate as we have little intuition as to what to expect. However, there are some meaningful tests we can perform to validate our simulations.
\begin{enumerate}[(i)]
\item If we perform a simulation using an economic model where all risk-factors are constant, a flat-accrual CDC fund exactly hits the target level of indexation.
\item The total area under the graph in Figure \ref{fig:pnlSingleQMeasure} is close to zero.
\item The plots for the pooled annuity fund scheme and the DC + Annuity scheme are as one would expect.
\item Plots generated using the ESG and using a Black--Scholes model with the same parameterisation are close.
\item Our plots of the trajectories for a flat-accrual CDC scheme demonstrate the properties such as smoothing one expects of a CDC fund. Our plots of the intergenerational cross-subsidies occurring in a flat-accrual CDC scheme have the qualitative properties one expects and closely match the predictions of Theorem \ref{thm:analyticFormula}.
\item The funds remaining at the end of a simulation are approximately zero.
\item Our Python codebase was developed independently from an R
codebase for modelling flat-accrual CDC funds created for \cite{catherineDonnelly}. It reproduces the results of that paper.
\item The results for the steady-state of the flat-accrual CDC simulation and the results for the annuity simulation have been
checked using a spreadsheet which repeats
the calculations, but in a recursive formulation. These also match the analytic formulae in Theorem \ref{thm:analyticFormula}.
\item When a pricing chart similar to Figure \ref{fig:pnlSingleByAge} is created, but with all investment in the risky asset, the results match
the analytic formulae in Theorem \ref{thm:analyticFormula}.
\end{enumerate}
We have, as far as possible, written our code in a modular fashion so that each of these tests increases confidence in the codebase overall.

\section{Relationship to regulations}
\label{sec:regulations}

We briefly describe how our equations correspond to the UK CDC regulations \cite{cdcRegulations}.
Equation \eqref{eq:cdcDefining} captures clause 17.4 (c), that benefit adjustments ``must be applied to all the members of the scheme without variation''.
Equation \eqref{eq:actuarialvaln} is essentially a mathematical formulation of
the remainder of section 17 (1-6). These rules specify the general principles that: the assets should match the liabilities (17.4(e));
that central estimates should be used throughout the calculations (17.3);
that the same increase should be applied each year (17.4 (a)) and should
include the projected change in inflation (17.5 (b)).

The regulators' interpretation of the regulations, \cite{tprWhatIsCDC},  clarifies that ``All qualifying benefits in a section must: be provided by reference to the same rate or amount for all members; have the same rate or amount of contributions paid by all members; have the same rate or amount of contributions paid by the employers; have the same normal pension age for all members''. These requirements are captured by equations \eqref{eq:defBeta}, \eqref{eq:defAlpha} and \eqref{eq:ageAndXi}.

Sections (8-13) of the regulations allow the scheme to make 
benefit cuts as necessary. There are complex rules on how benefit cuts should be made if they are spread over a number of years. We have for simplicity assumed benefit cuts and bonuses are made immediately.

The required relationship between the contribution rate and the level of benefits is hard to identify from the regulations. However, \cite{tprWhatIsCDC} states that: ``A qualifying benefit is subject to periodic adjustment, in accordance with the scheme rules, to achieve a balance between the value of the available assets and the required amount (i,e. the amount expected to be required to provide benefits collectively)''. While most of this idea is already expressed by equation \eqref{eq:actuarialvaln}, considering the contribution rate is also an important part of how ``balance'' might be achieved ``collectively''. However, there is no mention in this statement of achieving any particular target rate.
Nevertheless, in our discussions with industry representatives there appears
to be a general understanding that flat-accrual schemes do attempt
to target a particular indexation level.
In addition, the first gateway test 11.4, requires that for the first ten years centrally estimated benefits at least keep up with CPI.

\clearpage 
\section{Performance of age-related accrual and statistically calibrated accrual}

\begin{figure}[!htbp]
\begin{centering}
\includegraphics[width=0.7\linewidth]{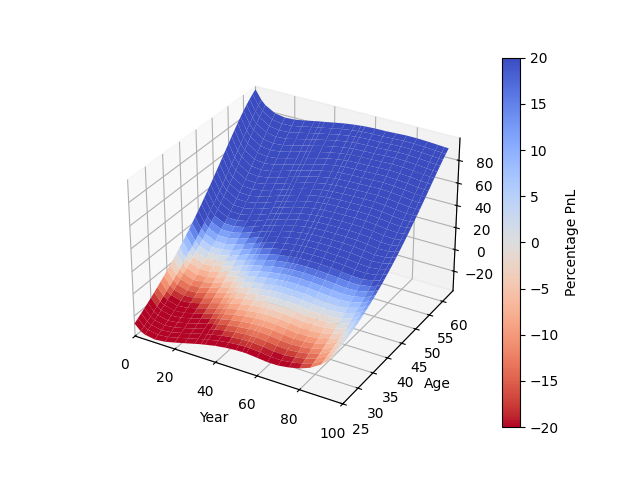}
\caption{Expected instantaneous profit for members in the age-based accrual CDC fund of Section \ref{sec:ageBased}.}
\label{fig:surfaceAge}
\end{centering}
\end{figure}

\begin{figure}[!htbp]
\begin{centering}
\includegraphics[width=0.7\linewidth]{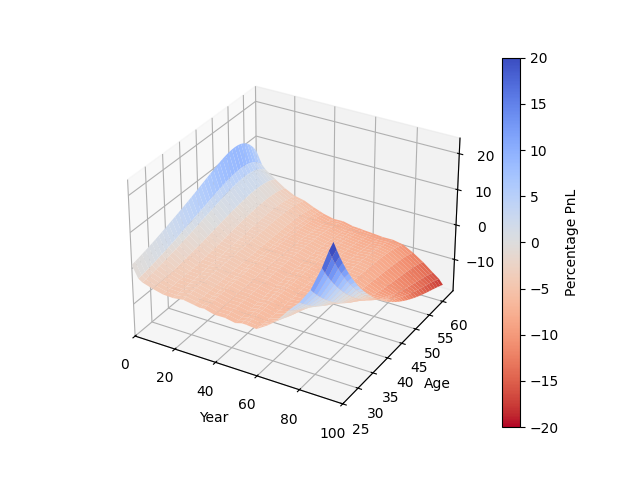}
\caption{Expected instantaneous profit for members of the statistically-calibrated dynamic-accrual CDC fund.}
\label{fig:surfaceStatistical}
\end{centering}
\end{figure}

\end{document}